\documentclass{aa}
\usepackage{amsmath}
\usepackage{graphicx}
\usepackage{dcolumn}
\usepackage{amssymb}
\usepackage{color}
\usepackage{url}
\usepackage{caption}
\usepackage{subcaption}
\usepackage{natbib}
\usepackage{hyperref}

\bibpunct{(}{)}{;}{a}{}{,}

\newcommand{\changeR}[1]{\textcolor{black}{#1}}

\newcommand{\changeRIV}[1]{\textcolor{black}{#1}}
\newcommand{\changeRV}[1]{\textcolor{black}{#1}}
 \newcommand{\changeRE}[1]{\textcolor{black}{#1}}
 \newcommand{\changeRR}[1]{\textcolor{black}{#1}}
 \newcommand{\changeRB}[1]{\textcolor{black}{#1}}
\newcommand{\ct}[1]{\textcolor{black}{#1}}
\newcommand{\kev}{~{\rm keV}}
\newcommand{\ergss}{~{\rm ergs/s}}

\begin{document}

\title {A thin diffuse component of the Galactic Ridge X-ray emission and
heating of the interstellar medium contributed by the radiation of Galactic X-ray binaries}

\author{Margherita Molaro\inst{\ref{inst1}}
\and
Rishi Khatri\inst{\ref{inst1}}
\and
Rashid A. Sunyaev\inst{\ref{inst1}{,}\ref{inst2}}
}

\institute{ Max Planck Institut f\"{u}r Astrophysik, Karl-Schwarzschild-Str. 1,
  85741, Garching, Germany \email{molaro@mpa-garching.mpg.de}\label{inst1}
\and
 Space Research Institute, Russian Academy of Sciences, Profsoyuznaya
 84/32, 117997 Moscow, Russia \label{inst2}}
\date{\today}
\authorrunning{Molaro,Khatri \& Sunyaev}
\titlerunning{Diffuse GRXE from Galactic XBs}

\abstract{We \ct{predict} a thin diffuse component \ct{of} the Galactic Ridge X-ray emission (GRXE) arising from the scattering of the radiation of bright X-ray binaries (XBs) by the interstellar medium. \ct{This scattered component has the same scale height as that of the gaseous disk ($\sim$ 80 pc) and is therefore thinner than the GRXE of stellar origin (scale height $\sim$ 130 pc).} The morphology of the scattered component is \ct{furthermore} expected to trace the clumpy molecular and HI clouds. We calculate this contribution to the GRXE from known Galactic XBs assuming that they are all persistent. The known XBs
sample is \ct{incomplete, however, because} it is \ct{flux limited} and spans the lifetime of X-ray astronomy ($\sim 50$ years)\ct{,} \ct{which is very short compared with} the characteristic time of 1000-10000 years that \ct{would have contributed} to the diffuse emission observed today due to time delays. We therefore also use a simulated sample of sources, to estimate the diffuse emission we should expect in an optimistic case assuming that the X-ray luminosity of our Galaxy is on average similar to that of other galaxies. In the calculations we also take into account the enhancement of the total scattering \ct{cross-section} due to coherence effects in the elastic scattering from multi-electron atoms and molecules.  This scattered emission can be distinguished from the contribution of {low
    X-ray luminosity 
    stars}  by the presence of narrow fluorescent K-$\alpha$ lines of Fe, Si\ct{,} and
    other abundant elements present in the interstellar medium {and by directly resolving the contribution of low
      X-ray luminosity stars.}  \changeRE{We
    find that within $1^{\circ}$ latitude of the Galactic plane the scattered emission contributes on average $10-30\%$ of
    the GRXE flux in the case of known sources and over $50\%$ in the case of simulated sources. In the latter case, the scattered component is found to even dominate the stellar emission in certain parts of the Galactic plane.}
    X-rays with energies $\gtrsim 1$ keV from XBs should also penetrate deep inside the HI and molecular clouds, \ct{where they are absorbed} and  heat  the interstellar
      medium.  We find that this heating rate dominates the heating by cosmic
      rays \changeRR{(assuming \ct{a} solar neighborhood energy density)} in a \ct{considerable} part of the Galaxy.
       }

\maketitle

\section{Introduction}
The unresolved or apparently diffuse X-ray sky is composed of an
   isotropic component originating in extragalactic sources such as active
   galactic nuclei, the  cosmic X-ray background (CXB) 
\citep{giac}, and a component confined mostly to the plane of our Galaxy 
known as the Galactic ridge X-ray emission (GRXE) \citep{worrall}. \changeRIV{The latter presents prominent emission lines at 6-7keV, which if
produced by highly \ct{ionized}
elements in a thermal plasma would imply a
temperature of 5-10keV
\citep{koyama1986,koyama1989,tanaka2002,muno2004,Revnivtsev2003}. A plasma with such
a high temperature could not be gravitationally bound and remain confined
to the Galactic plane
\citep{koyama1986,sunyaev1993,tanaka1999,tanaka2002}\ct{,} suggesting that the GRXE cannot be truly diffuse}. One solution that has
been explored is for
the GRXE to be composed of unresolved stellar sources with low X-ray luminosity. Evidence of this includes the
correlation of the emission with the stellar bulge and disk \citep{yako, Revnivtsev2003,Revnivtsev2006}, the resolution of over $80 \%$ of the emission in a small region of the sky \citep{Revnivtsev2009}, and the
similarity of the GRXE spectrum with the one produced by the superposition of the spectra of the low X-ray luminosity sources expected to contribute to
the emission 
\citep{ebi2001,ebi2005,Revnivtsev2006,morihana}. The nature of the GRXE is\ct{, however,} still not
conclusively resolved.

\citet{sunyaev1993} suggested that a part of the GRXE may arise from
diffuse gas and be composed of
radiation from the compact X-ray  sources scattered by the interstellar
medium \ct{(ISM). This} component of the GRXE would  
  also have \changeRR{a K-$\alpha$ line from neutral iron} in the \changeRIV{ISM}. \changeRE{
 \citet{Koyama1996}, using observations from ASCA, first resolved what was
initially thought to be a 6.7 keV  iron K-$\alpha$ line
from the Galactic center into a 7 keV and 6.7 keV lines from highly \ct{ionized}
H-like and He-like iron\ct{, respectively,} and
a 6.4 keV line from neutral iron, supporting  the
prediction of \citet{sunyaev1993}. This was later confirmed using Suzaku observations by \citet{Koyama2007} and \citet{Ebisawa2008}.}

 Our Galaxy is sprinkled with X-ray sources such as
\ct{low-mass and high-mass} binaries (LMXBs and HMXBs), supernova remnants,
pulsars\ct{,} and recurring novae. The X-ray radiation from these sources is
 scattered by free electrons, atomic and molecular hydrogen, helium\ct{,} and
heavier elements present in the interstellar medium. This reprocessed radiation
would be seen by us as a truly diffuse component emission, \ct{approximately tracing} the gas distribution in the Galaxy. This scattering of
X-rays by interstellar gas is different from the dust halos seen around
many X-ray sources
\citep{overbeck1965,trumper1973,rolf1983,predehl1995}. Scattering by dust
decreases very  sharply (approximately Gaussian) with the 
angle of scattering and is \ct{strong} only within $1^{\circ}$ of the
source \citep{MG1986,SmithDwek1998}. Scattering by atoms and molecules on the other hand is effective on
large angles and \changeRIV{would contribute up to $180^{\circ}$}.

\section{Expected contribution of scattered X-rays to GRXE} \label{contrXrays}
\par There have been a number of attempts to test the hypothesis that the
GRXE is composed of discrete {\ct{low-luminosity} stellar sources} \citep{ebi2001, muno2004,ebi2005,
  Revnivtsev2006,Revnivtsev2009,morihana} using Chandra and  \changeRV{X-ray
Multi-Mirror Mission (XMM-Newton)}
observatories.  \citet{Revnivtsev2009} succeeded in
resolving $\sim 80 \%$ of the  emission  in a narrow band around
6.7 keV into point sources in a
small region of the \ct{sky} of $16\times16$ arcmin centered at $l= 0.08^{\circ},b = -1.42^{\circ} $.  The
resolution of the GRXE into point sources in the Galactic plane has proven
more difficult because of the difficulty \changeRIV{in} separating out extra-galactic sources and insufficient sensitivity and angular resolution
to resolve the weak but densely populated Galactic sources. \changeRIV{\changeRV{Indirect}
evidence that weak X-ray sources such as 
cataclysmic variables and coronally active binaries form a significant component of GRXE came
from correlating the GRXE with the \ct{near-infrared} Galactic
emission, expected in this case since these sources
trace the old stellar population}. \cite{Revnivtsev2006} used data from
the \changeRV{Rossi X-Ray Timing Explorer - Proportional Counter Array (RXTE/PCA)} to show
that the 3-20keV component of the emission traces the stellar mass
distribution and is well fitted by models of the Galactic
stellar bar and disk from \ct{\citet{dwek}, \citet{bahcallsoneira}, \citet{kent}, \citet{freudenreich}, and \citet{denbinac}}. \changeRR{Using results from \cite{Sazonov2006}, \cite{Revnivtsev2006}} also found that the
 GRXE broad-band spectrum is very similar to the superposition of the spectra
of low-luminosity X-ray sources in the solar neighborhood. These results agree with the studies of the hard (17-60keV) component using the IBIS telescope \ct{onboard}
of \changeRV{INTErnational Gamma-Ray Astrophysics Laboratory (INTEGRAL)} by \citet{Krivonos2007}. {We refer to the contribution of low
luminosity (in X-rays) stellar sources to the GRXE as \emph{stellar GRXE}.}

This seems to suggest that the nature of the GRXE has \ct{finally been} solved. \changeRIV{\ct{However, there is still plenty of room}, especially in the central degree
of the Galactic plane, for a contribution from a diffuse scattered component
to the GRXE, {which we refer to as \emph{scattered GRXE}.}
 The scattered X-rays will follow the distribution of gas in the Galaxy
 with \ct{a} scale height of 80 pc ($0.6^{\circ}$ at a distance of 8 kpc), \ct{in contrast to} a stellar emission characterised by a scale height
 of {130 pc ($1^{\circ}$ at a distance of 8 kpc)} \citep{denbinac,denbin}}. We
 therefore do not expect \ct{a} significant contribution from the scattered emission
 in the field of view of \citet{Revnivtsev2009} at latitude $b=-1.42^{\circ}$\ct{,} but it is possible, in
an  optimistic model of Galactic X-ray luminosity, for the scattered
component to dominate the stellar component at $|b| \lesssim 0.5$. A
significant contribution from scattered diffuse X-rays is therefore not ruled out
by current observations\ct{,} and it is an interesting question to ask \changeRIV{what its contribution from scattering of luminous X-ray sources might be.}

\changeRIV{An important feature of
the scattered component is that} since the scattered X-rays \ct{travel} a longer
path to reach us from the original source, the X-rays seen by us today have
contribution from the cumulative past X-ray activity of the
Galaxy. {This implies that even sources currently undergoing
  a low quiescent phase may contribute to the average X-ray Galactic
  output had they experienced a higher level of activity in the past. An
  important example is the supermassive black hole Sgr A$^{*}$ situated at
  the Galactic center, \ct{which currently radiates} at least eight orders of
  magnitude below its Eddington limit with a quiescent $2-10$ keV luminosity of
  $\sim 10^{33-34}$ ergs/s \citep[e.g.][]{narayan1998,Baganoff2003}. 
\changeRV{The scattering of the hard \ct{X-ray} continuum
radiation of Sgr A$^{*}$ by an individual molecular cloud (Sgr B2) was first
observed by \citet{Revnivtsev2004}.}
Studies of the
  recent history of Sgr A$^{*}$'s activity through the reflected X-rays
  from the massive molecular clouds in its vicinity 
  \changeRR{(first proposed by \cite{sunyaev1993} and followed by \citet{Koyama1996,sgr3,Revnivtsev2004,sgr5,sgr6,sgr7,sgr8,sgr10,sgr11,sgr12,sgr13})} suggest that the source experienced much
  more luminous phases in the past than currently observed (see
  \citet{Ponti2013} for a review). \ct{In this paper} we ignore the
  contribution of this source and focus \ct{only} on the contribution of XBs (see Appendix \ref{app_sgrA} for a discussion of its possible
  contribution \ct{compared to that of} XBs).}

\citet{Grimm2002} calculated the  $2-10\kev$ luminosity of Galactic
binary sources averaged over the period 1996-2000. They give the average
luminosity of LMXBs to be $2-3\times 10^{39}$ ergs/s and \changeRIV{that of HMXBs to be} 
$2-3\times 10^{38}$ergs/s from \changeRV{RXTE All-Sky Monitor (ASM)} data. {The
total luminosity of the GRXE is currently estimated at $\sim 1.39 \times 10^{38}$ ergs/s in the $3-20$ keV band \citep{Revnivtsev2006} and $3.7 \pm 0.2
\times 10^{37}$ ergs/s in the $17-60$ keV energy range
\citep{Krivonos2007} with a 
spectrum  in the 3-20keV band with \ct{a} photon index $\Gamma \sim
1.2$.} For \changeR{a Thomson scattering optical depth of $\sim
1\%$ in the ISM through the Galactic plane}, we would expect the contribution to the GRXE from
\changeR{the scattered radiation}
 to be $\sim 10^{37}\ergss$, or about $10\%$ of the observed
luminosity. {In contrast to the {stellar GRXE}\ct{,} this $\sim 10\%$ scattered
contribution to the luminosity is concentrated in a much narrower region in
the Galactic plane, giving \ct{a} higher relative contribution to the local
surface brightness.}
\changeR{The sample of bright X-ray binaries we \changeRIV{use in the calculations}} is \ct{incomplete, however,} as is clear from a significant deficit of sources \ct{farther out} than
the Galactic center in the \citet{Grimm2002} sample. There is also significant uncertainty in the distances
of most of the sources. Moreover, we expect many \ct{high-luminosity} transients in the Galaxy with luminosities $\gtrsim 10^{38}\ergss$\ct{,}
 which are active for only few $\%$ of the time. Such transients would
contribute to the observed scattered flux today but would not be \changeRIV{present} in
the \ct{current} sample if they were inactive at the time of observations. 

A 
better estimate of the average X-ray luminosity of the Milky Way is possible by
considering the relation of \ct{LMXB and HMXB} populations with other Galactic
properties such as the stellar mass \ct{($M_{\star}$)} and the star formation
rate \ct{(SFR)}. The long evolutionary timescales of the \ct{low-mass} donor star imply that LMXBs are expected to follow the older stellar population\ct{,} while the short
lifetime of the \ct{high-mass} donors in HMXBs means that they should trace the
younger stellar population of the Galaxy. \citet{Grimm2002} pointed out that
we should therefore expect a relationship between the 
LMXB luminosity and \ct{$M_{\star}$} on the one hand and between the HMXB
luminosity and the \ct{SFR} of a galaxy on the other.
 \citet{grimm} and \citet{gilfanov2004} extended these studies
to \changeRR{other} galaxies which were followed by
\ct{\citet{rcs2003}, \citet{chp2004}, \citet{pr2007}, \citet{lehmer2010} and \citet{mineo2012}} and up to redshift
$z\sim 1.3$ using 66 galaxies in
\citet{Mineo2013}. The study of these relations in \changeRR{other galaxies}
avoids issues of incompleteness and distance uncertainties\ct{,} which are
important in the study of local XBs. \changeRR{These studies assume a linear relationship between the total 2-10 keV luminosity $L_X^{2-10{\rm keV}}$ of LMXBs and HMXBs and the  total \ct{$M_{\star}$} and \ct{SFR} of the galaxy respectively, \ct{that is} 
\changeRB{
\begin{equation}
L_X^{2-10{\rm keV}}\sim \alpha \times M_{\star}+ \beta \times SFR.\label{lehmer}
\end{equation} 
}
Estimated values for the proportionality factors are $\alpha \sim 8 \times 10^{28}$ erg/(s M$_\sun$) \citep{gilfanov2004} and $\beta\sim 2.2-2.6\times 10^{39}$ erg/(s M$_\sun$ yr$^{-1}$) \citep{grimm,Shtykovskiy2005,lehmer2010,mineo2012}.}

{The Milky Way is
therefore \ct{subluminous} in LMXBs by a factor
of $\sim 2$, consistent with the original results of 
\citet{gilfanov2004}, and by a factor of $\sim 30$ in \ct{luminosity of HMXBs} compared \ct{with} the other
galaxies.  The large discrepancy for the HMXBs may be explained if we take
into account that there is a delay of $10^6-10^7$ years between the time of
starburst and appearance of HMXBs \citep{gilfanov2004,sg2007}, corresponding to the
evolutionary timescale of the donor star, \changeRIV{which implies that} the relevant SFR is not the
present SFR\ct{,} but the SFR \ct{of} a million years ago. However, if we assume,
optimistically for our purpose, that the
discrepancy is \ct{caused by} the incompleteness of the observed sample and
 the short lifespan of X-ray astronomy \ct{of about} $50$ years, it is possible that
 we happen to live \changeRIV{in a time where} the observed X-ray luminosity of our Galaxy is
 below average and \changeRIV{that} if averaged over a \ct{time-scale} of $10^3-10^4$ years our
Galaxy \changeRIV{would} turn out to have \changeRIV{an average luminosity
  consistent with that observed in other galaxies}. \par \changeRR{The X-ray
  luminosity of most galaxies including our own is dominated by a few
  extremely bright sources\ct{,} and different sources may be the main
  contributors at different times.} \changeRR{Existence in the past of a
  few ultra-luminous X-ray sources (ULXs) with luminosities of
  $10^{39}-10^{40}$ erg/s, as observed in some \ct{star-forming} galaxies, would
  significantly affect the scattered component\ct{,} making it significantly
  brighter (Khatri \& Sunyaev in prep.).} \changeRR{Therefore such a
  \ct{strong} temporal fluctuation is plausible. This is also supported by the \ct{scatter of} more than an order of magnitude in the luminosity-SFR relations
  \citep{Mineo2013}} and \ct{by the} broad probability distributions for the
    X-ray luminosity at low star formation rates \citep{ggs2004}. \par \ct{For
 a Milky Way stellar mass of $6\times 10^{10} M_{\sun}$ \citep{massmodel}
 and star formation rate of $\sim 1M_{\odot}/{\rm yr}$ \citep[see for
 example][]{sfr2010} we should expect, under this assumption and based on above studies}, an average
 LMXB luminosity of \ct{$5\times 10^{39}\ergss$} and HMXB luminosity of
 \ct{$2-3\times 10^{39}\ergss$}. These estimates are \ct{higher by} a factor of $\sim 2$ (for LMXBs) and a factor of \ct{$\sim 10$} (for HMXBs) than the values\ct{, respectively,} quoted by \cite{Grimm2002}. \par In this optimistic case, we expect that in the
 Galactic plane almost $50\%$ of  the contribution to the GRXE might come
 from the scattered radiation. The observation of this thin component of the
GRXE \changeRIV{would therefore allow \ct{one} to test this hypothesis.}

{ The true luminosity probably lies somewhere between these
two extremes. We consider both a Monte Carlo population of XB sources based
on \changeRB{Eq. \ref{lehmer}}  to study the most luminous optimistic case,} and a \ct{catalog} of XB sources with known distances and spectral parameters from RXTE/ASM and INTEGRAL surveys to place a lower bound on the contribution of this effect.  

We also take into account the coherent or Rayleigh scattering of the X-ray
radiation on multi-electron species such as H2, He\ct{,} and heavy
elements. Rayleigh scattering significantly increases the GRXE flux
especially when the scattering angles are small. This 
applies to  the  scattering from the HMXBs\ct{,} which lie mostly in the plane of
the Galaxy. {Details of the effect of Rayleigh scattering are discussed in
Appendix \ref{appa}.} Molecular
clouds are thus very important and would \ct{be detected} as regions of high X-ray
intensity in the GRXE. 
{The compact nature of molecular clouds \ct{additionally}
boosts the scattered intensity, with molecular clouds \ct{evident} as
luminous features in the maps.} These features are clearly seen in our maps of
diffuse scattered component of \ct{the} GRXE  {using the  data from Boston
University-Five College Radio Astronomy Observatory Galactic Ring Survey  \citep{grs2006}.}

{The stellar component of the GRXE will also be scattered in the
  ISM and will form part of the scattered GRXE. We expect this component to
be \ct{subdominant} and \ct{to }contribute only $\sim \tau\sim 1\%$ to the GRXE. For
completeness we explicitly calculate and include this component. The
scattering of isotropic CXB\ct{,} on the other hand\ct{, cannot be observed} if we ignore
the small change in the spectrum of X-rays due to loss of energy to electron
recoil in Compton scattering and does not need to be taken into account. We
therefore leave the detailed calculation of scattering of CXB on cold
atomic and molecular gas for a future publication (Molaro et al. in prep.).}

\section{Galaxy model}
{The gas in the interstellar medium in the disk of the Galaxy has a
  multiphase character \citep{MO1977} loosely classified into  atomic
  (neutral or ionized) gas
  and molecular gas, with the atomic gas further  classified into cold
  neutral medium (CNM), warm neutral medium (WNM), warm ionized medium
  (WIM) and  hot ionized medium (HIM) \citep[see][for a recent
  review]{Kalberla}\ct{,} while a significant part of the molecular gas is concentrated in
  giant molecular clouds (GMCs). Most of the mass in the neutral atomic gas is
  confined to the disk of the Galaxy and is \ct{clearly} traced by HI 21 cm emission.
The  maps of the HI 21 cm emission in the Galaxy 
\citep{lab2005,gass2009}  can only
provide the total column density in parts of the sky, for example \ct{toward} the Galactic
center, where the velocity information cannot be used to infer the
\ct{three}-dimensional gas distribution. The CNM, which contains most of the HI, is clumped into clouds \ct{that} have
a volume filling factor of $\sim 10-20\%$ in the plane of the Galaxy in the
solar neighborhood \citep{Kalberla}. Any given line of sight should
therefore intersect many clouds\ct{,} making the column density much
smoother. A smooth  disk model therefore suffices for our calculation
and we use the models of 
\citet{denbinac} and \citet{denbin} for the HI gas distribution.} 

{ The situation is slightly
different for molecular clouds. The GMCs, mainly concentrated near the
\ct{Galactic center} and in the spiral arms, have sizes of $\sim 10-100$ pc
and average densities \ct{on the order} of
 $10^2-10^3~{\rm cm^{-3}}$ with mean separation between the clouds 10 times
the mean size \citep{Blitz1993,McKee2007}\ct{,} giving a \ct{volume-filling} factor in the plane of the Galaxy of
\changeRV{$\sim 0.1\%$}. They
are therefore compact enough to \ct{significantly} influence the morphology of the X-ray signal. The \ct{full-sky} maps of the molecular gas distribution  using
CO lines \citep{dame2001} do not  resolve and provide distances to
individual clouds \citep{dame2001}, although \ct{high-resolution} studies of many important
GMCs are available.}  Recently\ct{, high-resolution} data with kinematic
distances \ct{have} been made available from the Boston
University-Five College Radio Astronomy Observatory Galactic Ring Survey (GRS)
\citep{grs2006,GRS2009,romdutw,romdu}\ct{,} which covers the very important molecular ring structure in the
inner Milky Way. We use \ct{these} data to predict the X-ray scattering signal in
the longitude range $18^{\circ}\le \ell \le 55.7^{\circ} $ and latitude
range $-1^{\circ}\le b \le 1^{\circ}$. 
Although GRS covers only a $10\%$ of the Galactic plane, the
  results are easily extendable to \ct{and are }valid for the full sky, and in particular clearly
  show the spatial morphology and features in the GRXE expected
  {from the molecular clouds.}

\subsection{ISM distribution}\label{sec:di}
\begin{figure}
\includegraphics[trim = 70 0 70 10,clip,height = 10cm, width = 10 cm]{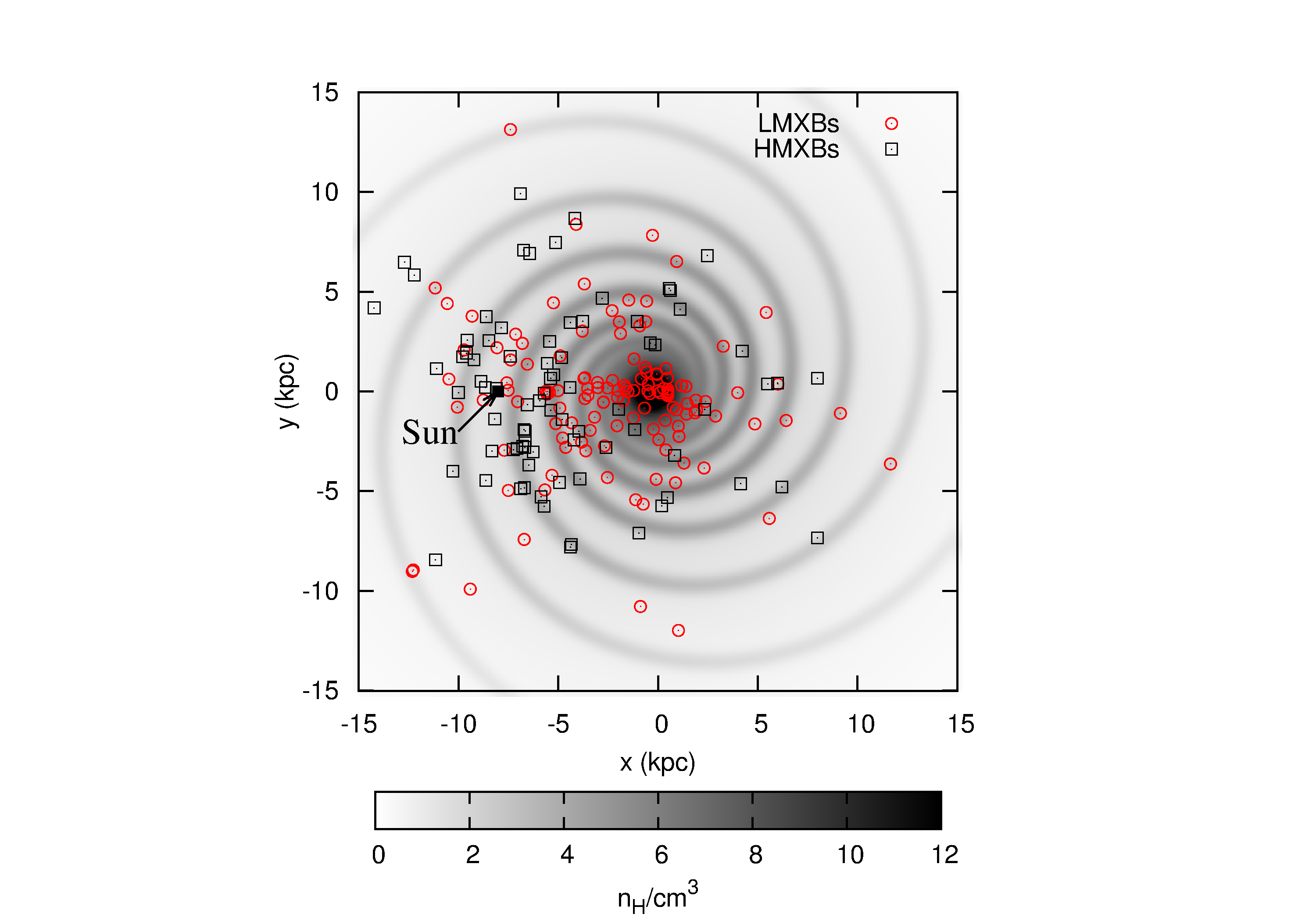}
\caption{Projected distribution on the Galactic plane of {
   persistent X-ray sources observed during the \ct{past} 40 years used in
          our calculations, shown against the spiral structure of the Galaxy.}
        \label{sourcesxy}}
\end{figure}
The total ISM mass in both atomic and molecular form within \ct{a} Galactocentric radius $R \lesssim 20$ kpc of the Galaxy is
  approximately $9.5 \times 10^{9}$ M$_{\sun}$  \citep{Kalberla}, with
  hydrogen {(HI and H2)} accounting for $71\%$ of this mass.  The ratio of average atomic
  to molecular gas is 4.9 \changeRR{within $R \lesssim 20$ kpc} \citep{Draine}, giving a total HI
  mass of $5.6 \times 10^{9}$ M$_{\sun}$ and \ct{a} total H2 mass of $1.15 \times
  10^{9}$ $M_{\sun}$ within this radius. \ct{Because of} the incompleteness of the GRS, we use a smooth disk model for the
mass distribution of HI and H2 for full Galactic plane calculations with
the above normalizations and use H2 data from the \ct{GRS} \citep{grs2006,romdutw,
   romdu} for
calculations limited to the 
longitude range covered by the survey. The latter case is used to study the effect of clumpiness and
small \ct{volume-filling} factor of
the molecular clouds on the observed X-ray intensity morphology (see Section
 \ref{clouds}).

The average density distribution of the interstellar HI gas in the Galactic
disk can be described by an exponential disk  \citep{denbinac,denbin}.  The ISM mass density
  distribution at a Galactocentric distance $R$ and Galactic plane height
  $z$ used in this model is given by
\begin{equation}
\rho_{{\rm HI}}(R,z) \propto \frac{1}{2z_{\rm d}}{\rm exp}\left(-\frac{R_m}{R}-\frac{R}{R_{\rm d}} - \frac{|z|}{z_{\rm d}}\right), \label{Eq:diskmodel}
\end{equation}
with parameters $R_m = 4 $kpc, $R_d = 6.4$ kpc and $z_d = 80$ pc.  
\par 
The lack of HI gas in the central region of the plane, due to the central
depression of radius 4 kpc in the HI disk\ct{,} is compensated \ct{for} by the concentration in
this region of most of the H2 component\ct{,} which  gives  a significant fraction of the scattering at low longitude values.
The average H2 distribution in the Galaxy is given by \citet{Misiriotis2006} as
\begin{equation}
\rho_{{\rm H2}}(R,z) \propto {\rm exp}(-R/R_{{\rm H2}}-|z|/z_d) \label{Eq:diskmodelh2}
\end{equation}
with parameters $R_{{\rm H2}} = 2.57$ kpc and $z_d = 80$ pc as in
  Eq. \ref{Eq:diskmodel}.
\begin{figure*}[!ht]
\centering
     \includegraphics[trim = 10 70 10 70, height = 5cm, width = 16 cm]{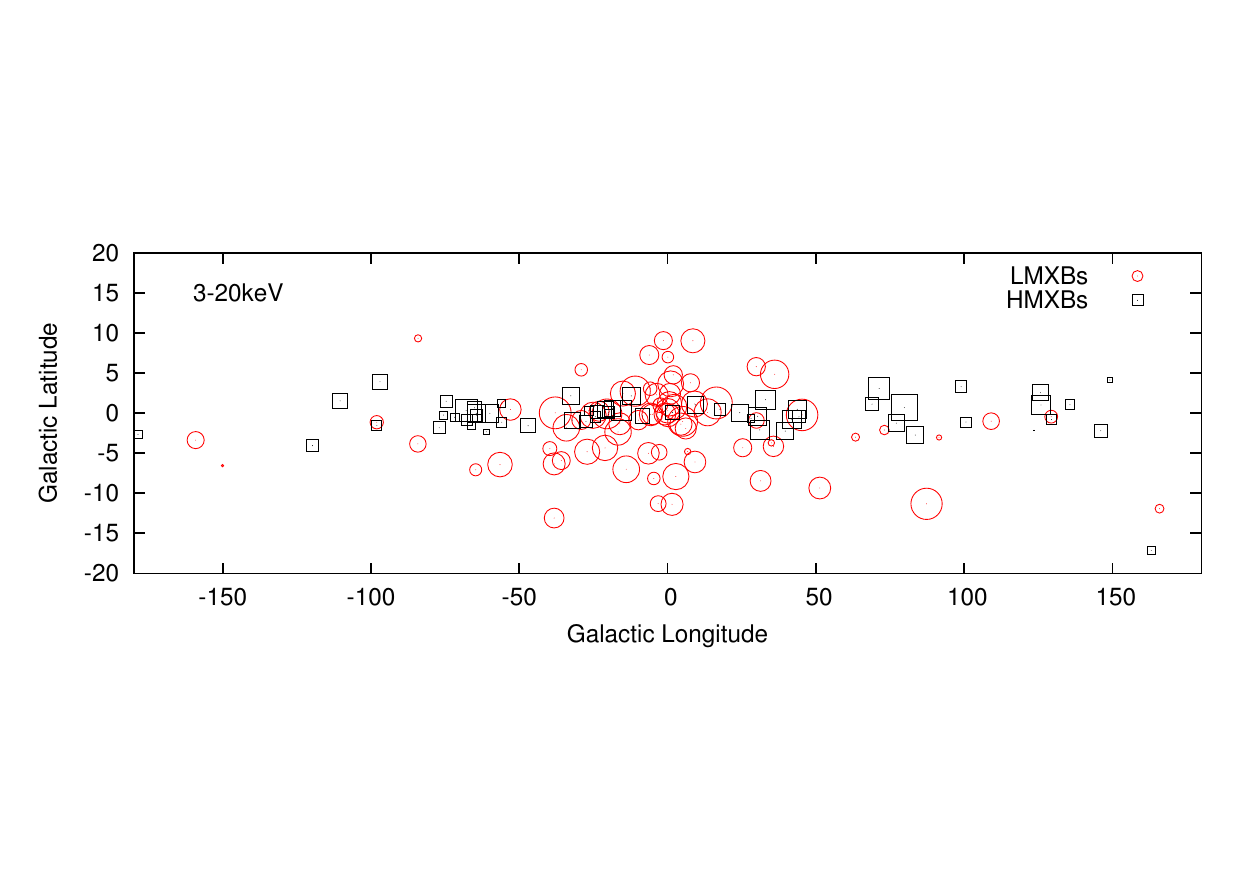} \\
     \includegraphics[trim = 10 70 10 60,height = 5cm, width = 16 cm]{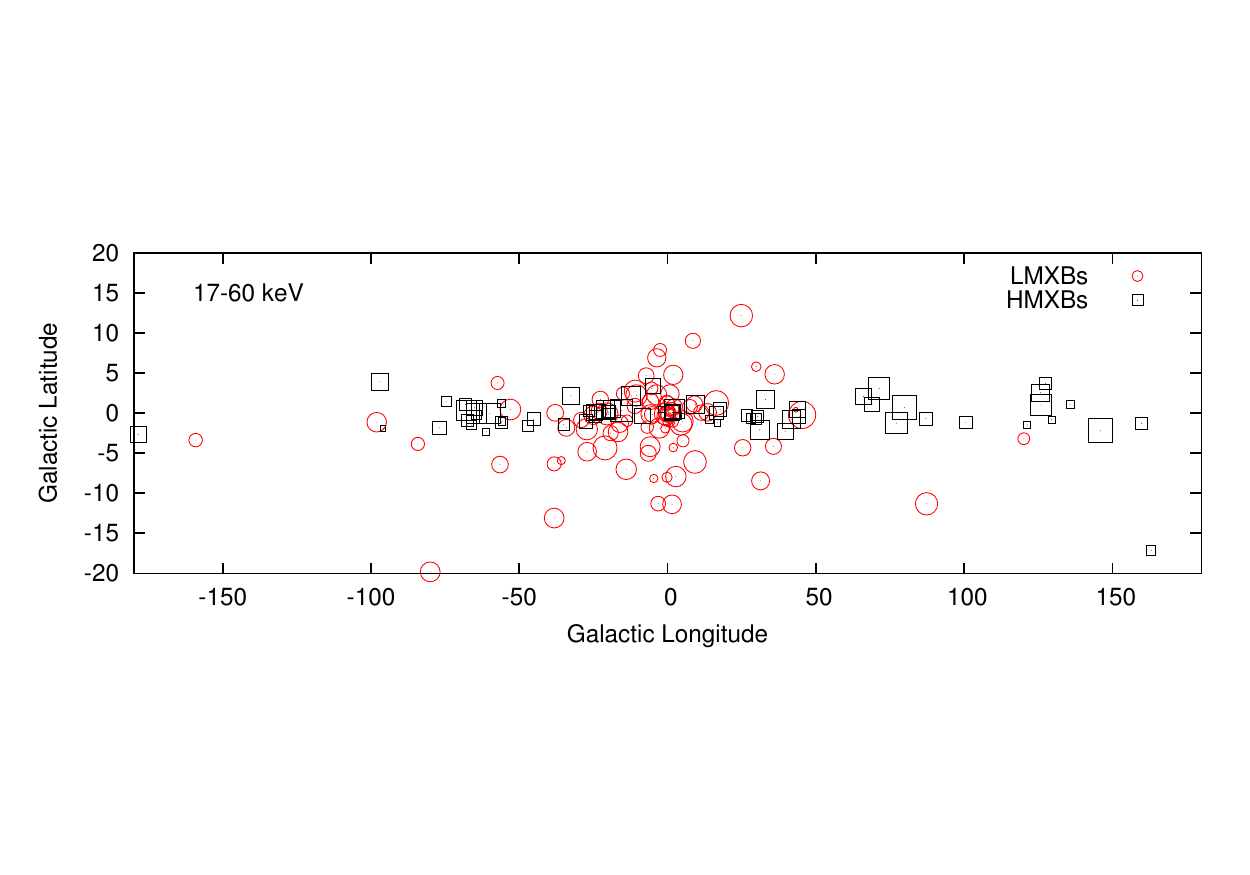} \\
        \caption{Distribution on the sky of observed LMXBs and HMXBs
          sources used in the calculations for energy ranges 3-20keV and
          17-60keV with \ct{the pointsize} proportional to their luminosity in the
          indicated energy range. \ct{A concentration} of HMXBs can be
            seen \changeRB{along the lines of sight} tangential to the spiral arms.}
        \label{sourceslb}
\end{figure*}
 
 We \ct{also} account for the concentration of matter in the Galactic disk into spiral arms.
This is particularly important \ct{because} HMXBs are also concentrated in
the  Galactic spiral arms: the combined effect of concentration of HMXBs as
well as atomic and molecular clouds \ct{in these structures} means that we \ct{expect} an enhancement of scattered X-ray intensity in the directions
tangential \ct{to the spiral arms}. 
 We describe the spiral structure of the Galaxy following the  prescription given in \citet{Vallee1995}, where the location of the density wave maxima of the $n$th spiral arm is given by
\begin{equation}
m\bigg[\theta - \theta_0 -{\rm ln}(R/R_0){\rm tan}(p)^{-1}\bigg] = (n-1)2\pi,
\end{equation}
where $m = 4$ is the total number of spiral arms, $p = 12^\circ$ (inward)
is the pitch angle and $R_0 = 2.5$ kpc. We have verified that the line of sights tangential to the spiral arms in the model are consistent with the location of the tangential directions based on different observations, as \ct{summarized} in \citet{Vallee2008}. 
We model the density around the spiral maxima as a Gaussian probability distribution with respect to the projected distance from the spiral arms' maxima on the Galactic plane $d$:
\begin{equation}
\rho_{Spiral} \propto \sum\limits_{n=1}^m {\rm exp}(-(d_n/w)^2),
\end{equation} 
where the typical width $w$ of the spiral arm is assumed to be 500 pc for the gas component.
The density of the wave is assumed to be three times higher than the
inter-arm density \citep{Levine}. The overall density distribution
including the spiral structure is therefore given by
\begin{equation}
\rho_{{\rm H}}(R,z,\theta) \propto  \left[1 + 3 \times \rho_{Spiral}(R,\theta)\right] \rho_{{\rm H}}(R,z)
\end{equation}
for both $\rho_{{\rm H}} = \rho_{{\rm HI}}$ and $\rho_{{\rm H}} = \rho_{{\rm H2}}$. The distribution of gas in the Galaxy in our model is
shown in Fig. \ref{sourcesxy}.
We assume constant abundances for  heavier elements throughout the Galaxy
with  abundances taken to be same as  the present-day solar photosphere from \citet{abundances}.

\subsection{Distribution of X-ray binaries in the Galaxy}
\textbf{Observed XBs distribution}
We use time-averaged X-ray flux measurements from different surveys to
compile a \ct{catalog} of X-ray \ct{binary} sources with known flux and position
in different energy bands (2-10keV and 17-60keV)\footnote{The full
  \ct{catalog} of sources used can be accessed from \url{http://www.mpa-garching.mpg.de/~molaro}}. 
References for distance estimates for each source are \ct{taken from} the SIMBAD database. 

For the energy range $17-60$keV we combine different INTEGRAL surveys \citep{Krivonos2007, Krivonos2012, Lutovinov2013} from which we select 86 \ct{LMXB and 70 HMXB} Galactic sources for a total 17-60keV luminosity of $1.97\times 10^{38}$erg/s and $6.2\times 10^{37}$erg/s\ct{,}  respectively, for which distance and \ct{best-fit} spectral parameters data are available. We also consider the contribution in this range of \ct{four} very luminous extra-galactic HMXB sources (SMC X-1, LMC X-1, LMC X-4, IGR J05007-7047), which is\ct{, however, close to negligible}. 
 \par For the lower energy range (2-10keV) we use the list of sources given in
 \citet{Grimm2002}\footnote{\changeRV{See \url{http://edoc.ub.uni-muenchen.de/1279/1/Grimm_Hans-Jakob.pdf} for the
 full list of sources and associated data.}}  for which distance and flux measurements from RXTE/ASM,
 which provides all-sky flux measurements averaged over $\sim 5$ \ct{yr}, are
 available. We also use the relation $L_{2-10{\rm keV}} \sim 0.5
 L_{17-60{\rm keV}}$ valid for NS binary sources \citep{Filippova2005} to
 infer the 2-10keV luminosity from the harder X-ray range \ct{catalog} for
 20 HMXB sources for which low-energy data are not available. In total\ct{,} the
 number of sources considered in this {energy range} is 61 for the HMXBs, for a
 total 2-10keV luminosity of $5.5 \times 10^{37}$ erg/s, and 81 for the
 LMXBs, with a total luminosity of $2.56\times 10^{39}$ erg/s.  The energy output of these sources in the 3-20 keV range is
 modeled using a model spectrum (described below  for the Monte Carlo
 sources) with a flat intensity spectrum and an exponential cutoff at 4.6
   keV and 20 keV for the LMXBs and HMXBs\ct{,} respectively. In the 17-60 keV range we use the \ct{best-fit} photon index from
 Integral surveys \citep{2007A&A...475..775K} and in some cases from \ct{the} literature on individual sources when the \changeRV{former} is not available. In
 general\ct{,} we expect variations in the photon spectrum in the 2-10 keV range
 from source to source and also with time. Given the uncertainties in the
 \ct{time-averaged} spectral properties of Galactic X-ray binaries, our
 description of \ct{the} average spectrum should be adequate. To turn the problem
 around, if the diffuse component of GRXE can be separated from point
 sources using high angular resolution observations, then the scattered
 GRXE measurement will directly give us the average spectrum of X-ray
 sources in the Galaxy, after correcting for the effect of energy
 dependence of Rayleigh scattering\ct{,} which makes the scattered spectrum
 softer \ct{than} the incident spectrum  (see Appendix \ref{appa}) {and the effect of X-ray absorption.}
The distribution of the \ct{catalog} sources used in the calculations is shown
in Fig. \ref{sourcesxy} {as a projection on the Galactic plane} and
in Fig. \ref{sourceslb} in the
Galactic angular coordinates. A list of  some of the sources responsible
for the peaks {in the scattered X-ray radiation} is given in Appendix \ref{appb} in Table \ref{tablepeaks}.

\noindent\textbf{Monte Carlo distribution}

{ We simulate the optimistic case discussed in Section \ref{contrXrays}}  by  populating the Galaxy with a Monte Carlo simulated population of HMXBs and LMXBs. The probability density function (pdf) for the
distribution of LMXBs in the Galaxy is taken to be proportional
to the mass density of stars, whereas the pdf of HMXBs is taken
to be proportional the mass density of the ISM; both the stars and ISM
density distributions are modelled using the { mass models
in \citet{denbinac} and  \citet{denbin} (Eq. \ref{Eq:diskmodel})} with stellar disk
parameters $z_{{\rm d}}=325$ pc, $R_{{\rm d}}=2.5$ kpc \ct{and} 
$R_{{\rm m}}=0$. HMXBs are therefore expected to be concentrated in
 the plane of the Galaxy, a fact that becomes important when
 we include the Rayleigh scattering at small angles on multi-electron atoms
 and molecules. We populate the Galaxy with 300 LMXBs and 100 HMXBs
{ using Monte Carlo sampling } from these distributions, { which is
consistent with observations taking into account the
incompleteness of \ct{catalog} at the \ct{low-flux}} end \citep{caton, cattw,
  catth}. The
sources are assigned luminosities sampled randomly from the luminosity
functions of \citet{gilfanov2004} for LMXBs and \citet{grimm} for HMXBs. The
total $2-10$keV luminosity of the Galaxy in our calculation is $\sim 4.7
\times 10^{39}$ ergs/s for \changeRR{simulated} LMXBs and $\sim 1.7 \times 10^{39}$ ergs
/s for \changeRR{simulated} HMXBs\ct{,} which
is consistent with the expectations from observations of other galaxies
{\citep{Grimm2002,grimm,gilfanov2004,lehmer2010,mineo2012,Mineo2013}.} As discussed in the previous section, we use the
following typical  spectrum  for all the sources\ct{,} but  differentiating between
LMXBs and HMXBs \ct{by assuming a} softer spectrum for LMXBs and a harder spectrum for
HMXBs in the 3-20 keV range:
\begin{equation}
\label{spectrum}
 F(E)\propto  E^{- \alpha}{\rm e}^{-\frac{E}{\beta}}\:\:\: {\rm  s}^{-1}{\rm keV}^{-1},
\end{equation}
with parameters $\alpha=1,\: \beta=20\:{\rm keV}$ for \changeRR{simulated} HMXBs
\citep{luto} and  $\alpha=1, \: \beta=4.6\:{\rm keV}$ for \changeRR{simulated} LMXBs, which
is a fit to the observed spectrum of GX 340+0 \citep{Gilfanov2003}.

\section{Scattering of X-rays on atoms and molecules}
\subsection{Scattering cross-sections}
\label{sec:crosssec}
Interaction of X-ray photons with light atoms {is  described} by
the following expression \changeRR{\citep[see e.g. ][for detailed discussion]{EP1970,prof,sunchu,suny2}:}
\begin{equation}
\frac{{\rm d}\sigma}{{\rm d}\Omega {\rm d}\omega_2} = \frac{r_e^2}{2}(1+{cos(\theta)}^2)\frac{\omega_2}{\omega_1}|\langle f|\sum_{j} e^{i\vec{q}\vec{r}_j} | i \rangle|^2 \times \delta(\Delta E_{if} - \Delta \omega),
\end{equation}
where $\omega_1$ and $\omega_2$ are the photon energy before and after the scattering, $\Delta E_{if}$ and $\Delta \omega = \omega_1 - \omega_2$ denote the energy change in the atom or molecule and in the photon\ct{,} respectively, $i$ and $f$ are the initial and final states of the system, $\vec{q}$ is the change in the photon momentum and $\vec{r}_j$ is the coordinate of the jth electron.
\par Depending on the final state of the system after the interaction, the
processes are classified as Rayleigh or elastic scattering, involving a
change in the direction of the incoming photon at constant frequency
($i=f$), Raman scattering, resulting in the excitation of the bound
electron ($f=$ {excited} state)\ct{,} and Compton scattering, resulting in the
\ct{ionization} of the atom or molecule ($f$= continuum state). \par These
cross-sections, which involve the scattering of radiation off electrons
bound in atomic or molecular hydrogen{,  helium\ct{,} and other heavy
  atoms and molecules}, include additional effects compared to scattering
off free electrons due to the discrete energy levels that they occupy and
their distribution over momentum from quantum mechanical effects
\citep{EP1970,sunchu,Vainshtein1998}. The latter in particular implies that the energy of the
scattered photon is not unambiguously constrained by the scattering angle,
and therefore that the Klein-Nishina (KN) formulation for the singly
differential scattering cross-section is no longer sufficient. One of the
most commonly used methods in computing the interaction is the impulse
approximation (IA) applied to the doubly differential cross-section, which
is relevant in the case of a change in the photon energy much larger than
the binding energy \citep{EP1970,sunchu}. The singly differential case\ct{,} on the other hand\ct{,} can be computed by implementing the corrections to the KN formulation via an incoherent scattering factor $S(x)${  \citep{bergsu},} such that  
\begin{equation}
\label{comp}
\left(\frac{{\rm d}\sigma}{{\rm d}\Omega} (\theta)\right)_{{\rm Compt}} = S(x(\theta)) \left(\frac{{\rm d}\sigma}{{\rm d}\Omega} (\theta)\right)_{{\rm KN}},
\end{equation}
where $\theta$ is the scattering angle and $x(\theta)$ is the momentum
transfer. \ct{For} heavier atoms or molecules, the most relevant
differences with respect to HI scattering are found at small angles due to
elastic scattering. In the presence of $Z$ electrons in the system\ct{,} these
can coherently scatter the photon, enhancing the scattering at
{small angles, where the change in energy of the photon is \ct{weak},}
by a factor of $Z^2$. Furthermore, the range of angles and energies at which
the elastic scattering remains dominant increases with additional electrons due
to the decreasing characteristic size of the electron distribution, $D$. The
condition that the elastic scattering dominates is in fact determined by
$\vec{q}\vec{r}_j \lesssim 1$, which in {the} \ct{small-angle}
approximation can be rewritten as $\theta 2 \pi D/ \lambda \lesssim 1$,
where $\lambda$ is the photon's wavelength \citep{suny2}. 
{The range of scattering angles over which Rayleigh scattering dominates\ct{,} however\ct{,} rapidly decreases with increasing energy. The enhancement of the total scattering \ct{cross-section} due to coherence effects therefore mostly contributes to the scattering in the softer energy band (see Appendix \ref{appa}).}

 The computation of the
elastic and inelastic cross sections for atoms is computed using the
routines  {from the publicly available library}, xraylib
\citep{xraylib}, which uses results from \citet{prof}. \ct{For} molecular hydrogen, which has a characteristic size of the electron distribution similar to that of HI, we approximate the scattering \ct{cross-section} by including a factor $Z^2 = 4$ for elastic scattering and a factor of 2 for inelastic scattering to the HI cross section, which was shown \ct{by \citet{suny2}} to be \ct{a} reasonable approximation. 
Heavier elements included in the calculations are chosen based on their
fractional contribution to the elastic scattering \ct{cross-section}. Ranking
heavier elements by their maximal contribution to Rayleigh scattering
(proportional to $Z^2 \times n_Z/n_H$), we include contributions from He, O, Fe, C, Ne, Si\ct{,} and Mg in addition to H (see Table \ref{elements}). Scattering on H2 and helium is especially
important for HMXBs,  \ct{whose} X-rays have paths to us with small
scattering angles\ct{,} resulting in significant enhancement of X-ray flux
\ct{than for} HI. \ct{The c}ontribution from all other elements combined would be
\ct{lower} than $\sim 1\%$. In principle, the scattering from molecules such
as CO \ct{and water} would be\ct{, in addition, }enhanced in the Rayleigh scattering
regime because of coherent scattering from all atoms in the molecules\ct{,} and
should be taken into account in a precise calculation. For example, in CO
the cross section in the Rayleigh limit will be $196\times$ Thomson in the
molecule\ct{,} compared \ct{with} $100\times$ Thomson if the two atoms were separate. Given the
uncertainties in \ct{the} fraction of elements in molecules and \ct{the} depletion of
elements in dust, in addition to \ct{the} uncertainties in metallicities in
different regions of the Galaxy, such a precise treatment is not warranted
at this stage\ct{,} and we ignore these otherwise important effects.

\subsection{ISM volume emissivity}
\begin{figure*}[!ht]
\centering
     \includegraphics[trim = 20 23 10 20, width=9cm, height=7cm]{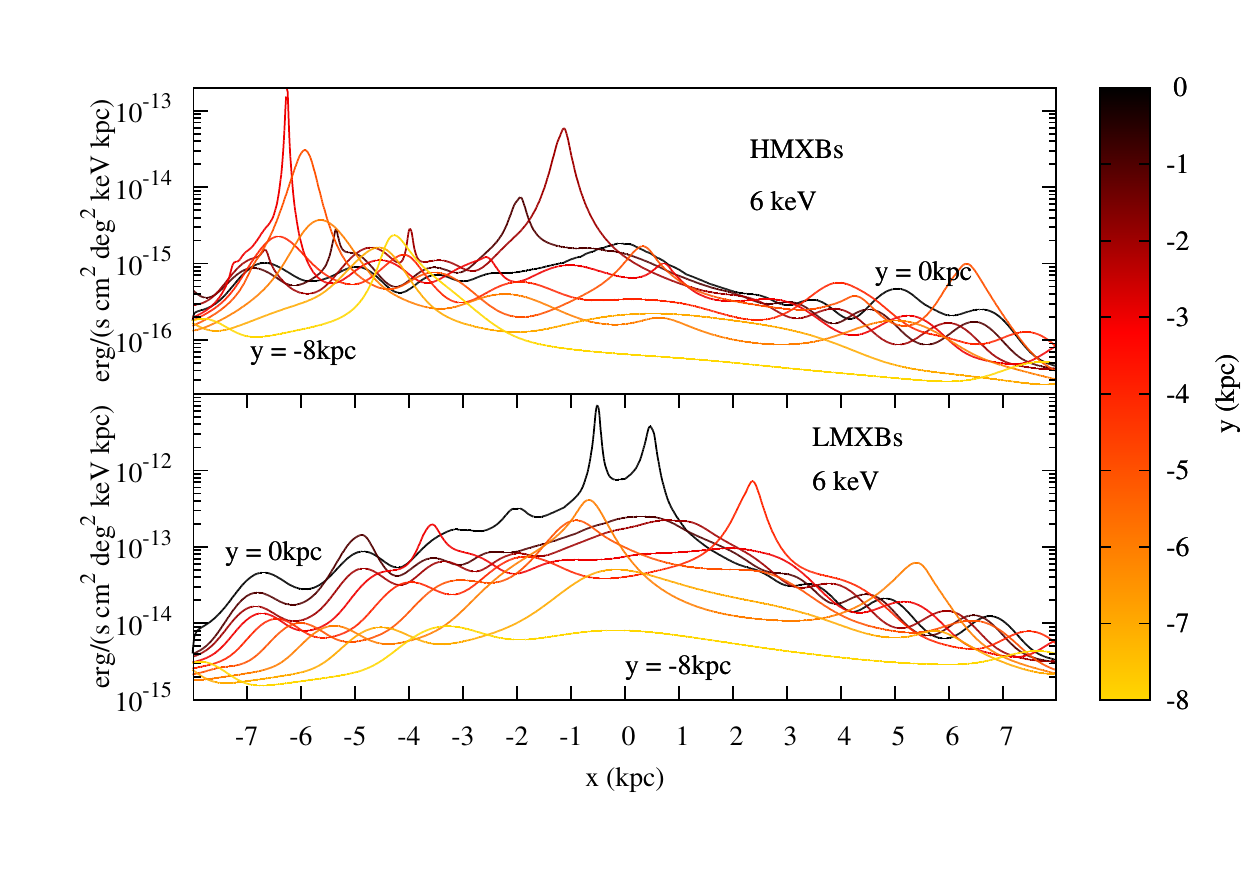}\includegraphics[trim = 18 23 10 20,width=9cm, height=7cm]{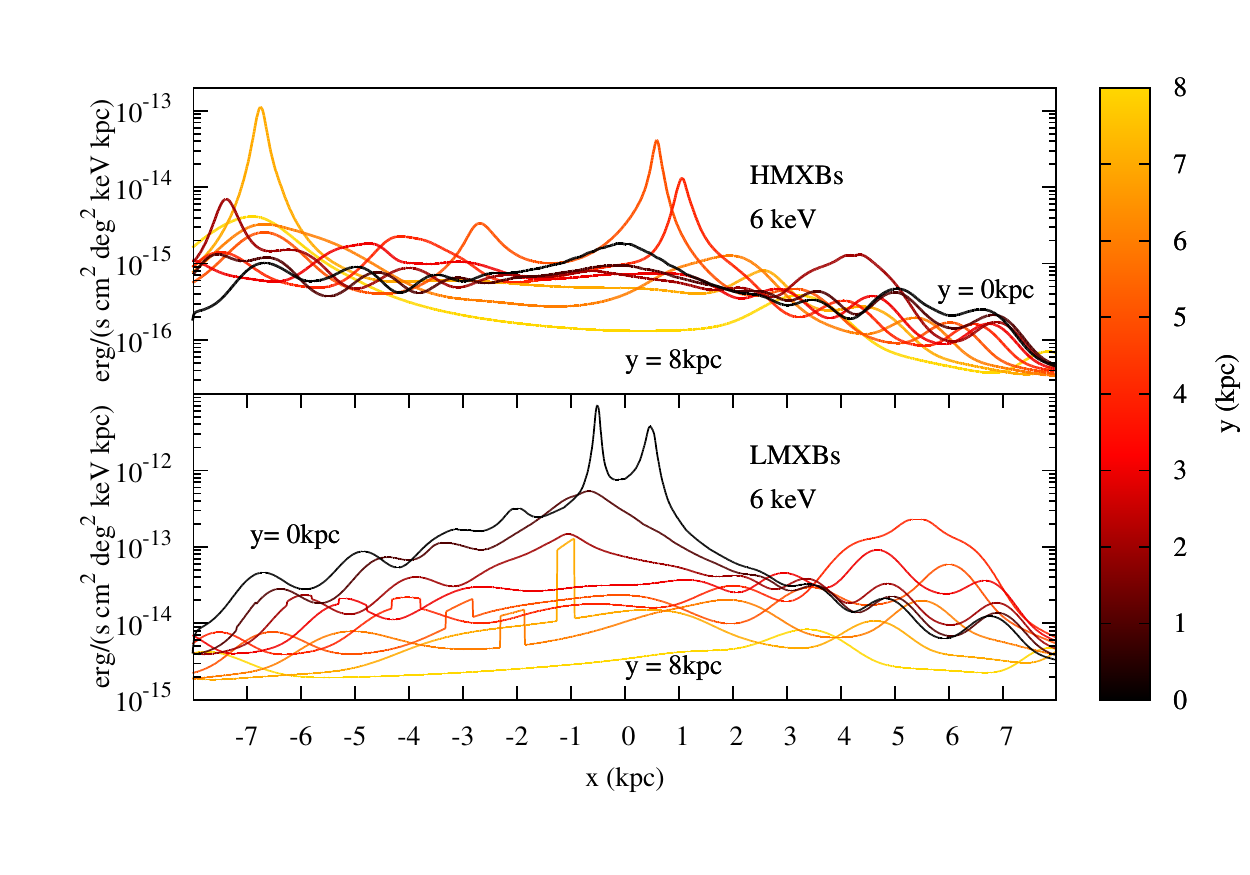} \\ 
       \caption{Profiles of the volume emissivity per deg$^2$ in our direction on the Galactic plane due to the observed sources used in our calculations, at different distances from the Galactic center (Sun's position (-8 kpc, 0, 0.015 kpc)). Units were chosen so that if this emissivity is integrated over 1 kpc it is numerically equal to the observed flux.  Proximity effects in the
         vicinity of sources are clearly visible. The concentration of
         LMXBs in the Galactic bulge results in a steady decrease in the
         overall profile of the volume emissivity from these sources away
         from the Galactic center. This is not the case for HMXBs \ct{because of}
         their distribution along the Galactic disk. Discontinuities in the
         LMXBs profiles in the positive y-range are \ct{caused by} temporal constraints in the illumination
         of its surroundings \ct{by} the very bright source GRS 1915+105 (located at $(-0.27{\rm kpc}, 7.83{\rm kpc}, -27{\rm pc})$)\ct{,}
         which was not observed during 1970-1992 and is assumed
           to \ct{have been} absent before 1992.}
       \label{volume_emissivity}
 \end{figure*}

The total volume emissivity per steradian, $\epsilon_{{\rm XBs}}$, of the gas due to illumination by Galactic X-ray binary sources at a distance $s$ from the observer along the line of sight $(l,b) $ is given by
\begin{equation}
\epsilon_{{\rm XBs}} (s, \nu) =  \sum_{Z}  \sum_{i} \frac{L_{i}(\nu)}{4\pi R^2_i(s)} \bigg(\frac{{\rm d}\sigma}{{\rm d}\Omega}(s, \nu) \bigg )_{Z}  n_{Z}(s) \label{emissivity},
\end{equation}
where the sum is \ct{computed} over the  X-ray sources $i$ and elements $Z$ contributing
to the scattering, $R_i$ is the distance of source with luminosity $L_i$
from the scattering point, $\left({\rm d}\sigma/ {\rm d}\Omega \right )$
is the  differential scattering cross-section including Rayleigh
  and Compton scattering, and $n_{Z}$ is the number density distribution of \ct{the} element $Z$. The total differential \ct{cross-section} is computed as a function of the position along the line of sight $s$ as well as of the radiation energy, since the angle at which radiation from a given source is to be scattered to reach the observer will depend on the position at which the scattering occurs. 
 Because of the low optical depth, the contribution of radiation
\ct{that undergoes} multiple scatterings before reaching the {observer is
  assumed} to be negligible. {The volume emissivity per unit solid
  angle in the Galactic
  plane is plotted in
  Fig. \ref{volume_emissivity} using observed data.}
{From {Eq.} \ref{emissivity}} we expect proximity zones near individual sources where the flux peaks sharply, as clearly seen in Fig. \ref{volume_emissivity}. Scattering of
X-rays in these proximity zones will result in X-ray halos around each
source, similar to the dust halos observed around many X-ray sources
\citep{overbeck1965,trumper1973,rolf1983,predehl1995} at angles
  $<1^{\circ}$ but
  extending to much larger angles. The profile is sharply peaked if the
  source is in the Galactic plane and the slice through the plane crosses
  the source. If\ct{,} on the other hand\ct{,} the source is above or below the plane,
  the profile is broader, as expected from  the geometry. LMXBs
 dominate in the center of the Galaxy \ct{because} they are concentrated in the
 bulge, while {in the disk and near the spiral arms  HMXBs are dominant.}

%
%

\section{Results}
The contribution of the scattered intensity is expected to be significant at \ct{low} latitudes where the interstellar gas is
concentrated. We therefore present the results in the form of longitude profiles of the total scattered
 intensity along the Galactic plane ($b =0^{\circ}$), in energy ranges 3 - 10, 10-20, 17-25 and 25-60 keV, chosen
  to allow  comparison with the RXTE and Integral observations of GRXE
  \changeRV{and future observations of Nuclear Spectroscopic Telescope
    Array (NuSTAR) \citep{Nustar} and hard X-ray
    telescopes onboard Astro-H \citep{astroh} and
     ART-XC \ct{onb}oard Spectrum-RG \citep{srg} satellites}. We \ct{furthermore distinguish between the contribution of }HMXBs and LMXBs for both real (Fig. \ref{Long_prof_real_sources}) and Monte Carlo sources (Fig. \ref{Long_prof_MC_sources}). 

%

\par  From the profiles it is easy to distinguish the contribution of
HMXBs along the Galactic disk and of LMXBs around the Galactic bulge; for
the \ct{hard-energy} range the contribution of LMXBs, {with \ct{a} softer
 average  spectrum \ct{than for} HMXBs,} significantly decreases. 
The halos of scattered radiation in the proximity of bright
  sources are also visible: the sources contributing the most to the scattering component are marked and \ct{are clearly detectable} as peaks in
the profile, with sharper peaks for sources in the Galactic plane and
broader peaks for sources significantly above or below the plane. \ct{When the line of sight crosses a source the}
sharpness of peaks  is limited by
the angular resolution of the calculation\ct{,} which is $5'$.  The LMXBs make a broad feature \ct{toward} the Galactic center from
superposition of several luminous sources\ct{,} while  the HMXBs contribute mainly in the
Galactic disk. At the \ct{low-energy} end the LMXBs dominate over the HMXBs, while
at the \ct{high-energy} end the \ct{contribution of the HMXBs} to the GRXE becomes \ct{similar} to
that of the LMXBs, \ct{because} the total GRXE emission in each band is proportional to
the total luminosity of \ct{the }corresponding X-ray sources in that band. The profile drops
sharply \ct{toward} the Galactic anti-center\ct{,} reflecting the decrease in ISM column
density.\footnote{\changeRR{We do not include the contribution of the Crab nebula,
  which \ct{because it is} only $\sim 1000$ years old would only \ct{contribute} a small peak near the Galactic anti-center}.}  Current observations of the GRXE do not have a high enough angular
  resolution to allow a direct comparison with the scattered profile.  We therefore make use of the relation between the GRXE
  emission and the Galactic stellar distribution first studied in
  \citet{Revnivtsev2006} to compare the diffuse scattered component with the
  GRXE of stellar origin.
\begin{figure*}[!ht]
\centering
     \includegraphics[trim = 40 100 40 50,height = 5.5cm, width = 14 cm]{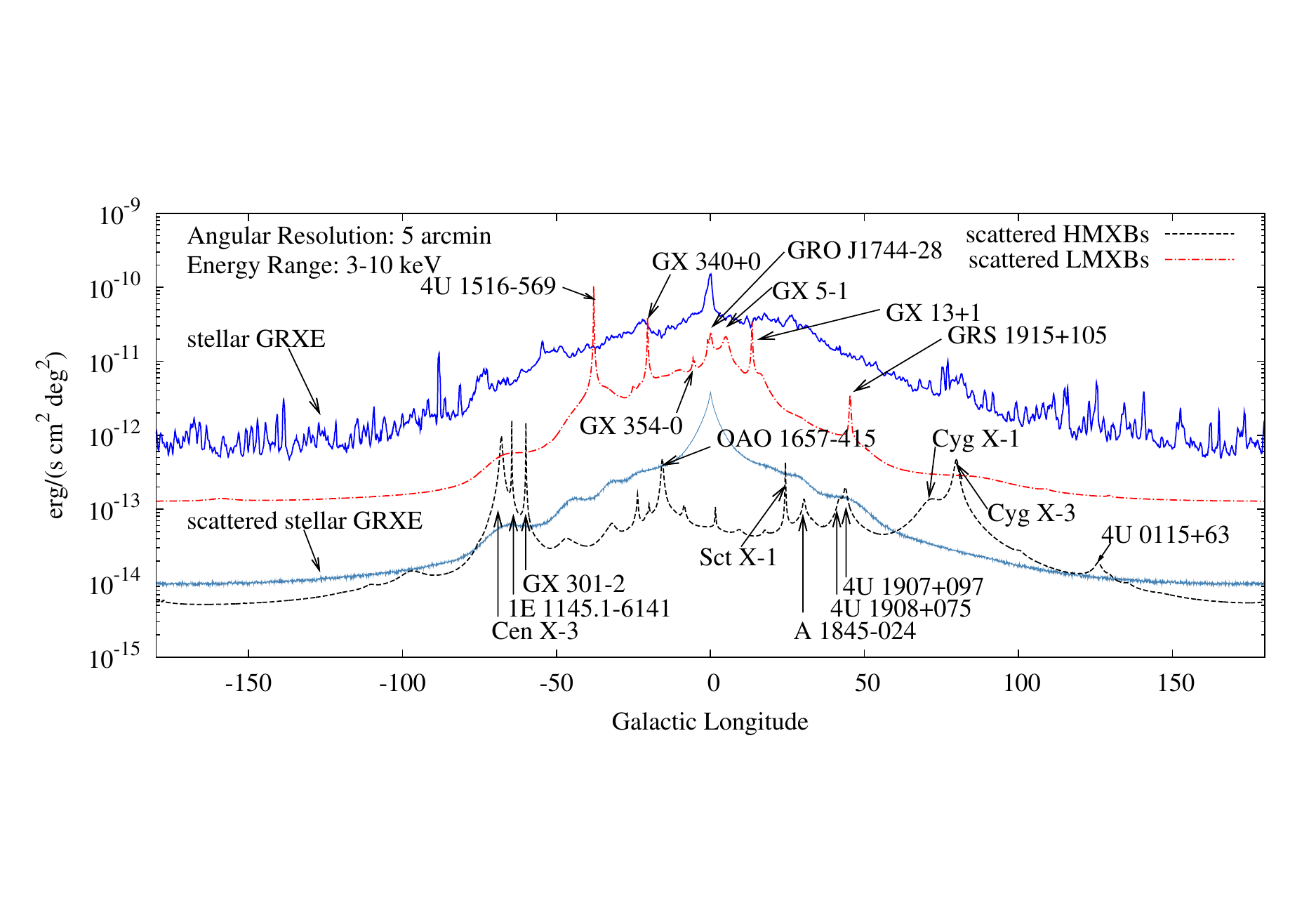} \\
\includegraphics[trim = 40 100 40 50,height = 5.5cm, width = 14 cm] {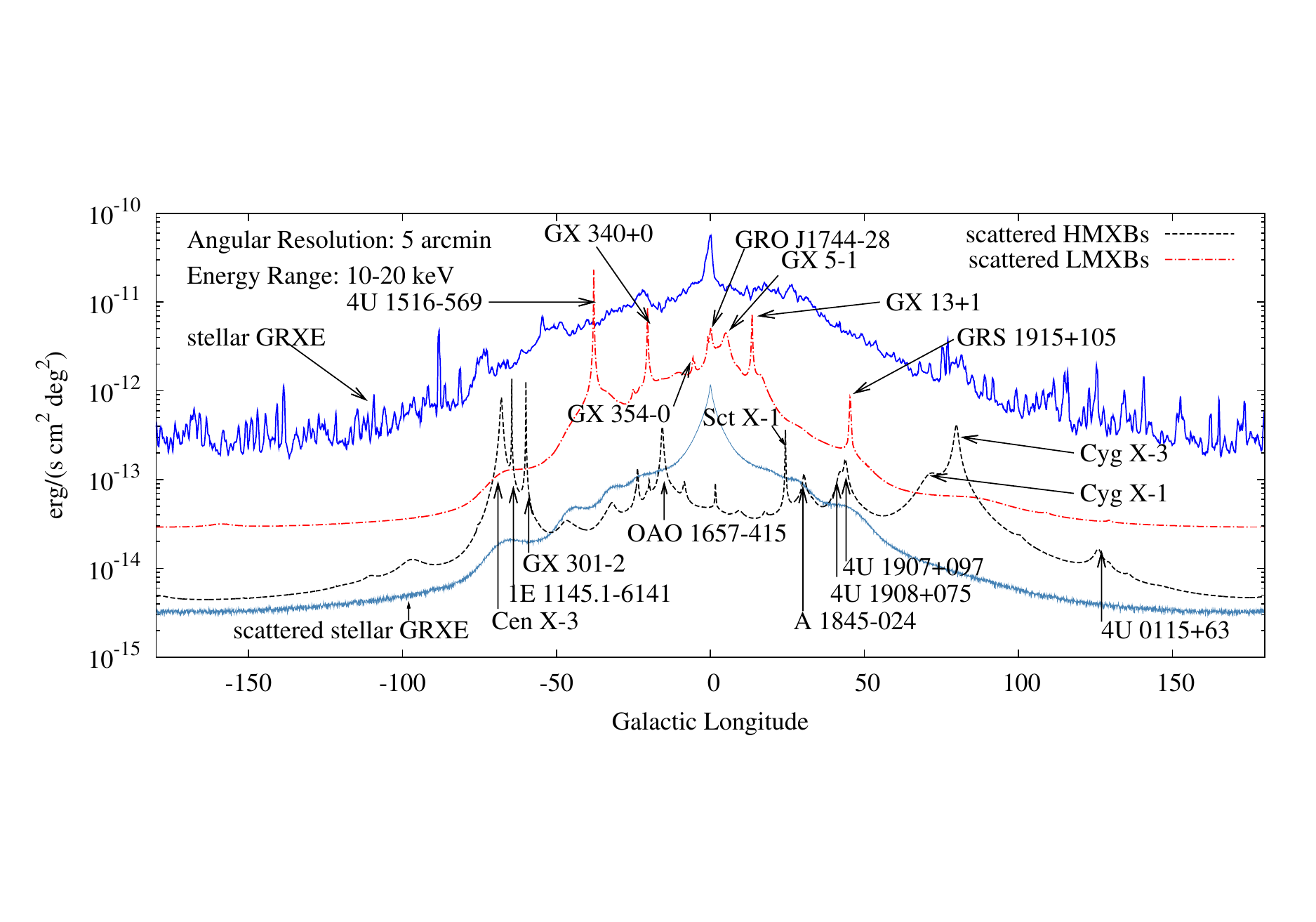} \\
     \includegraphics[trim = 40 100 40 50,height = 5.5cm, width = 14 cm]{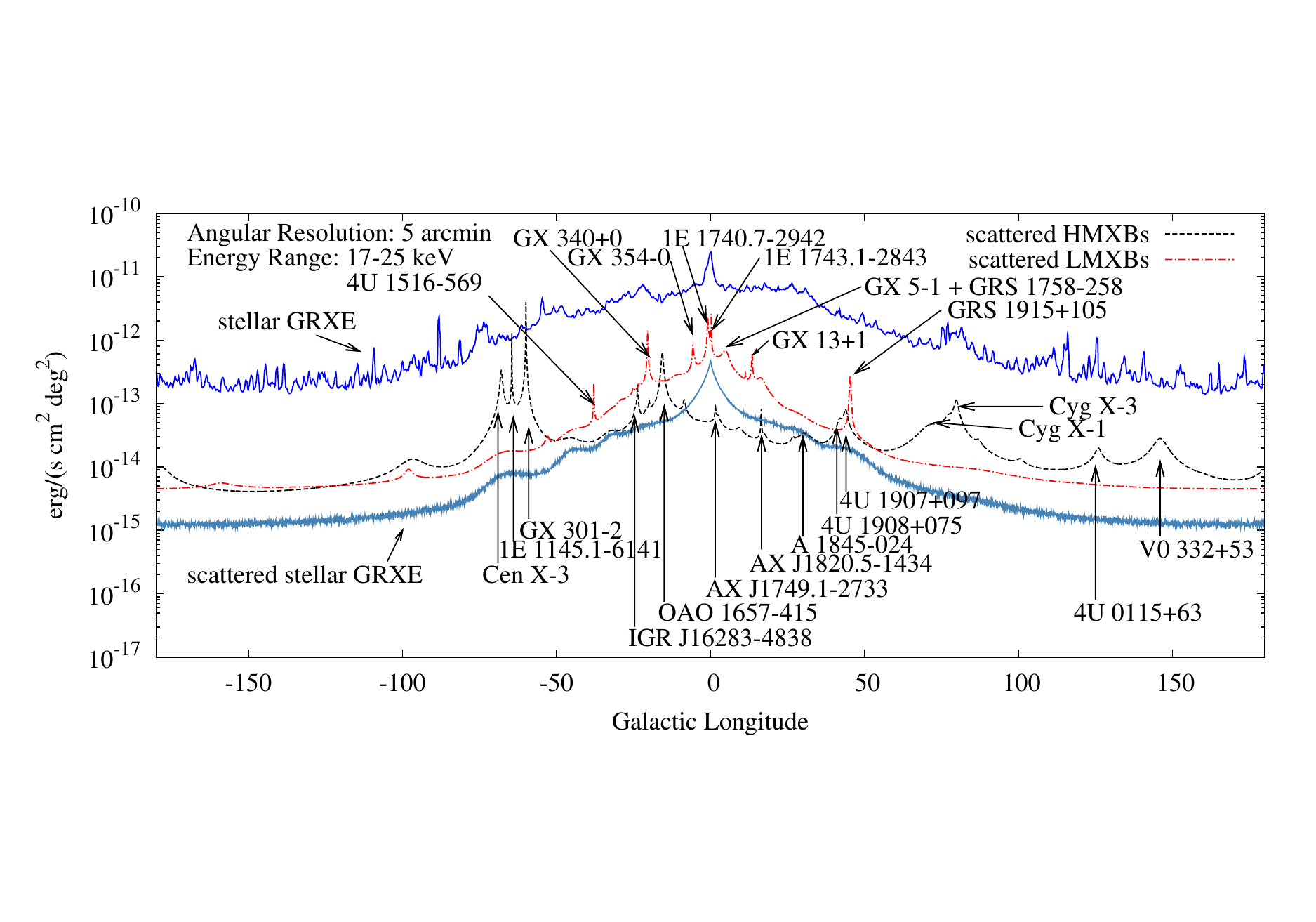}\\ 
\includegraphics[trim = 40 80 40 50,height = 5.5cm, width = 14 cm] {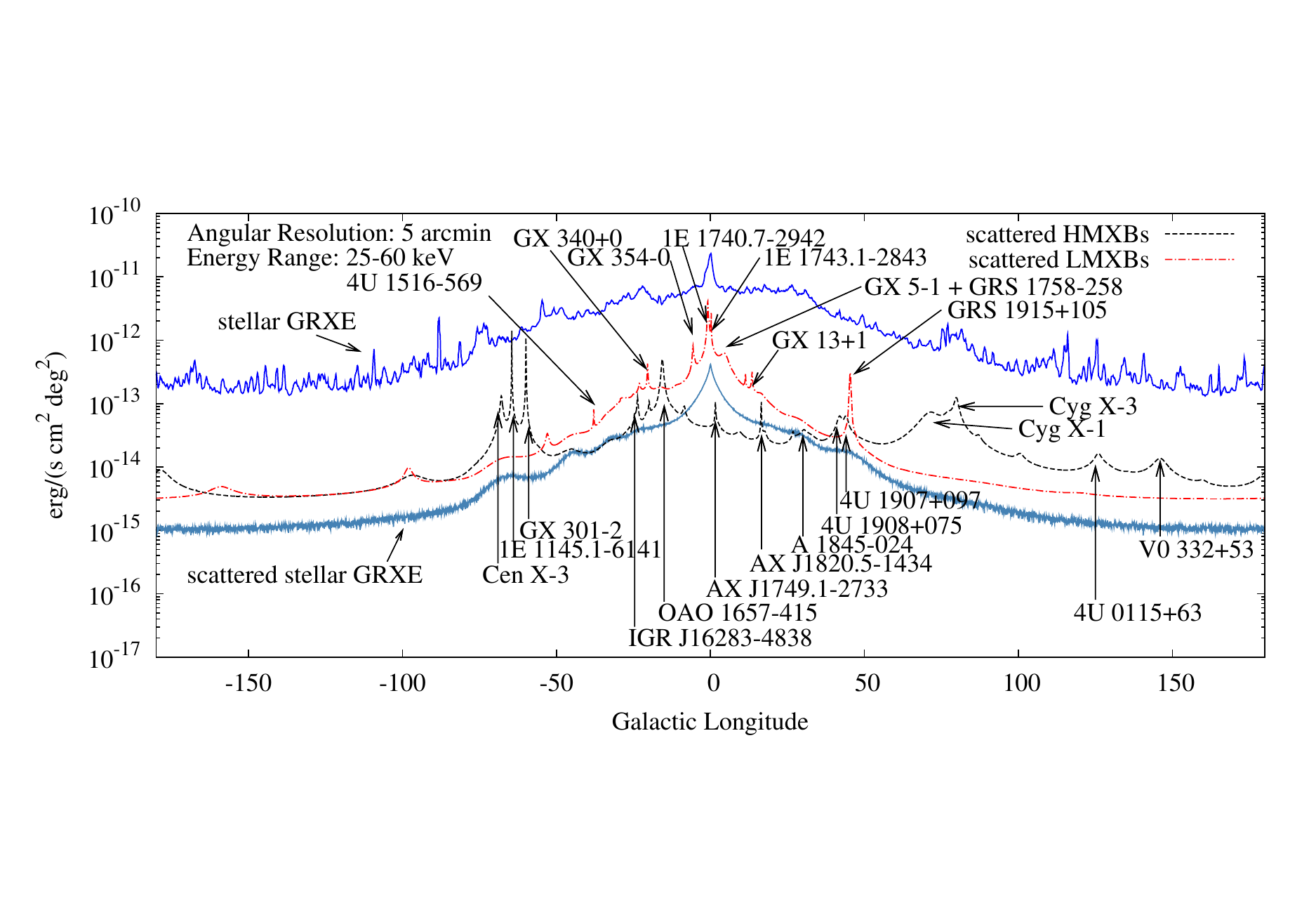} \\
       \caption{Longitude profiles of scattered \ct{XB} radiation in comparison with \ct{a near-infrared}-inferred GRXE profile in energy ranges 3-10, 10-20, 17-25\ct{,} and 25-60keV for real sources. The sum of the scattered component is 10-30\% of the stellar contribution to the GRXE. Data for the sources responsible for the labeled peaks are listed in table \ref{tablepeaks}.}
       \label{Long_prof_real_sources}
\end{figure*}
\begin{figure*}[!ht]
\centering
     \includegraphics[trim = 50 70 50 70, height = 4.5cm, width = 13 cm]{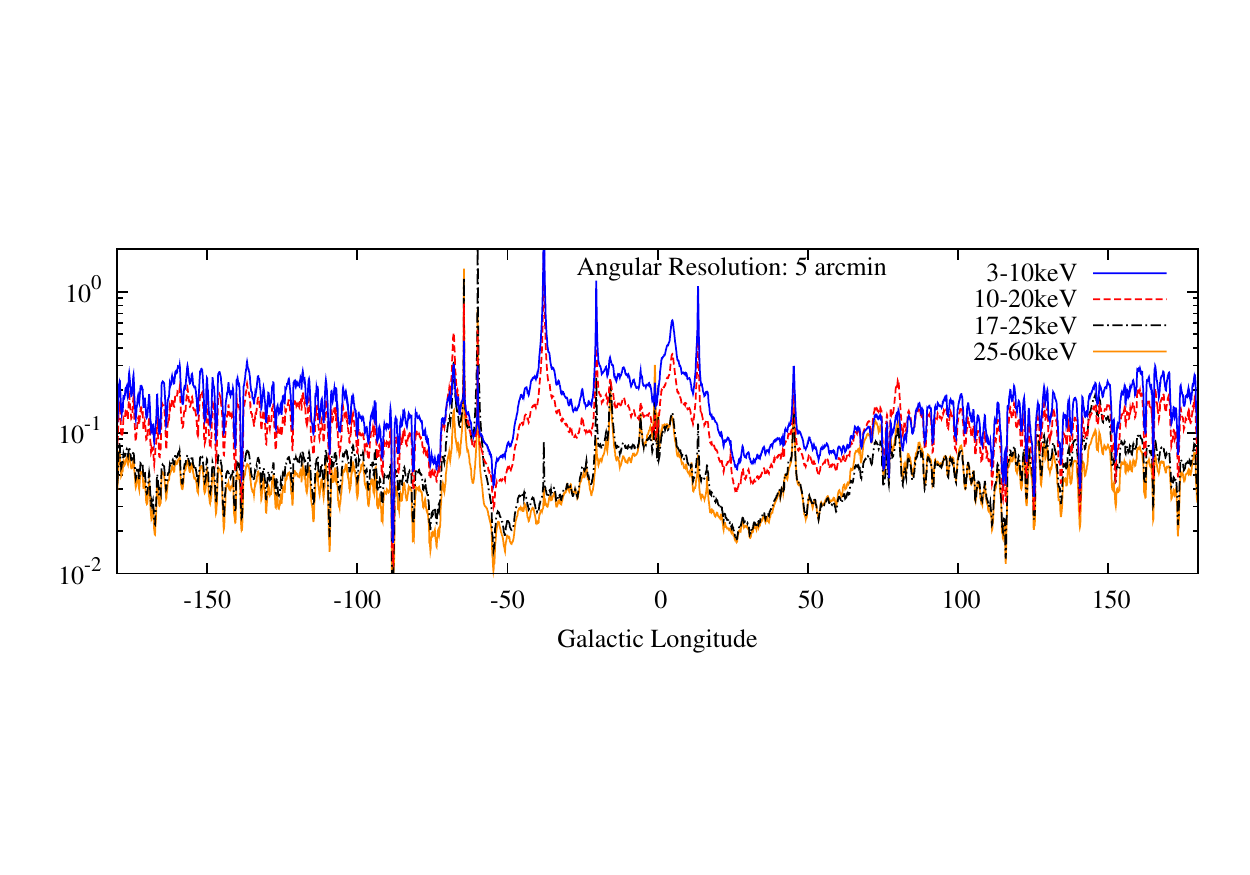}
       \caption{Ratio of scattered X-ray radiation (including HMXBs, LMXBs\ct{, and near-infrared-inferred} low-luminosity sources) to the \ct{near-infrared}-inferred GRXE
         profile for real sources for the energy ranges (from top to
         bottom) 3-10 keV, 10-20keV, 17-25keV\ct{,} and 25-60 keV. \changeRV{The average spectrum of the LMXB sources in our
           calculations is  significantly softer than the
GRXE spectrum observed by RXTE and Integral \citep{Revnivtsev2006,Krivonos2007}. As a result\ct{,} the contribution of the scattered component to GRXE 
decreases at energies above 10 keV.}}
       \label{Long_prof_ratios}
\end{figure*}

\begin{figure*}[!ht]
\centering
  \includegraphics[trim = 20 0 20 20, height = 6cm, width = 8cm]{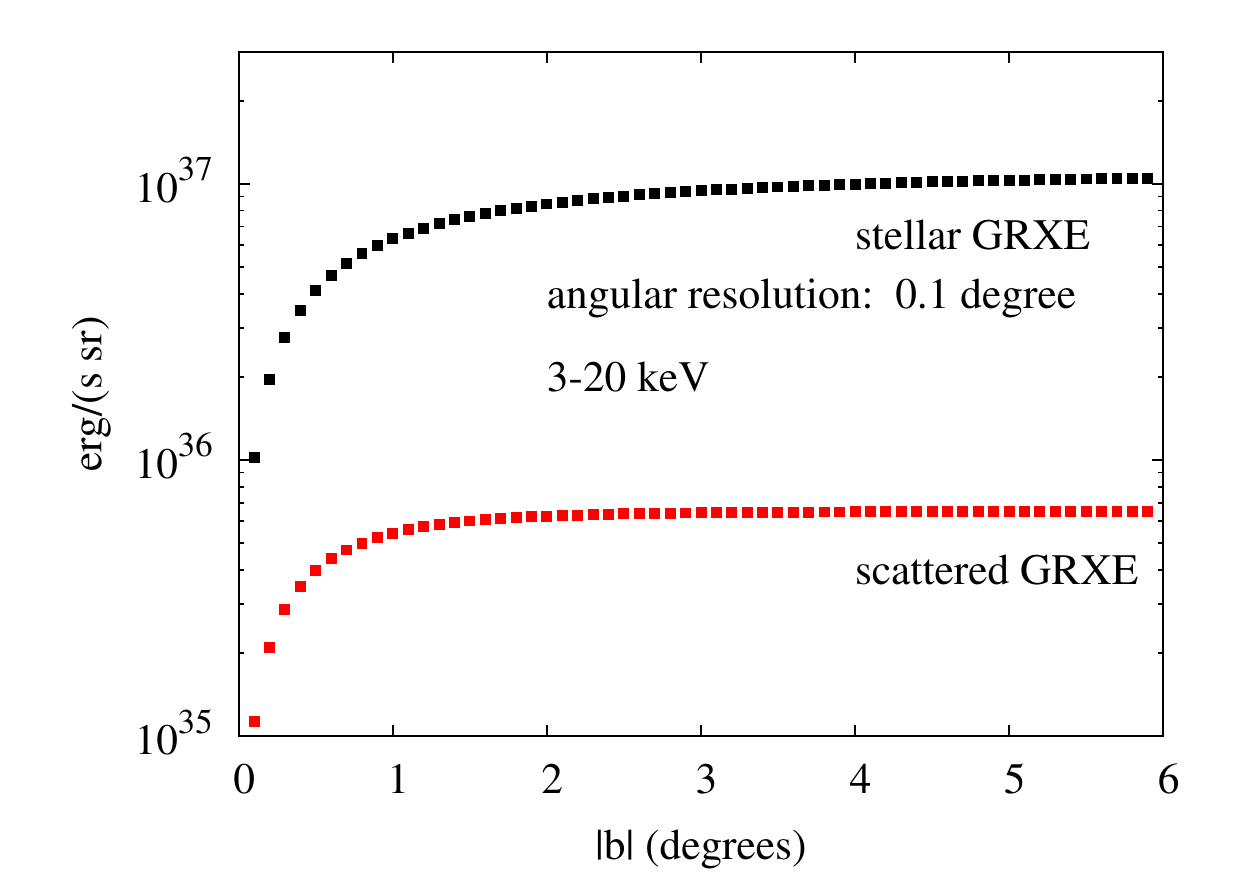}   \includegraphics[trim = 20 0 20 20, height = 6cm, width = 8cm]{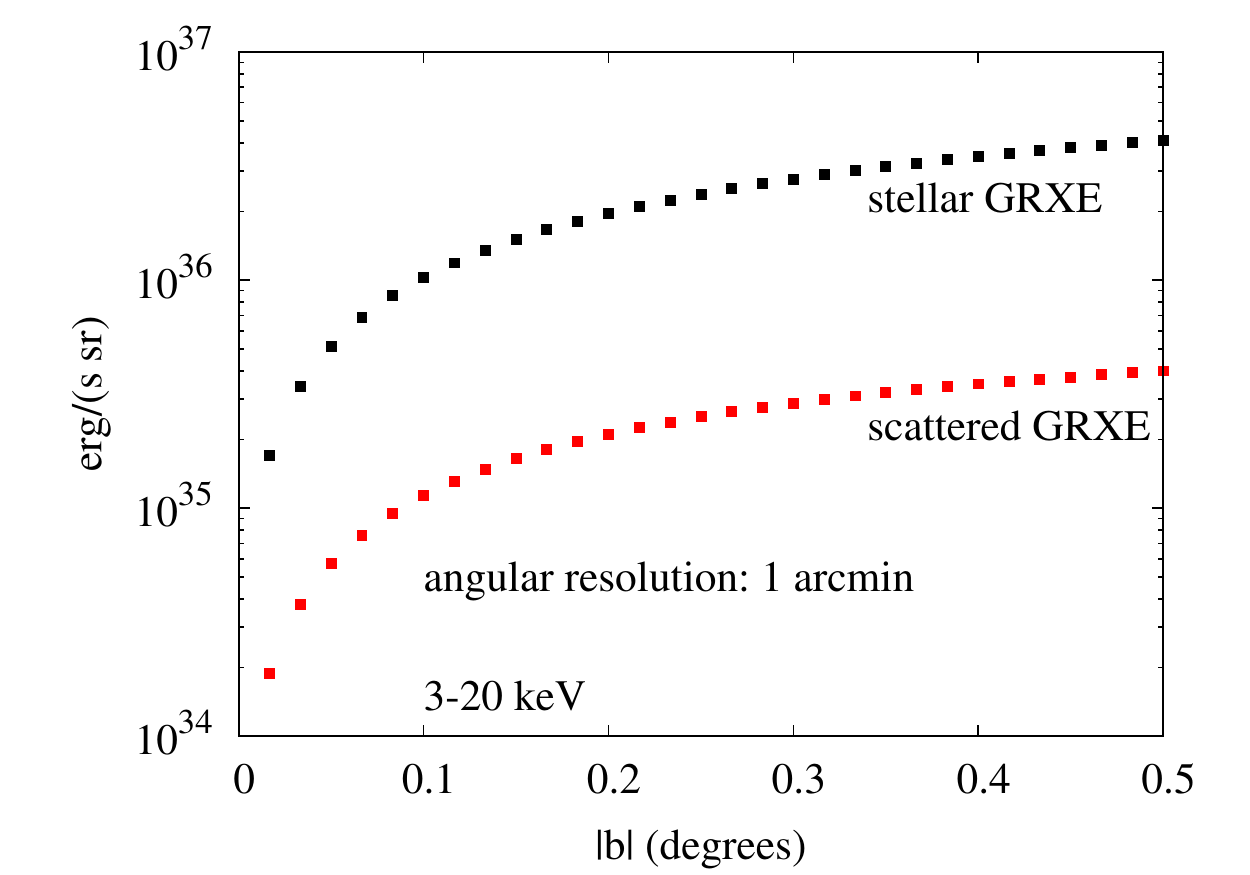}  \\
   \includegraphics[trim = 20 0 20 10, height = 6cm, width = 8cm]{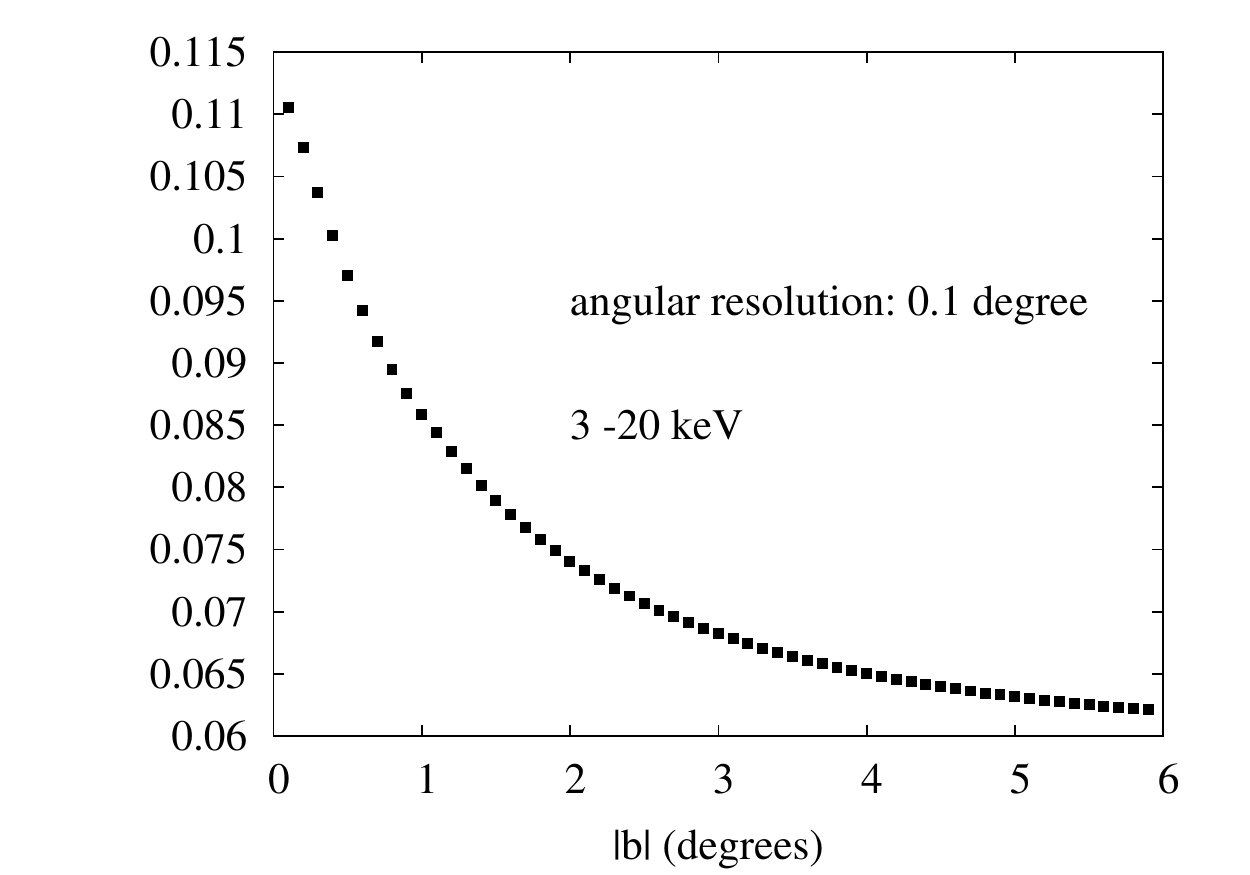}  \includegraphics[trim = 20 0 20 10, height = 6cm, width = 8cm]{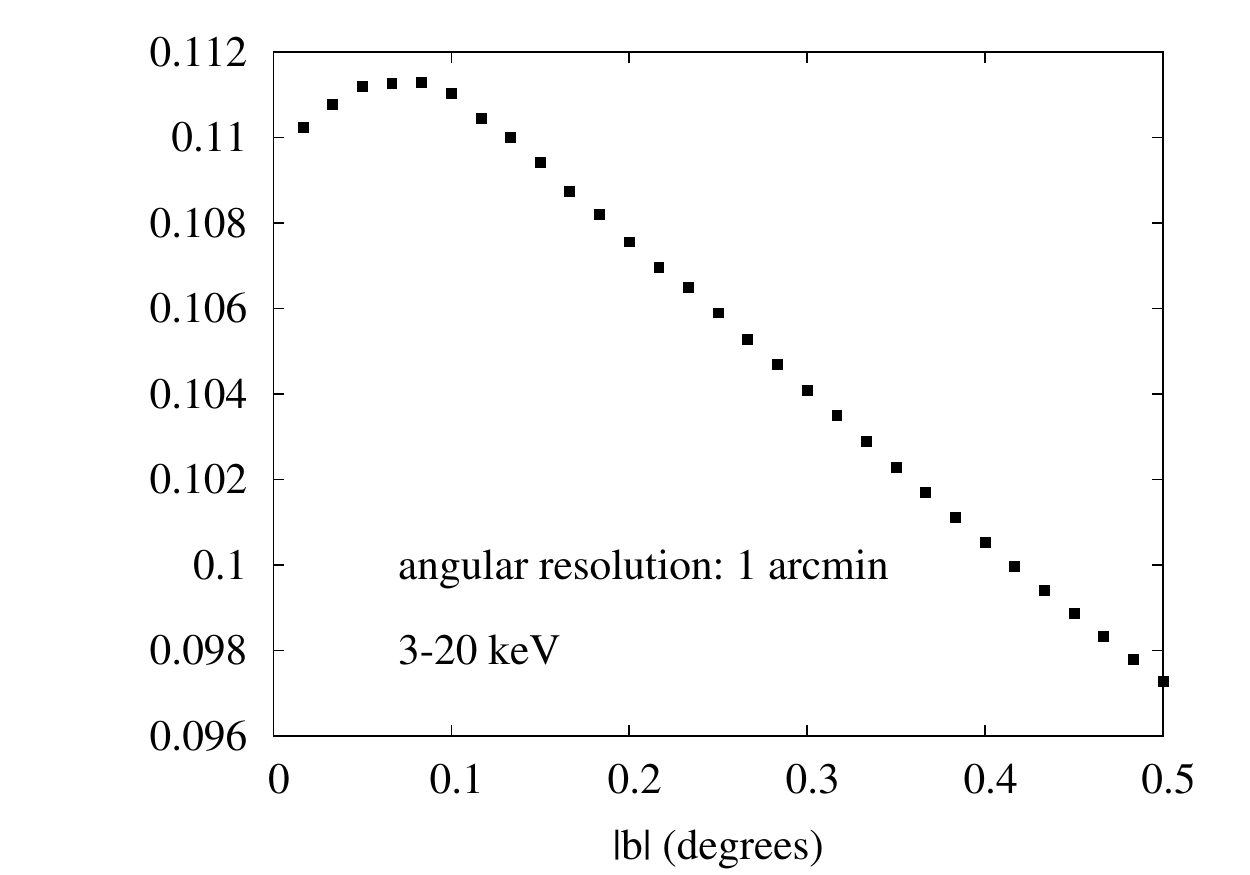}  \\
  \caption{\textit{Top \ct{panels}}: Luminosity/sr
radiated in our direction from the stellar GRXE and the scattered GRXE 
component (including stellar and XBs contributions) in the 3-20keV range 
integrated over the longitude range ($l < |180^{\circ}$|) and latitude 
range $-b$ to $b$.  The difference in the scale height of the scattered 
component compared \ct{with} a GRXE of stellar origin is clearly visible. 
\textit{Bottom \ct{panels}}: Ratio of the two  profiles. The peak in the 
ratio profile is \ct{caused by} the elevation of the \ct{Sun} above the Galactic 
plane by 15 pc.}
       \label{lum_profile}
\end{figure*}

\subsection{Contribution of the \changeRR{scattered} component compared \ct{with} the stellar emission}
 \ct{To} quantify the relative contribution of the ISM
scattering {with respect to the stellar emission,} we  use  the relation between the GRXE's {surface brightness} and the \ct{extinction-corrected} near-infrared emission found by \citet{Revnivtsev2006} ($I_{3-20{\rm keV}}$ ($10^{-11}$ erg s$^{-1}$cm$^{-2}$deg$^{-2}$) = 0.26$\times I_{3.5\mu{\rm m}}$(MJy/sr)) and \citet{Krivonos2007} ($F_{17-60}$keV = (7.52 $\pm$ 0.33)$\times 10^{-5}F_{4.9 \mu{\rm m}} $) for the 3-20 and 17-60keV ranges\ct{,} respectively.
The \ct{full-sky} infrared maps are available from the \changeRV{Cosmic Background
Explorer's Diffuse Infrared Background Experiment (COBE/DIRBE)}
experiment. We use the zodi-subtracted mission average map available on the LAMBDA archive at \url{http://lambda.gsfc.nasa.gov}.
\ct{To} account for interstellar extinction in the infrared data, we
use the method followed in \citet{Krivonos2007} based on the  assumption that the ratio of
intrinsic surface brightness in different infrared bands is uniform over
the sky. This in effect is equivalent to
the assumption that the average spectrum of the \ct{infrared} sources \ct{is the}
same in all directions, or that the average relative population of different
types of stars in different regions of the Milky Way is the same. The 'true
value' of the ratio {under this assumption can be}
estimated from the lines of sight at which the extinction is negligible and
the contribution from the  extragalactic sources is not important. We use this method {because} most of the publicly available extinction maps account
for the total extinction in our Galaxy and are useful only in the
case of extra-Galactic sources.  For the Galactic sources the extinction, in addition to direction, would also depend upon how far \ct{away} the sources are from the observer.
The extinction coefficient for emission at wavelength $\lambda$ can  be
calculated, following \citet{Krivonos2007}, as
\begin{equation}
A (l,b) = \frac{-2.5}{A_{1.25\mu{\rm m} }/A_{\lambda}-1}\bigg[{\rm log}_{10}\frac{I_{1.25\mu {\rm m}}}{I_{\lambda}} - {\rm log}_{10}\frac{I^0_{1.25\mu{\rm m}}}{I^0_{\lambda}} \bigg],
\end{equation}
where the "true" value of the ratio is estimated by averaging all measurements for $1^\circ$ around $b\sim 30^\circ$ and $A_\lambda/A_V$ values are taken from \citet{Rieke}. 
 The stellar component of the GRXE
  estimated in this way is also shown in Figs. \ref{Long_prof_real_sources} and \ref{Long_prof_MC_sources}. 

\subsection{Scattering of the stellar component}
The X-ray emission from the \ct{low-luminosity} stellar sources will
  also be scattered by the ISM, adding to the diffuse scattered X-rays from
  X-ray binary sources.  This effect can be accounted for by
  considering the scattering by the ISM of the observed GRXE, assumed to be of stellar origin.
We use the models for the \ct{three-dimensional} luminosity of the bulge and disk components of the GRXE based on \ct{infrared} distribution of the Galaxy given in \citet{Revnivtsev2006} for the 3-20 keV range and apply it to the 17-60keV range by \ct{normalizing} the distribution to the GRXE luminosity in this range as measured by \citet{Krivonos2007}. For the harder range we also assume the same ratio between the disk and bulge luminosity of $\sim 2.5$ measured in the 3-20 keV range. This ratio is consistent with the disk-to-bulge stellar mass ratio and therefore implies a uniform GRXE emissivity per unit stellar mass throughout the Galaxy. The bulge component is described by
\begin{equation}
\rho_{{\rm GRXE}, bulge} \propto r^{-1.8} {\rm exp}[{-r^3}],
\end{equation}
where $$r = \bigg[(y/y_0)^2 + (z/z_0)^2+ (z/z_0)^2 \bigg]^{1/2} $$ with $x_0 = 3.4 \pm 0.6$ kpc, $y_0 = 1.2 \pm 0.3$ kpc, $z_0 = 1.12 \pm 0.04$ kpc with the $x-y$ axes on the Galactic plane rotated by an angle $\alpha = 29 \pm 6^{\circ}$. The disk component\ct{,} on the other hand\ct{,} is described as
\begin{equation}
\rho_{{\rm GRXE}, disk} \propto {\rm exp}\bigg[ -(R/R_m)^3 - R/R_{disk} - z/z_{disk} \bigg],
\end{equation}
with parameters $R_{disk} = 2$ kpc, $z_{disk} = 0.13$ kpc and fixed $R_m
= 3$ kpc. The total bulge+disk luminosities considered are $\sim 1.39
\times 10^{38}$ erg/s and $\sim 4.2 \times 10^{37}$ erg/s in the 3-20
and 17-60 keV bands\ct{,} respectively. \par The volume {emissivity per unit solid
angle} due to scattering of this radiation at a distance $s$ from the observer in the line of sight $(l,b)$ is then given by
\begin{align}
\nonumber
\epsilon_{{\rm stellar}}(s, \nu) = &\sum_{Z} \iiint\limits_V \frac{\rho_{{\rm GRXE}}(x,y,z)f(\nu)}{4\pi d^2(x,y,z,s)} \\
& \times \bigg(\frac{{\rm d}\sigma}{{\rm d}\Omega}(s, \nu) \bigg )_{Z}  n_{Z}(s) {\rm d}V,
\label{scatt_grxe_em}
\end{align}
where $\rho_{{\rm GRXE}}$ is given in units of erg/cm$^3$, $d(x,y,z,s)$ is
the distance from any point $(x,y,z)$ in the Galaxy to the scattering
point, and $f(E)$ is the normalized spectral energy density taken
from the GRXE spectrum given in \citet{Revnivtsev2006} and \citet{Krivonos2007}.

We show the contribution of this effect as an additional component to the
longitude profiles given in Figs. \ref{Long_prof_real_sources} and \ref{Long_prof_MC_sources}. In the 3-20
keV energy range, the luminosity of GRXE is $\approx 1.4\times 10^{38}\ergss$
\citep{Revnivtsev2006}\ct{,} which is \ct{lower} than the luminosity of LMXBs ($\approx 2.43\times
10^{39}\ergss$)\ct{,} but \ct{higher} than the luminosity of HMXBs ($\approx 9\times
10^{37}\ergss$)\ct{,} and this is \ct{clearly reflected} in the contribution to the scattered X-ray intensity in
Fig. \ref{Long_prof_real_sources}. \changeRE{In the 17-60 keV energy range, the
luminosity of the GRXE is $3.7\times 10^{37}\ergss$ \citep{Krivonos2007}\ct{,}
which is a factor of $\sim 5$ \ct{lower than} the luminosity of LMXBs
($1.9\times 10^{38}\ergss$) and a factor of $\sim 2$ \ct{lower than} the luminosity
of HMXBs ($6.2\times 10^{37}\ergss$)}. We therefore expect the scattering
of \ct{the} stellar component of the GRXE  from the ISM  to be subdominant\ct{,} and this
is evident in Fig. \ref{Long_prof_real_sources}.  The ratio of
scattered X-rays to the GRXE inferred from \ct{near-infrared} data is shown in
Fig. \ref{Long_prof_ratios}. The diffuse scattered X-rays contribute
$10-30\%$ over most of the Galaxy with a  contribution
as high as $40\%$ near the luminous  sources in the 3-20 keV energy
range. The diffuse component in the 17-60 keV range \ct{contributes, on the other hand,} $\sim 10\%$ or \ct{less.  This  is low} but still non-negligible. 
\par
{We calculate the luminosity/sr radiated in our
  direction from within a given latitude and longitude range in the sky ${\rm d}\Omega$ (as opposed to the isotropic luminosity) for both the stellar GRXE and the scattered GRXE component as follows:}
\begin{align}
L_{{\rm GRXE,stellar}} &= \frac{1}{4 \pi}\iiint f(\nu)\rho_{{\rm GRXE}}(x,y,z) s^2 {\rm d} \nu {\rm d} s {\rm d}\Omega \\ 
L_{{\rm GRXE,scatt}} &= \iint [\epsilon_{{\rm XBs}}(s,\nu)+\epsilon_{{\rm stellar}}(s,\nu)] s^2 {\rm d} \nu {\rm d}s {\rm d}\Omega, 
\end{align}
where the total volume emissivity due to scattering $\epsilon(s, \nu)$ includes the contribution from \ct{both XB} (Eq. \ref{emissivity}) and low-luminosity stellar \ct{source} (Eq. \ref{scatt_grxe_em}) components.
{The luminosity/sr profiles are plotted and compared in Fig. \ref{lum_profile}. The
scattered radiation is on average $\sim 10\%$ as luminous as the stellar
component of the GRXE in the inner $\sim 1^{\circ}$ of the Galactic plane.}

\subsection{Optimistic case: \ct{Monte Carlo} simulated sources}
\begin{figure*}[!ht]
\centering
     \includegraphics[trim = 40 100 40 50, height = 5.5 cm, width = 14 cm]{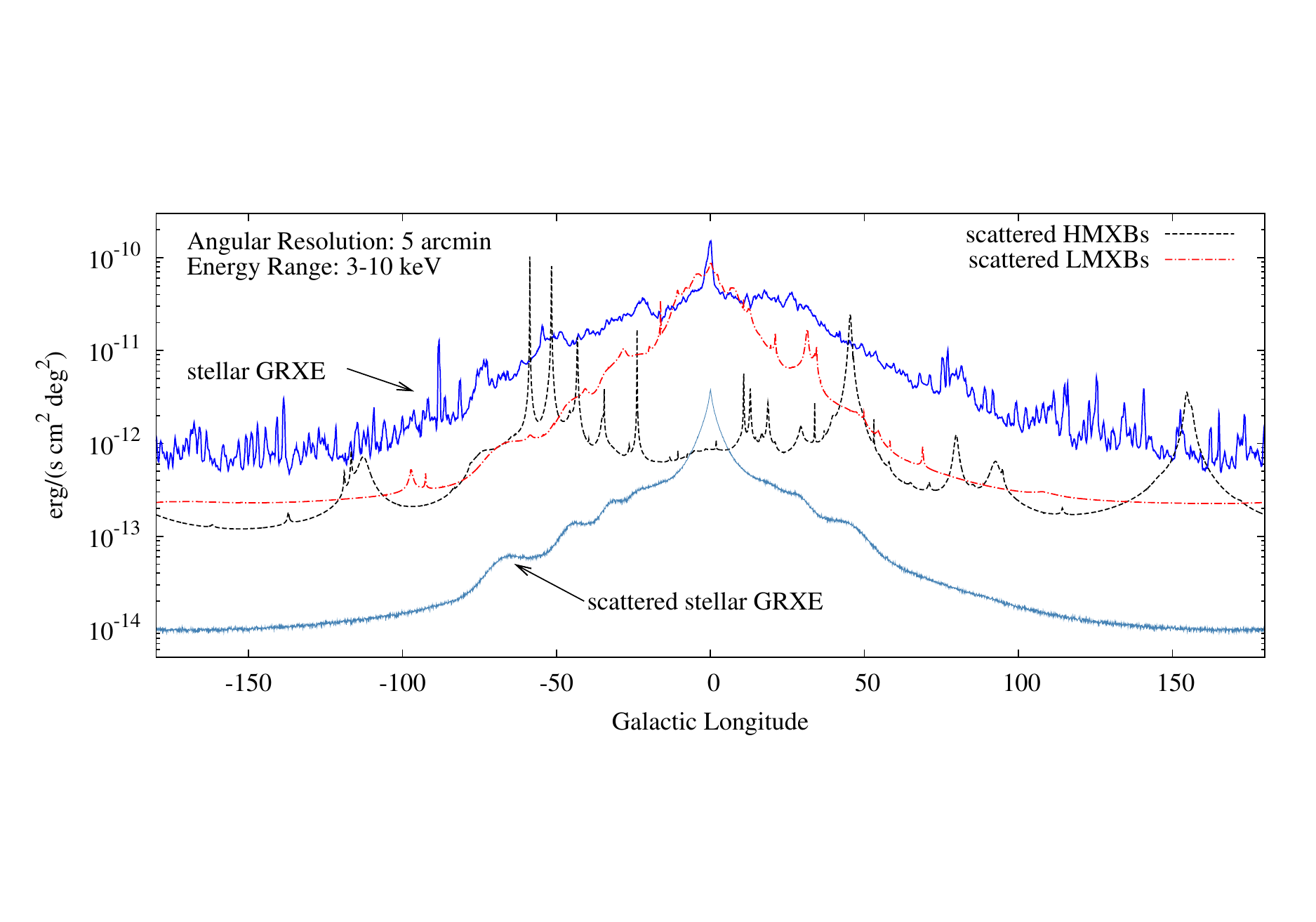} \\
\includegraphics[trim = 40 80 40 50,height = 5.5 cm, width = 14 cm] {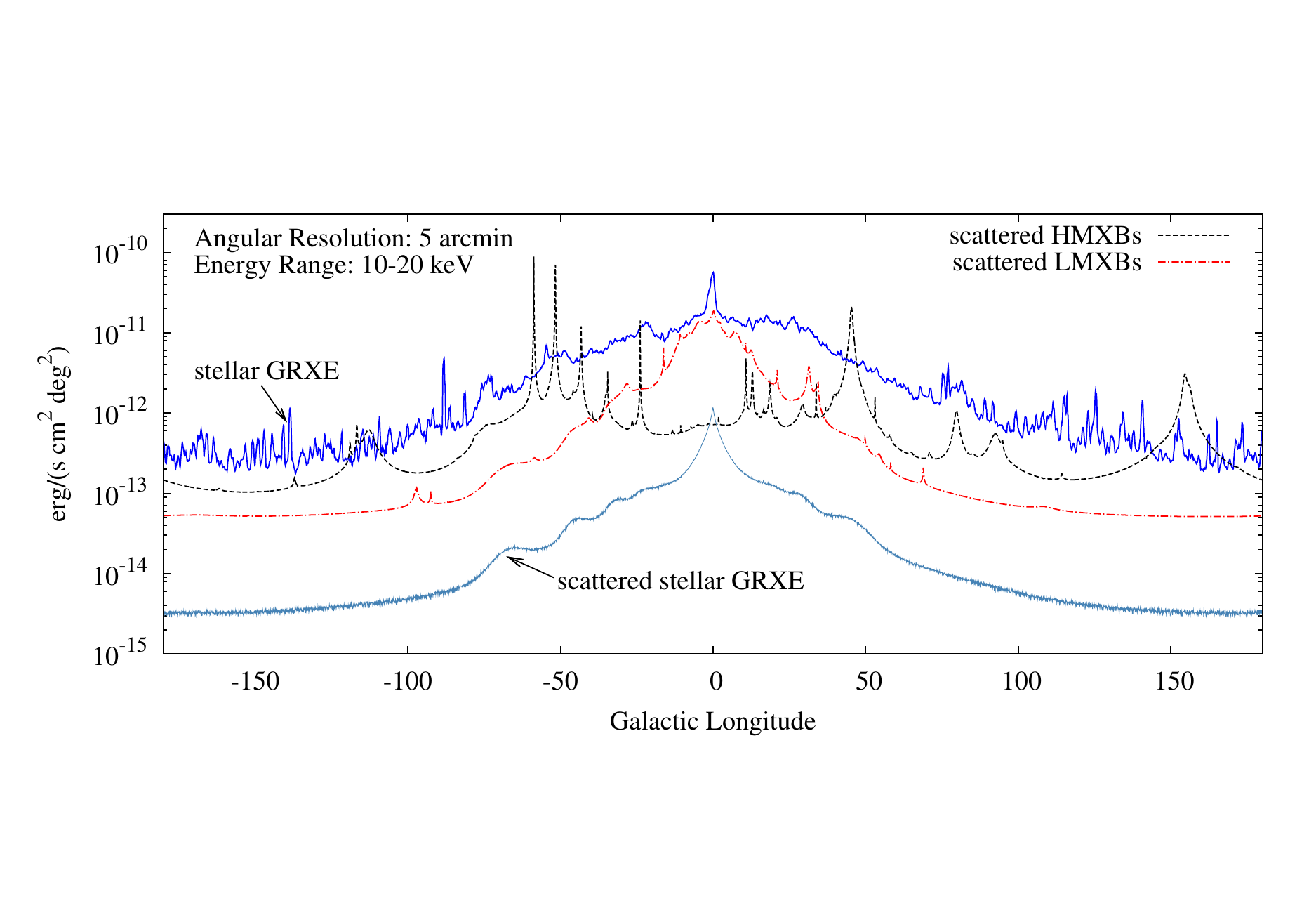} \\
       \caption{Longitude profiles of scattered \ct{XB} radiation in comparison
         with \ct{near-infrared-inferred} GRXE profile\ct{,} in energy ranges 3-10 and 10-20 keV
         for Monte Carlo sources. If our Galaxy was on average more
         luminous in X-rays \changeRV{during the last several thousand years
 compared \ct{with} the present\ct{,} then \changeRE{the scattered component of the
GRXE could be as bright as the stellar component}. In reality\ct{,} the scattered
component \ct{strongly depends} on the appearance of  just a few new bright sources
with luminosities close to the Eddington limit for a neutron star}.}
       \label{Long_prof_MC_sources}
\end{figure*}
\begin{figure*}[!ht]
\centering
\includegraphics[trim = 50 70 50 70, height = 4.5cm, width = 13 cm] {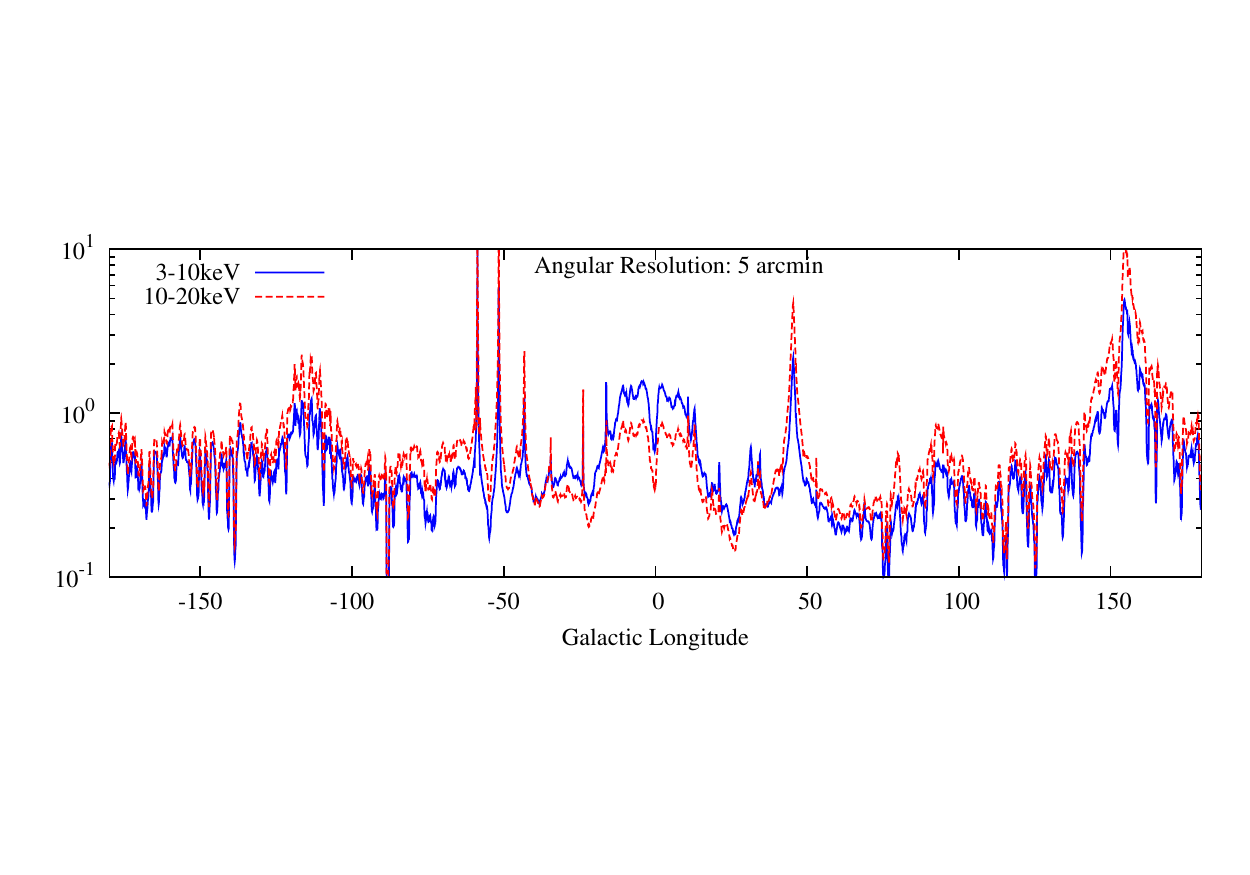}
       \caption{Ratio of scattered X-ray radiation (including HMXBs, LMXBs\ct{, and near-infrared-}inferred low-luminosity sources) to the \ct{near-infrared-inferred} GRXE
         profile for the Monte Carlo sources.}
       \label{Long_prof_ratios_MC}
\end{figure*}
To consider the possibility that \ct{the} current data significantly
  \ct{underestimate} the average total luminosity  of all the X-ray sources in
  the Galaxy, we use the Monte Carlo population of sources described below in
  the 3-20 keV range. The 2-10 keV luminosity of LMXBs, as discussed earlier, is
  $4.7\times 10^{39}\ergss$\ct{, which is a factor  of two higher than that of current}
data. The 2-10 keV luminosity of HMXBs is $1.7\times 10^{39} \ergss$\ct{,} which is
almost a factor of 50 \ct{higher} than what is currently observed. Compared \ct{with}
\changeRR{other} galaxies, our Galaxy is therefore significantly \ct{underluminous}
see e.g. \citet{grimm},\citet{lehmer2010},\citet{mineo2012} and \citet{Mineo2013}) even if we take a conservative star formation
rate of $1M_{\odot}/{\rm yr}$. The results for scattered emission \ct{are} shown
in Fig. \ref{Long_prof_MC_sources} in the form of longitude profiles at
latitude $b=0^{\circ}$. As expected, in this case the scattered X-rays from the XBs
account for almost all of the GRXE in the Galactic plane. This is
\ct{clearer} in the ratio
of total scattered to observed GRXE  shown in
Fig. \ref{Long_prof_ratios_MC}. \changeRE{Furthermore, \changeRB{recent} estimates of the SFR in the Galaxy \citep{Chomiuk2011} seem to suggest a higher SFR ($\sim 1.9 M_{\odot}/{\rm yr}$) than assumed in this work, which would imply an even higher discrepancy between the observed and expected luminosity of HMXBs. The contribution of scattered HMXBs could therefore be even higher than predicted by our simulated sample.}\par {The observation of the diffuse
  component of the GRXE therefore provides a test of the hypothesis that
  our Galaxy is on average significantly more luminous \ct{than} the
  current
  observations of the X-ray sources in the Galaxy suggest.}

\subsection{Comparison with the deep survey region of \citet{Revnivtsev2009}}
The results of previous sections may seem to contradict the results of
\citet{Revnivtsev2009}\ct{,} who resolved $80\%$ of the GRXE into point
sources. \ct{However, their field of view was located at latitude $b=-1.42^{\circ}$}. As
discussed earlier, the scale height of the ISM is much \ct{lower} ($\sim
0.6^{\circ}$)\ct{,} and therefore we expect the  diffuse component of the GRXE to
be significantly smaller at this latitude. We plot the latitude profile
at $l=0.08^{\circ}$ corresponding to the observation direction of
\citet{Revnivtsev2009} and compare the GRXE profile inferred from \ct{near-infrared}
data (see Fig. \ref{resolvedfov}). The scattered X-rays at $b=-1.42$ make up only $5\%$ of the
GRXE. Even in the optimistic case considered in the simulated population of sources, when the X-ray luminosity of the Galaxy
is much higher, the diffuse contribution is $\sim 26\%$. There is
  a coincidence here that a luminous \ct{Monte Carlo} source lies nearby\ct{,}
  giving \ct{a higher} than average flux in this direction: {the contribution} on the other side of the Galactic plane at $b = 1.42 ^\circ$ is $\sim 16\%$. This is consistent with the results of
  \citet{Revnivtsev2009}. Therefore a significantly higher X-ray luminosity
of Milky Way is allowed by current observations. \ct{The hypothesis} that the
average X-ray luminosity of our Galaxy is significantly higher \ct{than} indicated by current data can be tested by studies similar to \ct{that of}
\citet{Revnivtsev2009}\ct{,} but \ct{conducted} in the Galactic plane. \par Recently\ct{,}
  \citet{Morihana2012} and \citet{Iso2012} have claimed that the resolved
  fraction in this field of view may be closer to $50\%$\ct{,} allowing for a
  significantly larger contribution from the diffuse
  component. \ct{The a}nalysis of Suzaku satellite data by \citet{Uchiyama2013} also
  allows for a contribution from the scattered radiation to the GRXE consistent with our predictions for the simulated optimistic case.
\begin{figure}[!ht]
\centering
     \includegraphics[trim = 10 0 10 0,clip,height = 8cm, width = 9 cm]{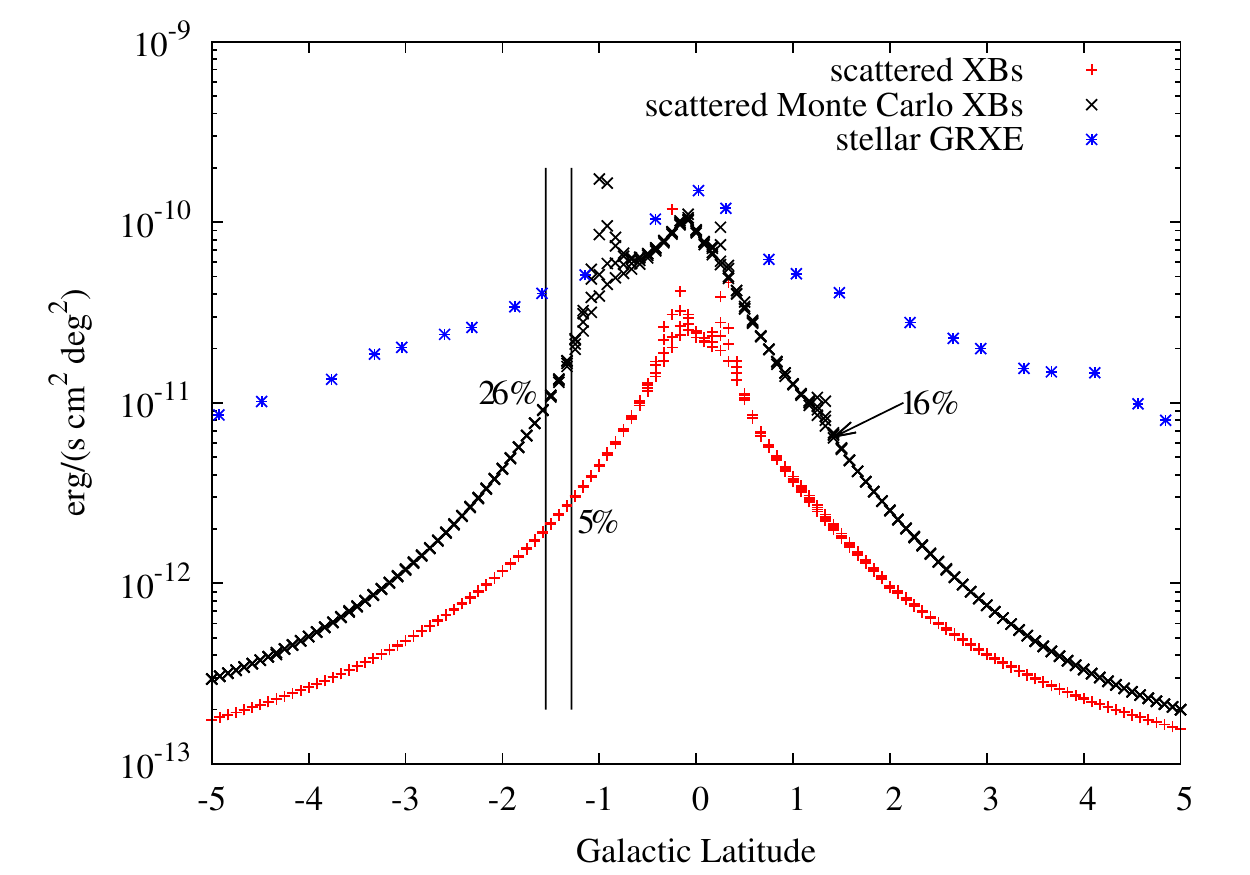} \\
       \caption{Latitude profiles of scattered \ct{XB} radiation in comparison with the \ct{near-infrared-inferred} GRXE profile in the energy range 3-10 keV for the field of view of $16 \times 16$ arcmin centered at $l = 0.08^\circ$, $b = -1.42^\circ$. In this region between 80 and 90 \% of the GRXE has been resolved into point sources \citep{Revnivtsev2009}. Percentages quoted show the fractional contribution of the diffuse GRXE component in each case. Because of the coincidental presence near this field of view of a luminous simulated source, we also quote the same value on the other side of the Galactic plane in the case of Monte Carlo sources.}
       \label{resolvedfov}
\end{figure}

\subsection{Effect of \ct{the} compactness of molecular clouds on the morphology
of \ct{the} diffuse GRXE}
\label{clouds}
\begin{figure*}[!ht]
 \centering
  \includegraphics[trim = 20 70 0 80, clip, width=16cm, height=5cm]{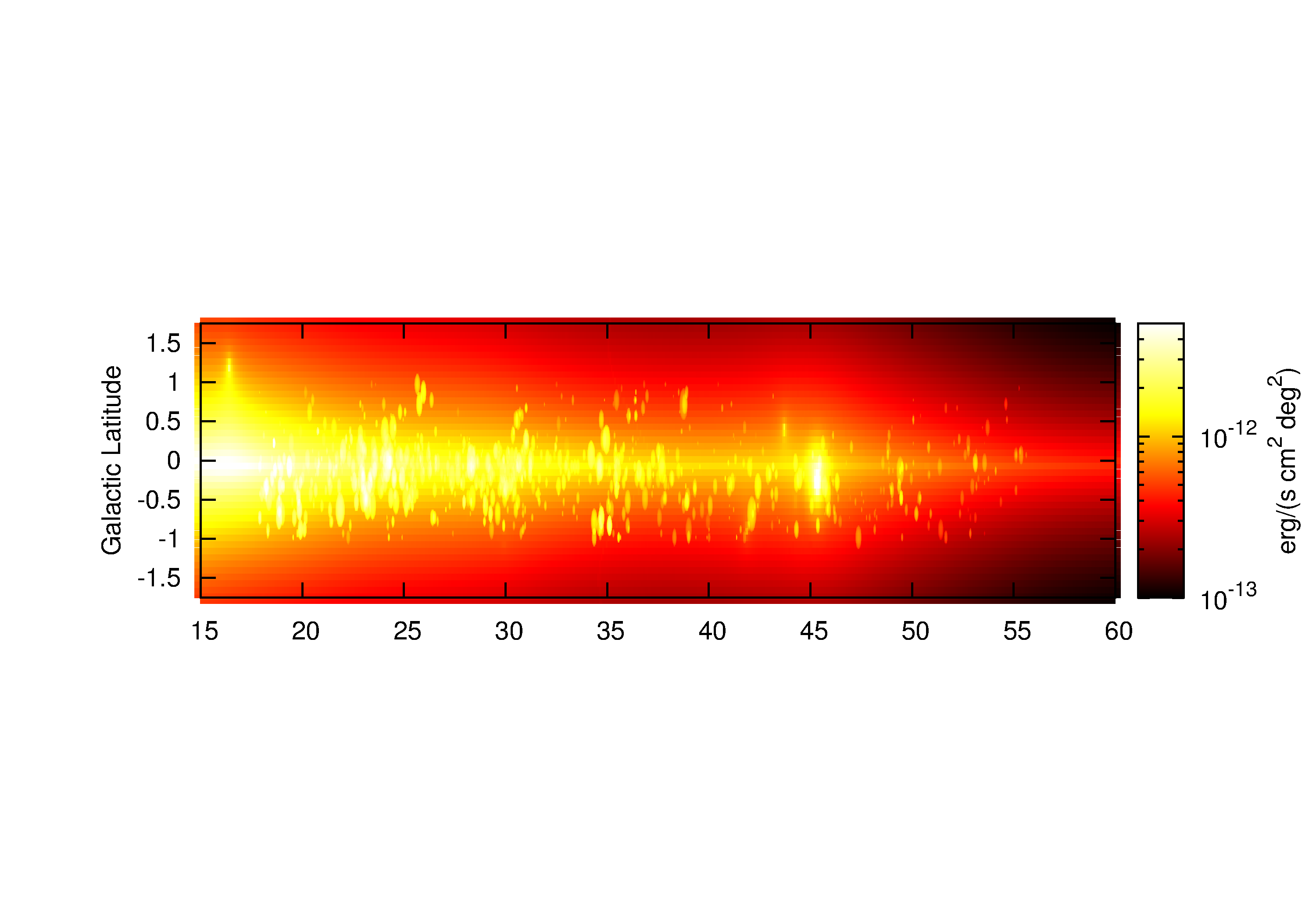}   
  \includegraphics[trim = 20 70 0 80, clip, width=16cm, height=5cm]{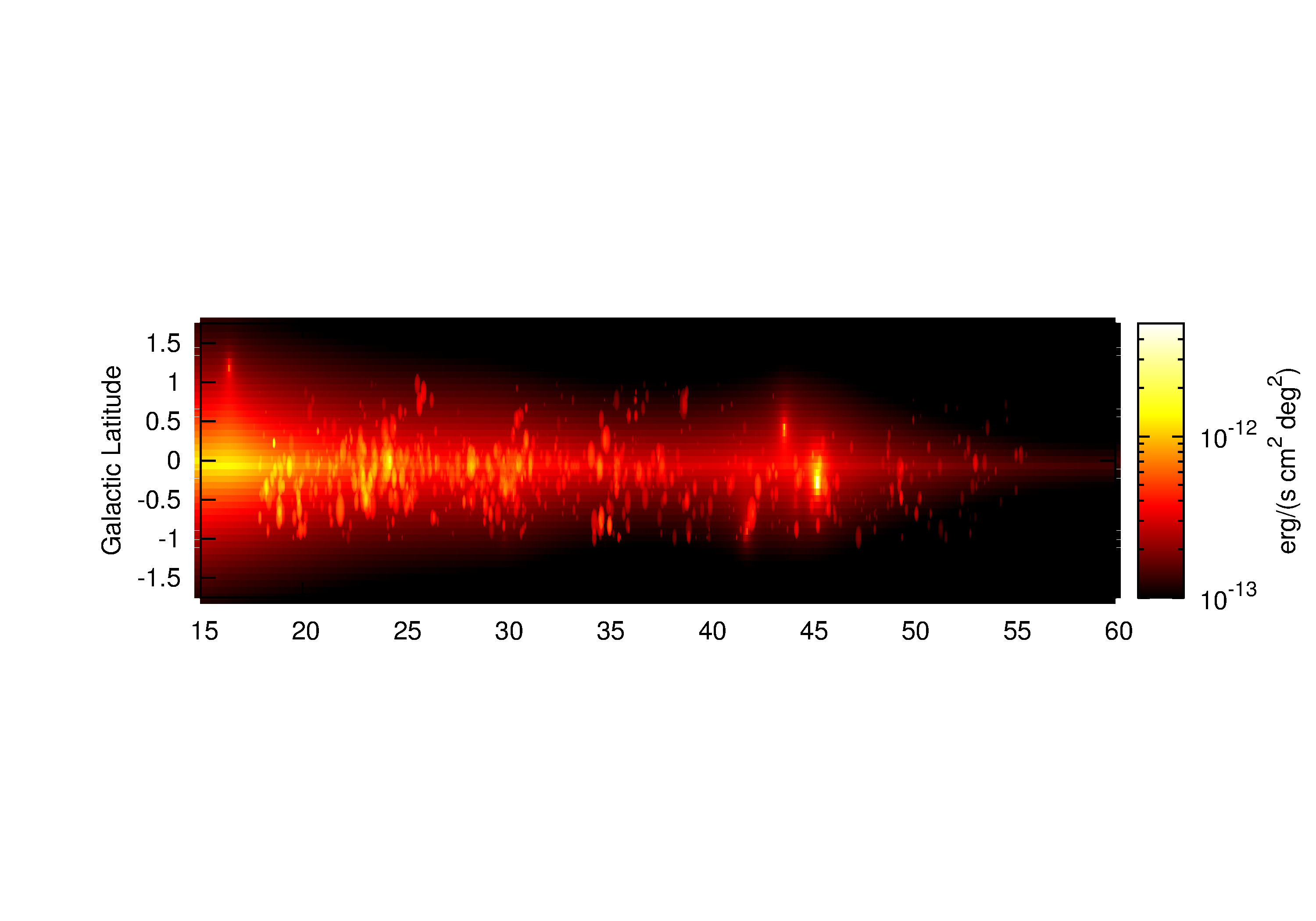} 
   \includegraphics[trim = 20 60 0 80, clip,width=16cm, height=6cm]{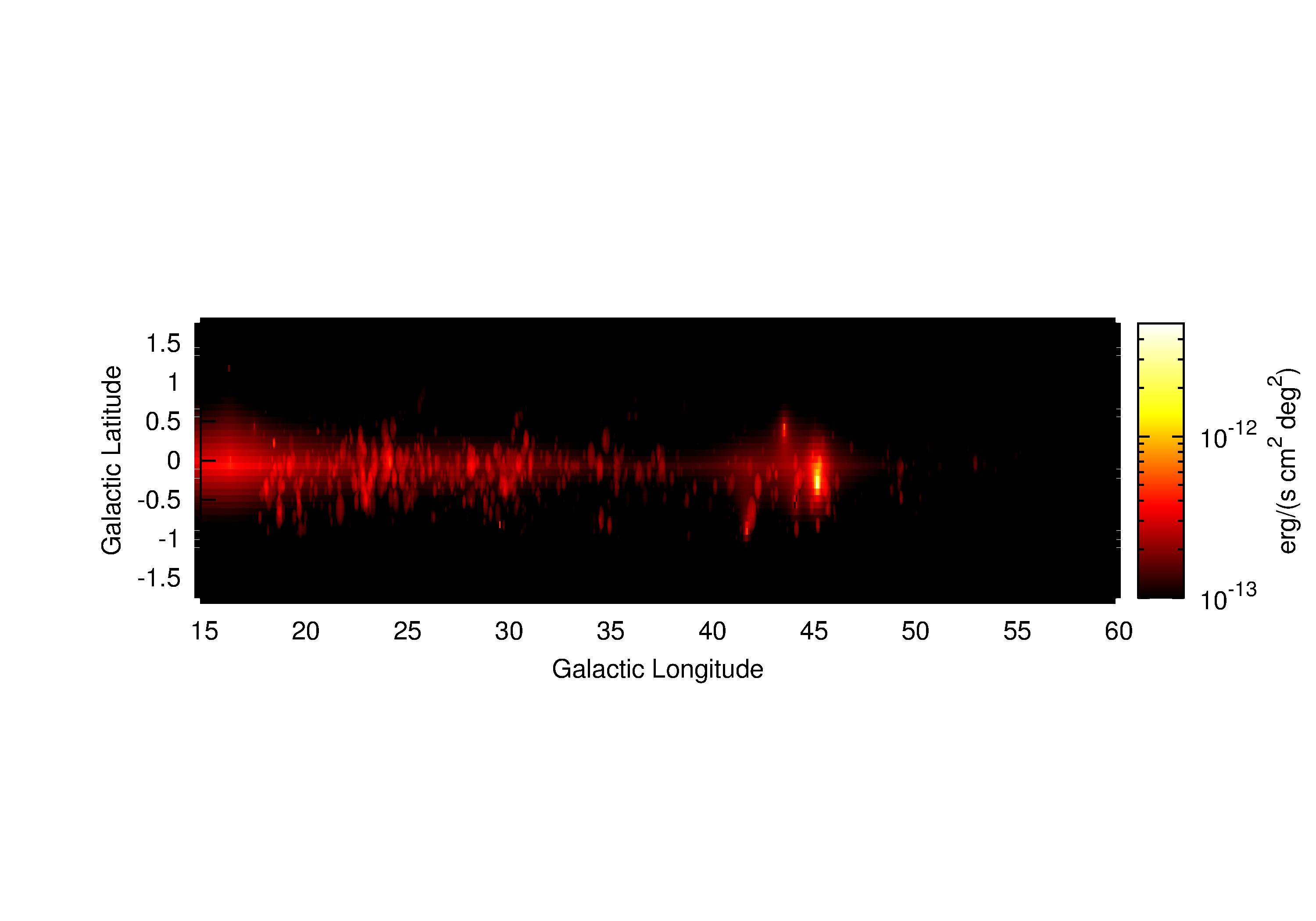} 
\caption{Maps of diffuse X-ray emission from the ISM with HI modeled as a
  smooth disk and H2 using data from the Galactic Ring Survey \changeRE{in the energy ranges 3-10 keV \textit{(top \ct{panel})}, 10-20 keV \textit{(middle \ct{panel})} and 17-60 keV \textit{(bottom \ct{panel})}}. Giant molecular
  clouds are immediately evident.}
        \label{clouds_map}
  \end{figure*}
The \ct{GRS} provides data for the spatial distribution, size and
 average density of $750$ molecular clouds in the Galactic plane in the
 latitude range $-1^{\circ}\le b \le 1^{\circ}$ and the longitude range
 $18^{\circ}\le \ell \le 55.7^{\circ} $. This survey covers a part
   of one of the 
   most significant \ct{concentrations} of molecular gas in the Galaxy \ct{that is }roughly
    distributed in a \ct{ring-like} structure about 4 kpc from the
   Galactic center \citep{stecker1975,cohen1977}\ct{,} although it may just be a
   spiral arm misinterpreted as a molecular ring \citep{db2012}. {We assume
   that the 
 density of each individual cloud is constant throughout the
 object.  The total H2 mass of the clouds in the GRS is  $3.5 \times 10^{7}$ M$_{\sun}$.}
 
 {We  account
for the absorption of X-rays in the molecular cloud  using photoabsorption \ct{cross-sections} from \citet{Morrison}.} \changeRR{Scattering of X-rays from XBs on molecular clouds, which
contain only a fraction of the total gas, gives a significant but patchy contribution to
the GRXE}. {Part of the reason \ct{for this }is that molecular clouds are dense and
occupy a small volume (and angular area on the sky). Locally,
along the lines of sight \ct{that intersect} molecular clouds, they dominate 
the much smoother contribution from HI.  In addition\ct{,} there is a} factor of 2 enhancement over atomic
hydrogen (with same number density of H atoms) because of Rayleigh
scattering {for small scattering angles.} We show the results after including the molecular clouds  in the \ct{GRS} instead of smooth
distribution of molecular gas in  Fig. \ref{clouds_map}. The molecular clouds are very obvious in these maps. This is
even \ct{clearer} in the   latitude profiles shown in       
 Fig. \ref{clouds_prof}  {along the direction tangent to the
   Scutum-Crux spiral arm of the Milky Way and the longitude profiles shown in Fig. \ref{long_clouds_prof}. Even at relatively high
   latitudes, the molecular clouds are able \ct{to significantly contribute, locally,} to the GRXE. } This suggests that at high angular resolution, where we can
resolve the molecular clouds, there should be significant fluctuations in
the diffuse component of the GRXE\ct{,} which would encode information about
the distribution of molecular clouds in the Galaxy. The comparison of their
contribution to the \ct{near-infrared-inferred} GRXE is given in
Fig. \ref{clouds_ratio}. {The molecular clouds are evident as 
narrow peaks on small scales, in addition to broader  peaks from the
presence of luminous X-ray sources, and locally they \ct{are responsible for an increase of a} factor of
more than 2
in the diffuse component of GRXE compared \ct{with} the average background.
\newline
We compare \ct{in Table \ref{tableGMC} }the results for the scattered flux from molecular clouds with
the \changeRV{Advanced Satellite for Cosmology and Astrophysics (ASCA)  \changeRR{\citep{Tanaka1994}}} measurement of X-ray flux in the field of view of giant molecular
clouds (GMC) \citep{cramphorn}. The ASCA
measurements provide an upper limit to the X-ray flux that \ct{might} be
produced by the scattering process.  The scattered X-ray component is well
below the limit set by ASCA. The X-ray telescopes Chandra and  XMM-Newton
  (and Athena+ in the future) are significantly more sensitive than ASCA\ct{,}
  which observed the giant molecular clouds in the Galactic plane for \ct{a relatively short time}.
}
\begin{figure}[!ht]
 \centering
  \includegraphics[trim = 10 0 0 20, width=9cm, height=8cm]{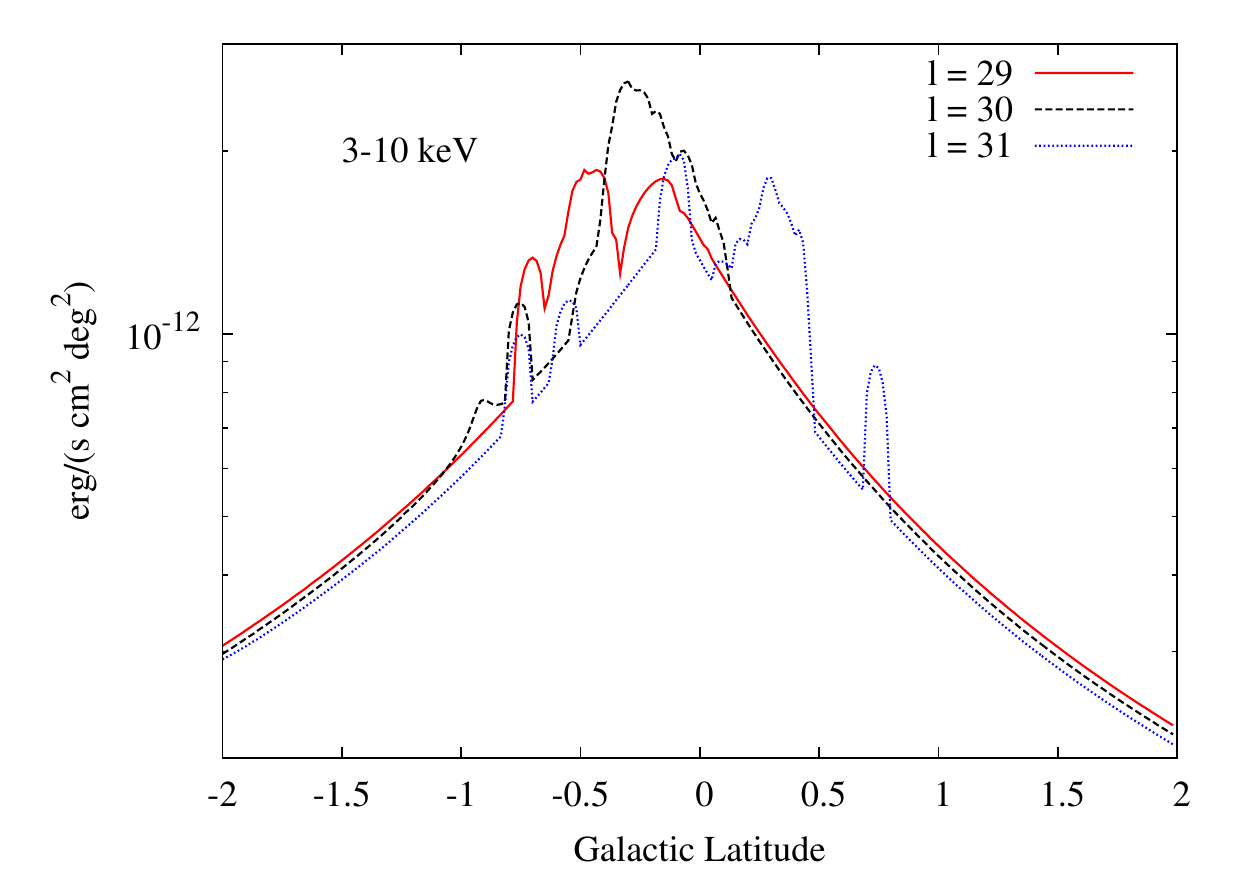}   
        \caption{Latitude profile of \ct{scattered XB} radiation near the line of sight
          tangential to one of the spiral arms {(Scutum-Crux). The
          lines of sight passing through giant molecular clouds can \ct{receive a
          significantly higher  contribution (than the average
          background)} from the scattered X-rays.}}        
\label{clouds_prof}
  \end{figure}

\begin{figure}[!ht]
 \centering
  \includegraphics[width=9cm, height=8cm]{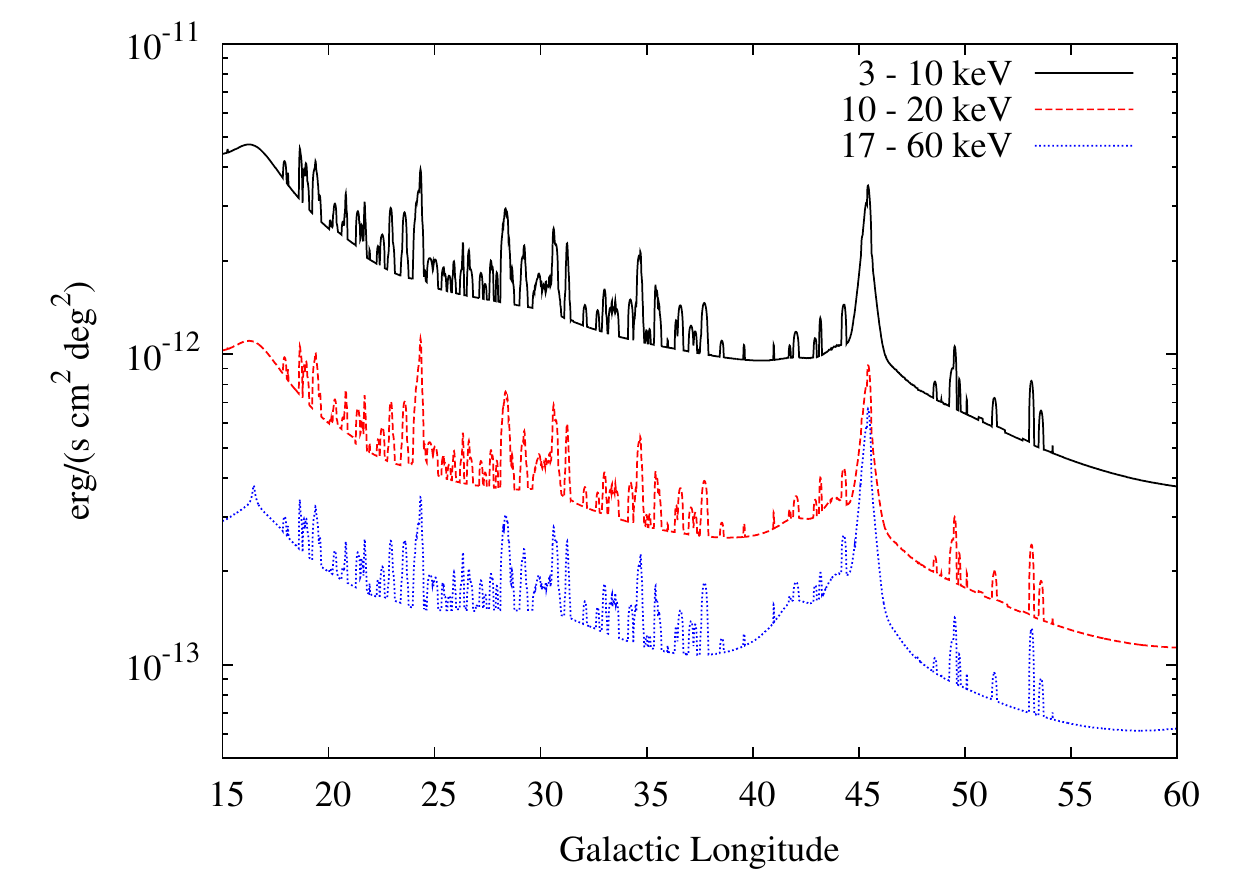}   
        \caption{Comparison of longitude profiles of scattered \ct{XB} radiation in different energy ranges including effects of molecular clouds \changeRR{(visible as narrow peaks in the profiles)} for the energy ranges (from top to bottom) 3-10 keV, 10-20keV, 17-25keV and 25-60 keV.}
        \label{long_clouds_prof}
  \end{figure}

\begin{figure}[!ht]
 \centering
  \includegraphics[trim = 10 0  2 20,width=9cm, height=8cm]{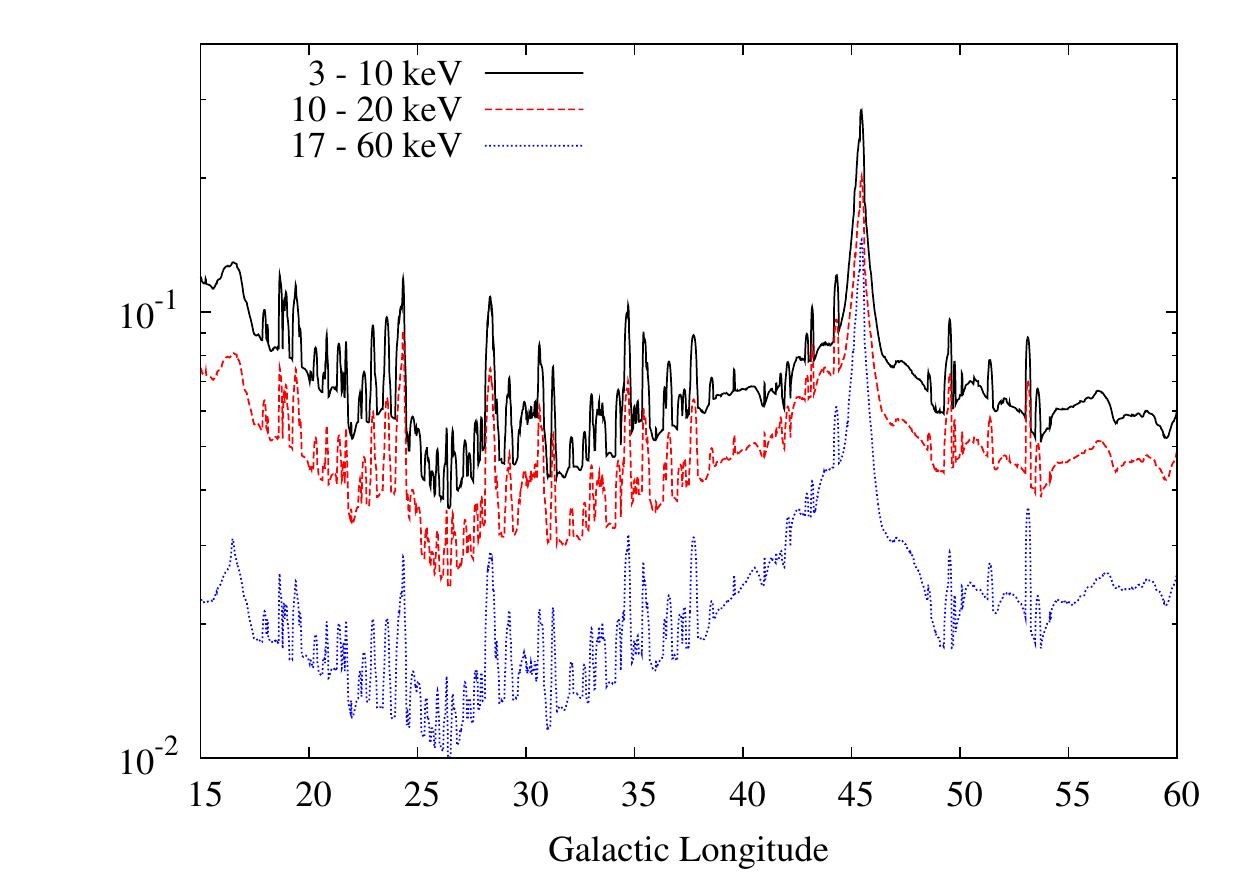}   
        \caption{Ratio of longitude profiles of scattered \ct{XB} radiation in different energy ranges including effects of molecular clouds \changeRR{(visible as narrow peaks in the profiles)} to the \ct{near-infrared}-inferred GRXE emission for the energy ranges (from top to bottom) 3-10 keV, 10-20keV, 17-25keV and 25-60 keV. }
        \label{clouds_ratio}
  \end{figure}

%
\begin{table}[!ht]
\caption{Comparison of XB scattered flux with ASCA measurements of 4-10keV flux in the field of view of 20 GMC}
\label {tableGMC}
 \begin{tabular}{ c c  c  c  }
GMC & F$_{{\rm ASCA}}$\tablefootmark{a}  & F$_{{\rm scatt}}$\tablefootmark{b} & F$_{{\rm scatt}}$/F$_{{\rm ASCA}}$ \\
& \textit{(erg/(s cm$^2$))}& \textit{(erg/(s cm$^2$))} & \\
\hline
\hline
 & & & \\
003 & 1.9$\times 10 ^{-11}$& 6.9$\times 10 ^{-13}$ &0.036 \\
014 & 1.5$\times 10 ^{-11}$& 2.1$\times 10 ^{-13}$ &0.014 \\
059 & 1.7$\times 10 ^{-11}$& 3.8$\times 10 ^{-13}$ &0.022 \\
080 & 1.5$\times 10 ^{-11}$& 9.9$\times 10 ^{-14}$ &0.0066 \\
085 & 1.6$\times 10 ^{-11}$& 1.8$\times 10 ^{-13}$ &0.011 \\
089 & 1.6$\times 10 ^{-11}$& 3.0$\times 10 ^{-13}$ &0.018 \\
116 & 1.5$\times 10 ^{-11}$& 6.1$\times 10 ^{-14}$ &0.0041 \\
122 & 1.8$\times 10 ^{-11}$& 2.0$\times 10 ^{-13}$ &0.011 \\
128 & 1.7$\times 10 ^{-11}$& 1.4$\times 10 ^{-13}$ &0.008 \\
151  & 1.4$\times 10 ^{-11}$& 1.6$\times 10 ^{-13}$ &0.011 \\
152  & 1.4$\times 10 ^{-11}$& 6.3$\times 10 ^{-14}$ &0.0045 \\
158  & 1.5$\times 10 ^{-11}$& 1.0$\times 10 ^{-13}$ &0.0067 \\
162  & 1.5$\times 10 ^{-11}$& 1.1$\times 10 ^{-13}$ &0.0071 \\
171  & 1.3$\times 10 ^{-11}$& 2.9$\times 10 ^{-13}$ &0.023 \\
191  & 1.0$\times 10 ^{-11}$& 7.2$\times 10 ^{-14}$ &0.0072 \\
193  & 1.0$\times 10 ^{-11}$& 2.3$\times 10 ^{-13}$ &0.023 \\
201  & 9.1$\times 10 ^{-11}$& 2.9$\times 10 ^{-13}$ &0.0031 \\
206  & 1.0$\times 10 ^{-11}$& 1.6$\times 10 ^{-13}$ &0.016 \\
214  & 7.7$\times 10 ^{-12}$& 3.1$\times 10 ^{-13}$ &0.04 \\
217  & 8.6$\times 10 ^{-12}$& 1.1$\times 10 ^{-13}$ &0.012 \\
\end{tabular}
\newline
\tablefoottext{a}{Flux as measured by ASCA \changeRIV{(upper limit)}}
\tablefoottext{b}{XB scattered flux in the field of view of the cloud}
\end{table}

\section{Contribution of LMXBs and HMXBs to the X-ray heating of the
  ISM}
\begin{figure*}[!ht]
\centering
\includegraphics[trim = 10 10 0 10,height = 8cm, width = 8 cm]{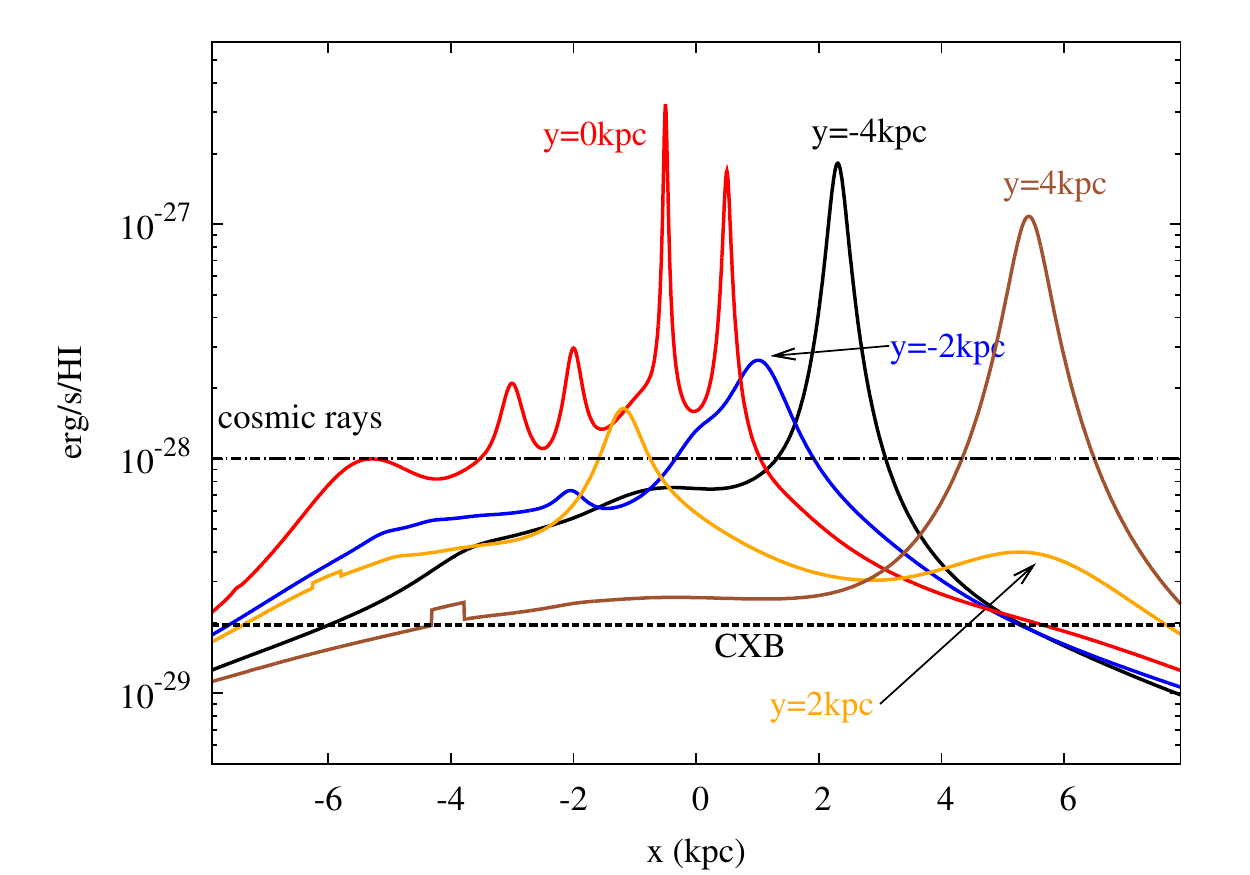}\includegraphics[trim = 10 10 0 10,height = 8cm, width = 8 cm]{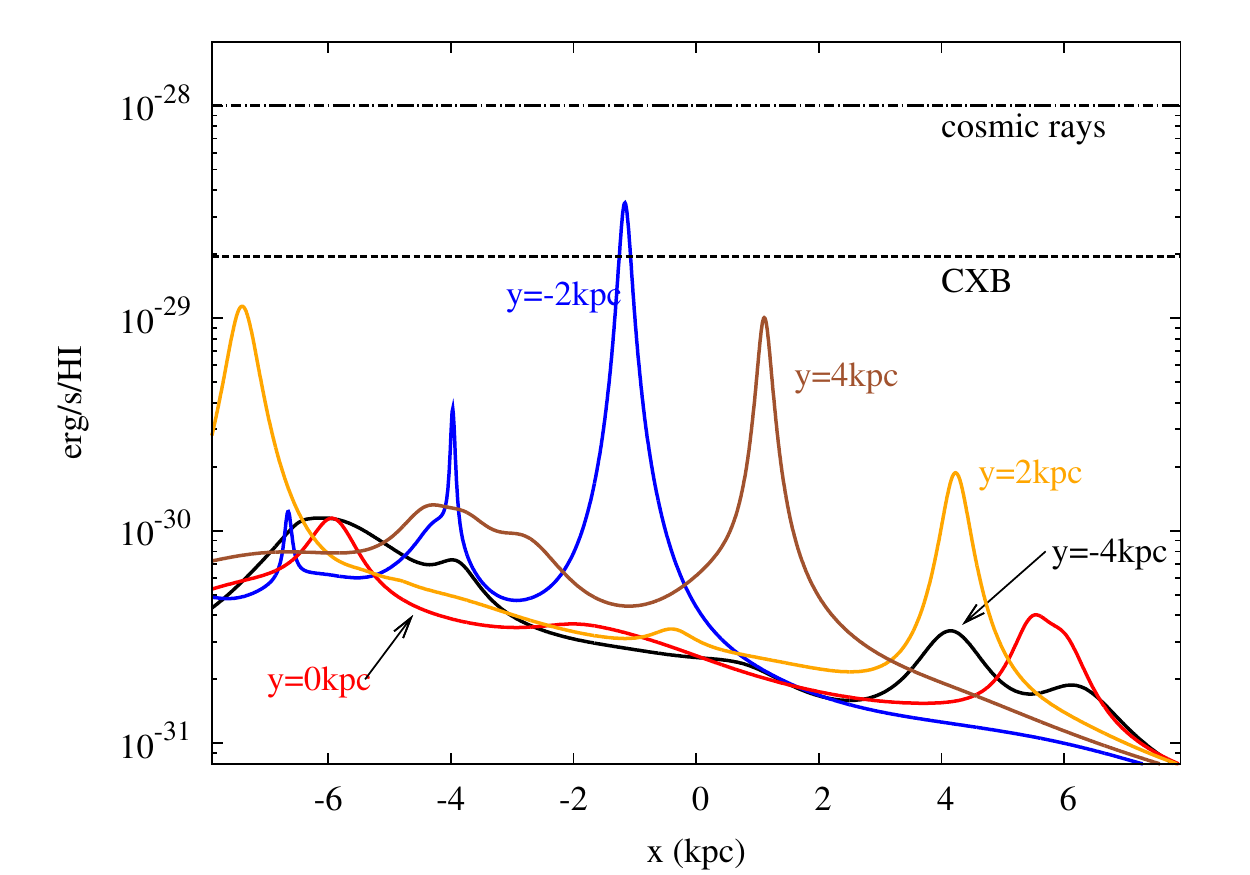}\\ 
\caption{Heating of the ISM per hydrogen atom from photoabsorption of
  X-rays from LMXBs \textit{(left panel)} and HMXBs  \textit{(right panel)}. For comparison the contribution from 
cosmic rays  following \ct{\citet{ppim} and \citet{Wolfire2003}} and CXB is also shown. \ct{The} X-ray
($\gtrsim 1$ keV)
contribution dominates compared \ct{with} the  cosmic rays (assuming \ct{the }same energy
density as the solar neighborhood) in a significant part of the Galaxy.}
\label{heating_map}
 \end{figure*}
\begin{figure}[!ht]
\centering
\includegraphics[trim = 20 40 30 50,clip,height = 7cm, width = 9 cm]{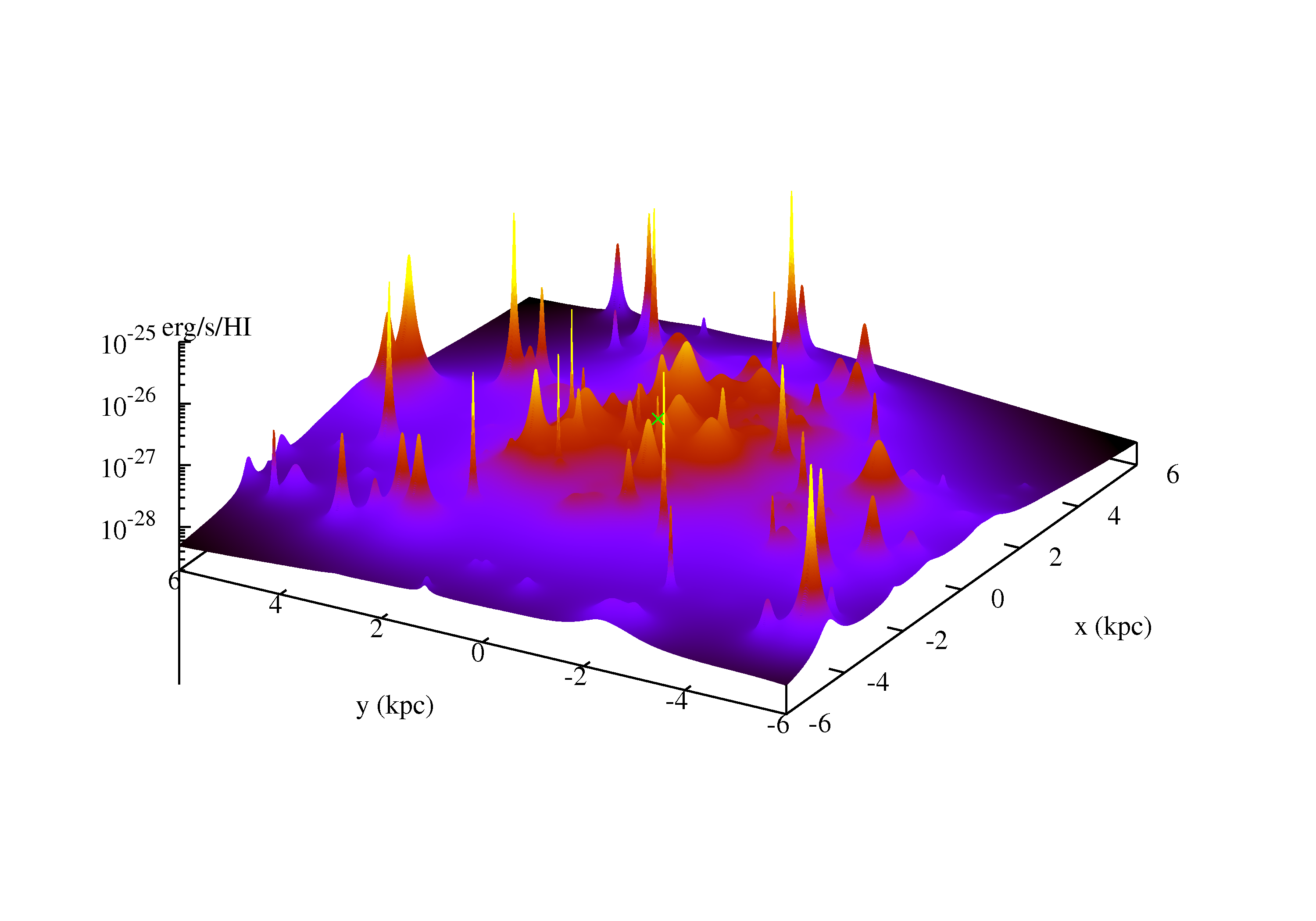} \\ \includegraphics[trim = 20 0 0 40,clip,height = 9cm, width = 10 cm]{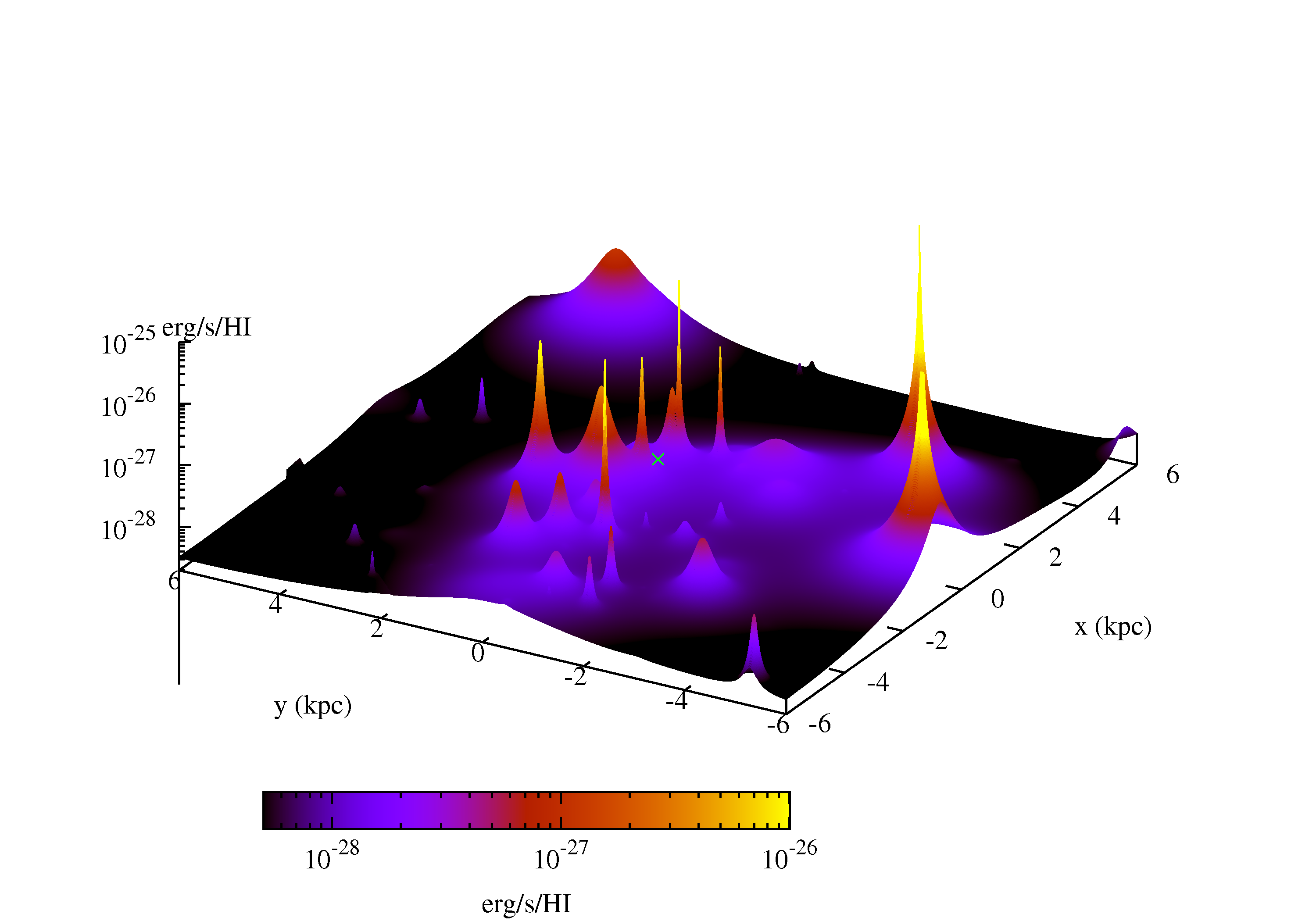} 
\caption{\ct{Three-dimensional} maps of the heating of the ISM per hydrogen atom from photoabsorption of
  X-rays from LMXBs, HMXBs and CXB on the Galactic plane for Monte Carlo sources (\textit{top \ct{panel}}) and real sources (\textit{bottom \ct{panel}}). \changeRR{In the outer regions of the 
Galaxy the heating rate approaches the lower bound of $\sim 1.9 \times 10^{-29} $
ergs/s/HI coming from absorbed CXB photons}. The green cross indicates the position of the Galactic center in each map.}
\label{heating_3dmap}
 \end{figure}
The competition between the different heating  (e.g. from irradiation,
cosmic rays, turbulent dissipation) and
cooling processes (e.g.  atomic lines such as CII,OI, Ly-{$\alpha$}) plays
a crucial role in determining the multiphase structure of the interstellar
medium, formation of molecular clouds and star formation
\changeRR{\citep{Pikelner1968,Field1969,MO1977,Wolfire1995,ppim,cox2005,snow2006,McKee2007}}. The contribution of soft X-rays
with energies $\lesssim 1$ keV
has been taken into account  in standard calculations
\citep{Wolfire1995,Wolfire2003} assuming a quasi-homogeneous (smoothly varying with
distance from the Galactic center) X-ray background. The 
heating from the soft X-rays\ct{, however, is} confined to the outer layers of the
clouds where they are almost completely absorbed. The harder X-rays at
energies $\gtrsim 1~{\rm keV}$ can\ct{, on the other hand,} penetrate the interiors of the
HI and molecular clouds. \changeRR{ X-ray binaries are the strongest
sources of X-rays at these energies in galaxies similar to the Milky way, and should therefore have a significant  effect on the
structure and chemistry of the ISM. The contribution to the ISM heating from these sources should be highly variable both in space, for the obvious reason of the rapid decrease of the source's X-ray flux with distance, and in time, \ct{because of }the transient nature of
most X-ray binaries. }

The energy deposition rate from X-rays at a point in the Galaxy is given by 
\begin{equation}
\label{Eq::heating}
\resizebox{.9\hsize}{!}{$\Gamma_x=  \sum_{i} \frac{L_{i}(\nu)}{4\pi R^2_i} \bigg[\sigma_{{\rm abs}}(\nu) -  \sum_{Z}\frac{n_Z}{n_H}\frac{E_{{\rm K-}\alpha,Z}}{E_{\nu}}Y_{Z}\times \sigma_{nl,Z}(\nu) \bigg]$,} 
\end{equation}
where the sum \ct{computed} is over the  X-ray sources $i$, $R_i$ is the distance of the source to the point of absorption and $L_i$ \ct{is} its luminosity, $\sigma_{{\rm abs}}$ is the photoelectric absorption \ct{cross-section} given in \citet{Morrison}, the sum $Z$ over heavy elements includes O, Fe, C, Ne, Si, Mg, $\sigma_{{\rm abs},Z}$ is the analytic fit to the partial photoionisation \ct{cross-section} of the $n=1,l=0$ state of element $Z$ as given in \citet{Verner}, $Y_Z$ is the K-$\alpha$ yield, $E_{{\rm K-}\alpha,Z}$ is the energy of the K-$\alpha$ line\ct{,} and $n_Z/n_H$ is the relative abundance for solar abundance estimates considered in \citet{Morrison}. Above 10 keV we use the partial K-shell \ct{ionization cross-sections} of Fe and Ni, which dominate 
the absorption \ct{cross-section} at these energies. \par
We plot the energy deposited by photoabsorption of X-rays in
Fig. \ref{heating_map} as profiles along the x-axes for different cuts
along the y-axes in the Galactic plane, with coordinates for the Galactic
center at $(x,y)=(0,0)$ and position of the Sun at $(x,y)=(-8,0)$ kpc. 
Not all the energy absorbed from the X-rays will go into heating the
clouds, \changeRR{as} some of it will be radiated away as \ct{low-energy} photons coming from
excitations and \ct{ionization} and subsequent recombinations of atoms
\citep{shull1985}. The conversion of the X-ray energy into heat will in
particular be most efficient if the \ct{ionization} fraction is large so that
most of the initially absorbed energy is dissipated via distant \ct{electron-electron}
collisions \ct{and not} through
excitations.  In addition, most of the initially absorbed X-ray energy
(minus the energy \ct{that escapes via the} fluorescent K-$\alpha,\beta$ lines of Fe and Ni)
 will  be
converted  \ct{in} heat if the hydrogen column density or the dust opacity is so \ct{high} that photons with
energy Ly-$\alpha$ or \ct{higher} ($\gtrsim \changeRV{10.2 ~{\rm eV}}$) cannot escape. For
the discussion below we assume, for the sake of comparison, that one or
more of these conditions holds and that most of the X-ray energy is absorbed. In
reality\ct{,} the fraction will vary significantly depending \ct{upon} the local
conditions \citep{shull1985}.

For
reference we plot the heating rate from CXB using the spectral fit
of \citet{cxbspe} and integrating up to 60 keV, although the contribution
from energies $\gtrsim 10~{\rm keV}$ is negligible. In addition to the
photoabsorption, hard X-rays also transfer energy to the gas when they
\ct{Compton-scatter}
on electrons, both free and bound in hydrogen and helium, through the recoil
effect. The average energy transferred by a photon of energy $E$ is $E\times
E/(m_{\rm e}c^2)$, where $m_{\rm e}$ is the mass of the electron and $c$ is
the speed of light in vacuum \citep[see e.g.][]{pss1983}.   The average
energy deposition  rate is therefore given by $\Gamma_{\rm recoil}=\int
{\rm d}
E F_{\rm X}\sigma_{\rm KN} E/(m_{\rm e}c^2)(1-2.2 E/(m_{\rm e}c^2))$, 
where $F_{\rm X}$ is the
specific X-ray flux in units of (${\rm ergs/s/cm}^2{\rm /keV}$) and 
$\sigma_{\rm KN}$ is the
Klein-Nishina \ct{cross-section;} we include the \ct{lowest-order} 
Klein-Nishina correction to the average recoil energy. For solar metallicity, 
taking $\log \text{Fe}=7.52$
normalized to $\log
\text{H}=12$ used in \citet{Morrison}, the heating from recoil
  dominates the heating from photoabsorption at $\approx 28 ~{\rm
    keV}$. For CXB the contribution of \ct{the} recoil effect to the heating rate is
  $\approx 4\times 10^{-32}\ergss/{\rm electron}$, counting free electrons
  as well as those bound in hydrogen and helium and in the outer shells of
  heavier elements. \ct{In general, therefore,} we expect the recoil effect to be
 negligible, \ct{but it might} become important in the vicinity of the hard
 X-ray sources. {The CXB, in contrast to X-ray binaries, \ct{uniformly heats
   the intergalactic medium  everywhere}. The  energy density in CXB peaks
   around $z\sim 0.7-1$ \citep{cxb1,cxb2,cxb3} and its contribution \ct{is}
   therefore more important in high redshift galaxies.} 

An
important point to note is that the area of influence of luminous sources
extends quite far, up to  a few kpc away from the source for the most luminous
sources. These heating rates \ct{need to} be compared with the heating rate from
the cosmic rays estimated to be, \changeRR{on average in the solar neighborhood},
$\Gamma_{CR}\approx 10^{-28}$ erg/s/HI and also with typical heating rates
from photoelectric heating by dust grains. The X-ray heating is
comparable \ct{with} and even stronger than the cosmic ray heating \changeRR{(assuming \ct{a} solar neighbourhood energy density)} in a significant
part of the Galaxy and may be \ct{higher than} other heating
 mechanisms \ct{typically} considered. {The optimistic case, where the
X-ray luminosity of the Milky Way is scaled to match the activity in the
other galaxies\ct{,} is shown in Fig. \ref{heating_3dmap} using the
Monte Carlo sources. The X-ray heating dominates over the solar neighbourhood energy density of cosmic rays in
almost the whole Galaxy and if Milky Way activity in the past was close to
this level, X-ray heating would have played an important role in the
energetics of the different phases of the ISM. The \ct{star-forming} galaxies have \ct{a} much higher
X-ray luminosity \ct{than} even this optimistic case for the Milky Way and the X-ray heating 
therefore \ct{very likely plays}
a crucial role in the evolution of their ISM.}

\section{Conclusions}
Weak X-ray sources such
as \changeRV{cataclysmic variables (CVs)} and coronally active binaries are currently believed to be the main
contributors to the GRXE. We investigate the scattering by the interstellar medium of X-rays
produced by Galactic X-ray sources as a possible truly diffuse component of
the GRXE emission.  {We include contributions from coherent
  scattering from molecular hydrogen, helium and heavy elements in addition
  to the atomic hydrogen. The scattering of X-rays from luminous sources
  results in bright regions around the sources, the proximity zones,
  similar to the dust halos\ct{,} but extending to much larger scattering angles
  \ct{than}  the  $\lesssim
  1^{\circ}$ for dust halos.  The scattered photons from the ISM originating in
  the known X-ray binary sources and stellar
  sources in the
  Galaxy are able to \ct{contribute}, on average, $10-30\%$ of the flux currently
  identified as GRXE in the Galactic plane. The profile of the scattered
  intensity follows the distribution of gas in the Galaxy with a scale
  height of 80 pc ($0.6^{\circ}$ at 8 kpc distance)\ct{,} which is much sharper
  than the profile of stellar distribution with a scale height of $\sim
  130$ pc. Therefore we expect the stellar point sources to dominate away
  from the Galactic plane\ct{,} while the scattered X-rays would be concentrated
  within the central $1^{\circ}$ of the plane. Molecular clouds \ct{locally
  enhance the scattered intensity}, imprinting their morphology on
  the GRXE. The coherent scattering by electrons in the molecular hydrogen
  operates on \ct{a comparatively wide} range of scattering angles at low energies
  \ct{than at} high energies, making the scattered spectrum softer \ct{than} the incident spectrum (see Appendix \ref{appa}). }

{We also find that the X-rays \changeRE{from} the luminous binary sources
  contribute significantly to the heating rate in the interstellar medium,
  especially in the interiors of the clouds where the UV or softer X-rays
  are unable to penetrate. This important effect has so far been ignored in
  the standard studies of the multiphase ISM
  \citep{MO1977,Wolfire1995,Wolfire2003}. This effect would be even \ct{stronger} at high redshifts ($z\sim 0.7-1$) where the CXB energy density
  peaks and is much higher
  \ct{than} today 
  as well as in \ct{star-forming} galaxies\ct{,} which have \ct{a} much higher X-ray luminosity
  \ct{than} the Milky Way.}

{We only calculate the X-ray flux integrated over particular
  X-ray bands \ct{because} this will be more readily observable in reasonable
  observation time with  telescopes in operation now or expected in the
  near future\ct{,} such as NuSTAR \citep{Nustar}, \changeRR{Spectrum-RG/ART-XC \citep{srg}} and Astro-H \changeRE{\citep{Takahashi2010}}. There is additional information in the spectrum of the
  scattered emission \ct{that} may help \ct{to }distinguish between the scattered GRXE
  and stellar GRXE. Of particular interest in this regard is the emission
  of fluorescent lines of neutral elements such as iron, sulphur and
  silicon. We \ct{have also assumed} that the sources are persistent. The
  transient nature of most X-ray binary sources as well as extreme events
  such as gamma-ray bursts and giant flares from magnetars will leave
  unique signatures in the morphology of the scattered GRXE. We will
  consider these effects in a separate publication (Khatri \& Sunyaev in prep.). }

Compared \ct{with} other galaxies, the Milky Way is
  \ct{underluminous} in X-rays according to current observations. It is
  possible that the Milky Way is just going through a phase of low X-ray
  activity at present but was more luminous in the past. {In particular, if the \ct{time-averaged} luminosity of Milky Way is the same as inferred from averaged relations between X-ray luminosity and star formation rate/stellar mass of \changeRR{other} galaxies, \ct{then the scattered component will on average} contribute to more than 50\% to the GRXE in the galactic plane and would dominate the stellar component in a significant part of the Galaxy.} This hypothesis is
  allowed by current observations and is testable by observations of the
  Galactic plane with high angular resolution and sensitivity, similar to
  the study by \citet{Revnivtsev2009} below the Galactic plane.

\begin{acknowledgements}
We would like to thank Mike Revnivtsev for discussions \changeRE{and Ken Ebisawa, Marat Gilfanov, Christopher Mckee, Jeremiah Ostriker\ct{,} and Scott Tremaine for comments on the manuscript}. This research has made use of the SIMBAD database, operated at CDS, Strasbourg, France.
\end{acknowledgements}

\bibliographystyle{aa}
\bibliography{references}

\begin{thebibliography}{196}
\expandafter\ifx\csname natexlab\endcsname\relax\def\natexlab#1{#1}\fi

\bibitem[{{Asplund} {et~al.}(2009){Asplund}, {Grevesse}, {Sauval}, \&
  {Scott}}]{abundances}
{Asplund}, M., {Grevesse}, N., {Sauval}, A.~J., \& {Scott}, P. 2009, \araa, 47,
  481

\bibitem[{{Baganoff} {et~al.}(2003){Baganoff}, {Maeda}, {Morris}, {Bautz},
  {Brandt}, {Cui}, {Doty}, {Feigelson}, {Garmire}, {Pravdo}, {Ricker}, \&
  {Townsley}}]{Baganoff2003}
{Baganoff}, F.~K., {Maeda}, Y., {Morris}, M., {et~al.} 2003, \apj, 591, 891

\bibitem[{{Bahcall} \& {Soneira}(1980)}]{bahcallsoneira}
{Bahcall}, J.~N. \& {Soneira}, R.~M. 1980, \apjs, 44, 73

\bibitem[{{Bandyopadhyay} {et~al.}(1999){Bandyopadhyay}, {Shahbaz}, {Charles},
  \& {Naylor}}]{1999MNRAS.306..417B}
{Bandyopadhyay}, R.~M., {Shahbaz}, T., {Charles}, P.~A., \& {Naylor}, T. 1999,
  \mnras, 306, 417

\bibitem[{{Barger} {et~al.}(2005){Barger}, {Cowie}, {Mushotzky}, {Yang},
  {Wang}, {Steffen}, \& {Capak}}]{cxb3}
{Barger}, A.~J., {Cowie}, L.~L., {Mushotzky}, R.~F., {et~al.} 2005, \aj, 129,
  578

\bibitem[{{Bassa} {et~al.}(2006){Bassa}, {Jonker}, {in't Zand}, \&
  {Verbunt}}]{2006A&A...446L..17B}
{Bassa}, C.~G., {Jonker}, P.~G., {in't Zand}, J.~J.~M., \& {Verbunt}, F. 2006,
  \aap, 446, L17

\bibitem[{{Baykal} {et~al.}(2002){Baykal}, {Stark}, \&
  {Swank}}]{2002ApJ...569..903B}
{Baykal}, A., {Stark}, M.~J., \& {Swank}, J.~H. 2002, \apj, 569, 903

\bibitem[{Bergstrom {et~al.}(1993)Bergstrom, Suri\'c, Pisk, \& Pratt}]{bergsu}
Bergstrom, P.~M., Suri\'c, T., Pisk, K., \& Pratt, R.~H. 1993, Physical Review
  A, 48, 1134

\bibitem[{{Bhattacharyya} {et~al.}(2006){Bhattacharyya}, {Strohmayer},
  {Markwardt}, \& {Swank}}]{2006ApJ...639L..31B}
{Bhattacharyya}, S., {Strohmayer}, T.~E., {Markwardt}, C.~B., \& {Swank}, J.~H.
  2006, \apjl, 639, L31

\bibitem[{{Binney} \& {Tremaine}(2008)}]{denbin}
{Binney}, J. \& {Tremaine}, S. 2008, {Galactic Dynamics: Second Edition}
  (Princeton University Press)

\bibitem[{{Blitz}(1993)}]{Blitz1993}
{Blitz}, L. 1993, in Protostars and Planets III, ed. E.~H. {Levy} \& J.~I.
  {Lunine}, 125--161

\bibitem[{{Capelli} {et~al.}(2012){Capelli}, {Warwick}, {Porquet}, {Gillessen},
  \& {Predehl}}]{sgr10}
{Capelli}, R., {Warwick}, R.~S., {Porquet}, D., {Gillessen}, S., \& {Predehl},
  P. 2012, \aap, 545, A35

\bibitem[{{Cartwright} {et~al.}(2013){Cartwright}, {Engel}, {Heinke},
  {Sivakoff}, {Berger}, {Gladstone}, \& {Ivanova}}]{2013ApJ...768..183C}
{Cartwright}, T.~F., {Engel}, M.~C., {Heinke}, C.~O., {et~al.} 2013, \apj, 768,
  183

\bibitem[{{Chaty} {et~al.}(2006){Chaty}, {Mignani}, \&
  {Israel}}]{2006MNRAS.365.1387C}
{Chaty}, S., {Mignani}, R.~P., \& {Israel}, G.~L. 2006, \mnras, 365, 1387

\bibitem[{{Chen} {et~al.}(2010){Chen}, {Zhang}, {Torres}, {Wang}, \&
  {Li}}]{2010A&A...510A..81C}
{Chen}, Y.-P., {Zhang}, S., {Torres}, D.~F., {Wang}, J.-M., \& {Li}, T.-P.
  2010, \aap, 510, A81

\bibitem[{{Chenevez} {et~al.}(2007){Chenevez}, {Falanga}, {Kuulkers}, {Walter},
  {Bildsten}, {Brandt}, {Lund}, {Oosterbroek}, \& {Zurita
  Heras}}]{2007A&A...469L..27C}
{Chenevez}, J., {Falanga}, M., {Kuulkers}, E., {et~al.} 2007, \aap, 469, L27

\bibitem[{{Chomiuk} \& {Povich}(2011)}]{Chomiuk2011}
{Chomiuk}, L. \& {Povich}, M.~S. 2011, \aj, 142, 197

\bibitem[{{Christian} \& {Swank}(1997)}]{1997ApJS..109..177C}
{Christian}, D.~J. \& {Swank}, J.~H. 1997, \apjs, 109, 177

\bibitem[{{Clavel} {et~al.}(2013){Clavel}, {Terrier}, {Goldwurm}, {Morris},
  {Ponti}, {Soldi}, \& {Trap}}]{sgr13}
{Clavel}, M., {Terrier}, R., {Goldwurm}, A., {et~al.} 2013, ArXiv e-prints

\bibitem[{{Cohen} \& {Thaddeus}(1977)}]{cohen1977}
{Cohen}, R.~S. \& {Thaddeus}, P. 1977, \apjl, 217, L155

\bibitem[{{Colbert} {et~al.}(2004){Colbert}, {Heckman}, {Ptak}, {Strickland},
  \& {Weaver}}]{chp2004}
{Colbert}, E.~J.~M., {Heckman}, T.~M., {Ptak}, A.~F., {Strickland}, D.~K., \&
  {Weaver}, K.~A. 2004, \apj, 602, 231

\bibitem[{{Coleiro} \& {Chaty}(2013)}]{2013ApJ...764..185C}
{Coleiro}, A. \& {Chaty}, S. 2013, \apj, 764, 185

\bibitem[{{Corbel} {et~al.}(2005){Corbel}, {Kaaret}, {Fender}, {Tzioumis},
  {Tomsick}, \& {Orosz}}]{2005ApJ...632..504C}
{Corbel}, S., {Kaaret}, P., {Fender}, R.~P., {et~al.} 2005, \apj, 632, 504

\bibitem[{{Cox}(2005)}]{cox2005}
{Cox}, D.~P. 2005, \araa, 43, 337

\bibitem[{{Cramphorn} \& {Sunyaev}(2002)}]{cramphorn}
{Cramphorn}, C.~K. \& {Sunyaev}, R.~A. 2002, \aap, 389, 252

\bibitem[{{Dame} {et~al.}(2001){Dame}, {Hartmann}, \& {Thaddeus}}]{dame2001}
{Dame}, T.~M., {Hartmann}, D., \& {Thaddeus}, P. 2001, \apj, 547, 792

\bibitem[{{Degenaar} {et~al.}(2010){Degenaar}, {Jonker}, {Torres}, {Kaur},
  {Rea}, {Israel}, {Patruno}, {Trap}, {Cackett}, {D'Avanzo}, {Lo Curto},
  {Novara}, {Krimm}, {Holland}, {de Luca}, {Esposito}, \&
  {Wijnands}}]{2010MNRAS.404.1591D}
{Degenaar}, N., {Jonker}, P.~G., {Torres}, M.~A.~P., {et~al.} 2010, \mnras,
  404, 1591

\bibitem[{{Dehnen} \& {Binney}(1998)}]{denbinac}
{Dehnen}, W. \& {Binney}, J. 1998, \mnras, 294, 429

\bibitem[{{Del Santo} {et~al.}(2007){Del Santo}, {Chenevez}, {Brandt},
  {Bazzano}, \& {Ubertini}}]{2007ATel.1207....1D}
{Del Santo}, M., {Chenevez}, J., {Brandt}, S., {Bazzano}, A., \& {Ubertini}, P.
  2007, The Astronomer's Telegram, 1207, 1

\bibitem[{{Dobbs} \& {Burkert}(2012)}]{db2012}
{Dobbs}, C.~L. \& {Burkert}, A. 2012, \mnras, 421, 2940

\bibitem[{{Doroshenko} {et~al.}(2007){Doroshenko}, {Doroshenko}, {Postnov},
  {Cherepashchuk}, \& {Tsygankov}}]{2007ESASP.622..495D}
{Doroshenko}, R.~F., {Doroshenko}, V.~A., {Postnov}, K.~A., {Cherepashchuk},
  A.~M., \& {Tsygankov}, S.~S. 2007, in ESA Special Publication, Vol. 622, ESA
  Special Publication, 495

\bibitem[{{Doroshenko} {et~al.}(2010){Doroshenko}, {Suchy}, {Santangelo},
  {Staubert}, {Kreykenbohm}, {Rothschild}, {Pottschmidt}, \&
  {Wilms}}]{2010A&A...515L...1D}
{Doroshenko}, V., {Suchy}, S., {Santangelo}, A., {et~al.} 2010, \aap, 515, L1

\bibitem[{{Dotani} {et~al.}(2000){Dotani}, {Asai}, \&
  {Wijnands}}]{2000ApJ...543L.145D}
{Dotani}, T., {Asai}, K., \& {Wijnands}, R. 2000, \apjl, 543, L145

\bibitem[{{Draine}(2011)}]{Draine}
{Draine}, B.~T. 2011, {Physics of the Interstellar and Intergalactic Medium}
  (Princeton University Press)

\bibitem[{{Dwek} {et~al.}(1995){Dwek}, {Arendt}, {Hauser}, {Kelsall}, {Lisse},
  {Moseley}, {Silverberg}, {Sodroski}, \& {Weiland}}]{dwek}
{Dwek}, E., {Arendt}, R.~G., {Hauser}, M.~G., {et~al.} 1995, \apj, 445, 716

\bibitem[{{Ebisawa} {et~al.}(2001){Ebisawa}, {Maeda}, {Kaneda}, \&
  {Yamauchi}}]{ebi2001}
{Ebisawa}, K., {Maeda}, Y., {Kaneda}, H., \& {Yamauchi}, S. 2001, Science, 293,
  1633

\bibitem[{{Ebisawa} {et~al.}(2005){Ebisawa}, {Tsujimoto}, {Paizis},
  {Hamaguchi}, {Bamba}, {Cutri}, {Kaneda}, {Maeda}, {Sato}, {Senda}, {Ueno},
  {Yamauchi}, {Beckmann}, {Courvoisier}, {Dubath}, \& {Nishihara}}]{ebi2005}
{Ebisawa}, K., {Tsujimoto}, M., {Paizis}, A., {et~al.} 2005, \apj, 635, 214

\bibitem[{{Ebisawa} {et~al.}(2008){Ebisawa}, {Yamauchi}, {Tanaka}, {Koyama},
  {Ezoe}, {Bamba}, {Kokubun}, {Hyodo}, {Tsujimoto}, \&
  {Takahashi}}]{Ebisawa2008}
{Ebisawa}, K., {Yamauchi}, S., {Tanaka}, Y., {et~al.} 2008, \pasj, 60, 223

\bibitem[{{Eisenberger} \& {Platzman}(1970)}]{EP1970}
{Eisenberger}, P. \& {Platzman}, P.~M. 1970, \pra, 2, 415

\bibitem[{{Falanga} {et~al.}(2005{\natexlab{a}}){Falanga}, {Bonnet-Bidaud},
  {Poutanen}, {Farinelli}, {Martocchia}, {Goldoni}, {Qu}, {Kuiper}, \&
  {Goldwurm}}]{2005A&A...436..647F}
{Falanga}, M., {Bonnet-Bidaud}, J.~M., {Poutanen}, J., {et~al.}
  2005{\natexlab{a}}, \aap, 436, 647

\bibitem[{{Falanga} {et~al.}(2005{\natexlab{b}}){Falanga}, {Kuiper},
  {Poutanen}, {Bonning}, {Hermsen}, {di Salvo}, {Goldoni}, {Goldwurm}, {Shaw},
  \& {Stella}}]{2005A&A...444...15F}
{Falanga}, M., {Kuiper}, L., {Poutanen}, J., {et~al.} 2005{\natexlab{b}}, \aap,
  444, 15

\bibitem[{{Field} {et~al.}(1969){Field}, {Goldsmith}, \& {Habing}}]{Field1969}
{Field}, G.~B., {Goldsmith}, D.~W., \& {Habing}, H.~J. 1969, \apjl, 155, L149

\bibitem[{{Filippova} {et~al.}(2005){Filippova}, {Tsygankov}, {Lutovinov}, \&
  {Sunyaev}}]{Filippova2005}
{Filippova}, E.~V., {Tsygankov}, S.~S., {Lutovinov}, A.~A., \& {Sunyaev}, R.~A.
  2005, Astronomy Letters, 31, 729

\bibitem[{{Freudenreich}(1996)}]{freudenreich}
{Freudenreich}, H.~T. 1996, \apj, 468, 663

\bibitem[{{Galloway} {et~al.}(2006){Galloway}, {Psaltis}, {Muno}, \&
  {Chakrabarty}}]{2006ApJ...639.1033G}
{Galloway}, D.~K., {Psaltis}, D., {Muno}, M.~P., \& {Chakrabarty}, D. 2006,
  \apj, 639, 1033

\bibitem[{{Gando Ryu} {et~al.}(2012){Gando Ryu}, {Nobukawa}, {Nakashima},
  {Tsuru}, {Koyama}, \& {Uchiyama}}]{sgr12}
{Gando Ryu}, S., {Nobukawa}, M., {Nakashima}, S., {et~al.} 2012, ArXiv e-prints

\bibitem[{{Giacconi} {et~al.}(1962){Giacconi}, {Gursky}, {Paolini}, \&
  {Rossi}}]{giac}
{Giacconi}, R., {Gursky}, H., {Paolini}, F.~R., \& {Rossi}, B.~B. 1962,
  Physical Review Letters, 9, 439

\bibitem[{{Gilfanov}(2004)}]{gilfanov2004}
{Gilfanov}, M. 2004, \mnras, 349, 146

\bibitem[{{Gilfanov} {et~al.}(2004){Gilfanov}, {Grimm}, \& {Sunyaev}}]{ggs2004}
{Gilfanov}, M., {Grimm}, H.-J., \& {Sunyaev}, R. 2004, \mnras, 351, 1365

\bibitem[{{Gilfanov} {et~al.}(2003){Gilfanov}, {Revnivtsev}, \&
  {Molkov}}]{Gilfanov2003}
{Gilfanov}, M., {Revnivtsev}, M., \& {Molkov}, S. 2003, \aap, 410, 217

\bibitem[{{Grimm} {et~al.}(2002){Grimm}, {Gilfanov}, \& {Sunyaev}}]{Grimm2002}
{Grimm}, H.-J., {Gilfanov}, M., \& {Sunyaev}, R. 2002, \aap, 391, 923

\bibitem[{{Grimm} {et~al.}(2003){Grimm}, {Gilfanov}, \& {Sunyaev}}]{grimm}
{Grimm}, H.-J., {Gilfanov}, M., \& {Sunyaev}, R. 2003, \mnras, 339, 793

\bibitem[{{Gruber} {et~al.}(1999){Gruber}, {Matteson}, {Peterson}, \&
  {Jung}}]{cxbspe}
{Gruber}, D.~E., {Matteson}, J.~L., {Peterson}, L.~E., \& {Jung}, G.~V. 1999,
  \apj, 520, 124

\bibitem[{{Haardt} {et~al.}(2001){Haardt}, {Galli}, {Treves}, {Chiappetti},
  {Dal Fiume}, {Corongiu}, {Belloni}, {Frontera}, {Kuulkers}, \&
  {Stella}}]{2001ApJS..133..187H}
{Haardt}, F., {Galli}, M.~R., {Treves}, A., {et~al.} 2001, \apjs, 133, 187

\bibitem[{{Harrison} {et~al.}(2013){Harrison}, {Craig}, {Christensen},
  {Hailey}, {Zhang}, {Boggs}, {Stern}, {Cook}, {Forster}, {Giommi},
  {Grefenstette}, {Kim}, {Kitaguchi}, {Koglin}, {Madsen}, {Mao}, {Miyasaka},
  {Mori}, {Perri}, {Pivovaroff}, {Puccetti}, {Rana}, {Westergaard}, {Willis},
  {Zoglauer}, {An}, {Bachetti}, {Barri{\`e}re}, {Bellm}, {Bhalerao},
  {Brejnholt}, {Fuerst}, {Liebe}, {Markwardt}, {Nynka}, {Vogel}, {Walton},
  {Wik}, {Alexander}, {Cominsky}, {Hornschemeier}, {Hornstrup}, {Kaspi},
  {Madejski}, {Matt}, {Molendi}, {Smith}, {Tomsick}, {Ajello}, {Ballantyne},
  {Balokovi{\'c}}, {Barret}, {Bauer}, {Blandford}, {Brandt}, {Brenneman},
  {Chiang}, {Chakrabarty}, {Chenevez}, {Comastri}, {Dufour}, {Elvis}, {Fabian},
  {Farrah}, {Fryer}, {Gotthelf}, {Grindlay}, {Helfand}, {Krivonos}, {Meier},
  {Miller}, {Natalucci}, {Ogle}, {Ofek}, {Ptak}, {Reynolds}, {Rigby},
  {Tagliaferri}, {Thorsett}, {Treister}, \& {Urry}}]{Nustar}
{Harrison}, F.~A., {Craig}, W.~W., {Christensen}, F.~E., {et~al.} 2013, \apj,
  770, 103

\bibitem[{{Hasinger} {et~al.}(2005){Hasinger}, {Miyaji}, \& {Schmidt}}]{cxb2}
{Hasinger}, G., {Miyaji}, T., \& {Schmidt}, M. 2005, \aap, 441, 417

\bibitem[{Hubbell {et~al.}(1975)Hubbell, Veigele, Briggs, Brown, Cromer, \&
  Howerton}]{prof}
Hubbell, J.~H., Veigele, W.~J., Briggs, E.~A., {et~al.} 1975, Journal of
  Physical and Chemical Reference Data, 4, 471

\bibitem[{{Hynes} {et~al.}(2004){Hynes}, {Steeghs}, {Casares}, {Charles}, \&
  {O'Brien}}]{2004ApJ...609..317H}
{Hynes}, R.~I., {Steeghs}, D., {Casares}, J., {Charles}, P.~A., \& {O'Brien},
  K. 2004, \apj, 609, 317

\bibitem[{{Iaria} {et~al.}(2005){Iaria}, {di Salvo}, {Robba}, {Lavagetto},
  {Burderi}, {Stella}, \& {van der Klis}}]{2005A&A...439..575I}
{Iaria}, R., {di Salvo}, T., {Robba}, N.~R., {et~al.} 2005, \aap, 439, 575

\bibitem[{{in't Zand} {et~al.}(2005){in't Zand}, {Cornelisse}, \&
  {M{\'e}ndez}}]{2005A&A...440..287I}
{in't Zand}, J.~J.~M., {Cornelisse}, R., \& {M{\'e}ndez}, M. 2005, \aap, 440,
  287

\bibitem[{{in't Zand} {et~al.}(2003){in't Zand}, {Kuulkers}, {Verbunt},
  {Heise}, \& {Cornelisse}}]{2003A&A...411L.487I}
{in't Zand}, J.~J.~M., {Kuulkers}, E., {Verbunt}, F., {Heise}, J., \&
  {Cornelisse}, R. 2003, \aap, 411, L487

\bibitem[{{in't Zand} {et~al.}(2002){in't Zand}, {Verbunt}, {Kuulkers},
  {Markwardt}, {Bazzano}, {Cocchi}, {Cornelisse}, {Heise}, {Natalucci}, \&
  {Ubertini}}]{2002A&A...389L..43I}
{in't Zand}, J.~J.~M., {Verbunt}, F., {Kuulkers}, E., {et~al.} 2002, \aap, 389,
  L43

\bibitem[{{Inui} {et~al.}(2009){Inui}, {Koyama}, {Matsumoto}, \&
  {Tsuru}}]{sgr6}
{Inui}, T., {Koyama}, K., {Matsumoto}, H., \& {Tsuru}, T.~G. 2009, \pasj, 61,
  241

\bibitem[{{Iso} {et~al.}(2012){Iso}, {Ebisawa}, \& {Tsujimoto}}]{Iso2012}
{Iso}, N., {Ebisawa}, K., \& {Tsujimoto}, M. 2012, in American Institute of
  Physics Conference Series, Vol. 1427, American Institute of Physics
  Conference Series, ed. R.~{Petre}, K.~{Mitsuda}, \& L.~{Angelini}, 288--289

\bibitem[{{Israel} {et~al.}(2001){Israel}, {Negueruela}, {Campana}, {Covino},
  {Di Paola}, {Maxwell}, {Norton}, {Speziali}, {Verrecchia}, \&
  {Stella}}]{2001A&A...371.1018I}
{Israel}, G.~L., {Negueruela}, I., {Campana}, S., {et~al.} 2001, \aap, 371,
  1018

\bibitem[{{Jackson} {et~al.}(2006){Jackson}, {Rathborne}, {Shah}, {Simon},
  {Bania}, {Clemens}, {Chambers}, {Johnson}, {Dormody}, {Lavoie}, \&
  {Heyer}}]{grs2006}
{Jackson}, J.~M., {Rathborne}, J.~M., {Shah}, R.~Y., {et~al.} 2006, \apjs, 163,
  145

\bibitem[{{Jonker} \& {Nelemans}(2004)}]{2004MNRAS.354..355J}
{Jonker}, P.~G. \& {Nelemans}, G. 2004, \mnras, 354, 355

\bibitem[{{Kalberla} {et~al.}(2005){Kalberla}, {Burton}, {Hartmann}, {Arnal},
  {Bajaja}, {Morras}, \& {P{\"o}ppel}}]{lab2005}
{Kalberla}, P.~M.~W., {Burton}, W.~B., {Hartmann}, D., {et~al.} 2005, \aap,
  440, 775

\bibitem[{{Kalberla} \& {Kerp}(2009)}]{Kalberla}
{Kalberla}, P.~M.~W. \& {Kerp}, J. 2009, \araa, 47, 27

\bibitem[{{Kennea} \& {Campana}(2006)}]{2006ATel..818....1K}
{Kennea}, J.~A. \& {Campana}, S. 2006, The Astronomer's Telegram, 818, 1

\bibitem[{{Kent} {et~al.}(1991){Kent}, {Dame}, \& {Fazio}}]{kent}
{Kent}, S.~M., {Dame}, T.~M., \& {Fazio}, G. 1991, \apj, 378, 131

\bibitem[{{Kinugasa} {et~al.}(1998){Kinugasa}, {Torii}, {Hashimoto}, {Tsunemi},
  {Hayashida}, {Kitamoto}, {Kamata}, {Dotani}, {Nagase}, {Sugizaki}, {Ueda},
  {Kawai}, {Makishima}, \& {Yamauchi}}]{1998ApJ...495..435K}
{Kinugasa}, K., {Torii}, K., {Hashimoto}, Y., {et~al.} 1998, \apj, 495, 435

\bibitem[{{Kong} {et~al.}(2000){Kong}, {Homer}, {Kuulkers}, {Charles}, \&
  {Smale}}]{2000MNRAS.311..405K}
{Kong}, A.~K.~H., {Homer}, L., {Kuulkers}, E., {Charles}, P.~A., \& {Smale},
  A.~P. 2000, \mnras, 311, 405

\bibitem[{{Koyama} {et~al.}(1989){Koyama}, {Awaki}, {Kunieda}, {Takano}, \&
  {Tawara}}]{koyama1989}
{Koyama}, K., {Awaki}, H., {Kunieda}, H., {Takano}, S., \& {Tawara}, Y. 1989,
  \nat, 339, 603

\bibitem[{{Koyama} {et~al.}(2007){Koyama}, {Hyodo}, {Inui}, {Nakajima},
  {Matsumoto}, {Tsuru}, {Takahashi}, {Maeda}, {Yamazaki}, {Murakami},
  {Yamauchi}, {Tsuboi}, {Senda}, {Kataoka}, {Takahashi}, {Holt}, \&
  {Brown}}]{Koyama2007}
{Koyama}, K., {Hyodo}, Y., {Inui}, T., {et~al.} 2007, \pasj, 59, 245

\bibitem[{{Koyama} {et~al.}(1996){Koyama}, {Maeda}, {Sonobe}, {Takeshima},
  {Tanaka}, \& {Yamauchi}}]{Koyama1996}
{Koyama}, K., {Maeda}, Y., {Sonobe}, T., {et~al.} 1996, \pasj, 48, 249

\bibitem[{{Koyama} {et~al.}(1986){Koyama}, {Makishima}, {Tanaka}, \&
  {Tsunemi}}]{koyama1986}
{Koyama}, K., {Makishima}, K., {Tanaka}, Y., \& {Tsunemi}, H. 1986, \pasj, 38,
  121

\bibitem[{{Krivonos} {et~al.}(2007{\natexlab{a}}){Krivonos}, {Revnivtsev},
  {Churazov}, {Sazonov}, {Grebenev}, \& {Sunyaev}}]{Krivonos2007}
{Krivonos}, R., {Revnivtsev}, M., {Churazov}, E., {et~al.} 2007{\natexlab{a}},
  \aap, 463, 957

\bibitem[{{Krivonos} {et~al.}(2007{\natexlab{b}}){Krivonos}, {Revnivtsev},
  {Lutovinov}, {Sazonov}, {Churazov}, \& {Sunyaev}}]{2007A&A...475..775K}
{Krivonos}, R., {Revnivtsev}, M., {Lutovinov}, A., {et~al.} 2007{\natexlab{b}},
  \aap, 475, 775

\bibitem[{{Krivonos} {et~al.}(2012){Krivonos}, {Tsygankov}, {Lutovinov},
  {Revnivtsev}, {Churazov}, \& {Sunyaev}}]{Krivonos2012}
{Krivonos}, R., {Tsygankov}, S., {Lutovinov}, A., {et~al.} 2012, \aap, 545, A27

\bibitem[{{Kuulkers} {et~al.}(2003){Kuulkers}, {den Hartog}, {in't Zand},
  {Verbunt}, {Harris}, \& {Cocchi}}]{2003A&A...399..663K}
{Kuulkers}, E., {den Hartog}, P.~R., {in't Zand}, J.~J.~M., {et~al.} 2003,
  \aap, 399, 663

\bibitem[{{Lehmer} {et~al.}(2010){Lehmer}, {Alexander}, {Bauer}, {Brandt},
  {Goulding}, {Jenkins}, {Ptak}, \& {Roberts}}]{lehmer2010}
{Lehmer}, B.~D., {Alexander}, D.~M., {Bauer}, F.~E., {et~al.} 2010, \apj, 724,
  559

\bibitem[{{Levine} {et~al.}(2006){Levine}, {Blitz}, \& {Heiles}}]{Levine}
{Levine}, E.~S., {Blitz}, L., \& {Heiles}, C. 2006, Science, 312, 1773

\bibitem[{{Lin} {et~al.}(2007){Lin}, {Homan}, {Remillard}, \&
  {Wijnands}}]{2007ATel.1183....1L}
{Lin}, D., {Homan}, J., {Remillard}, R., \& {Wijnands}, R. 2007, The
  Astronomer's Telegram, 1183, 1

\bibitem[{{Liu} {et~al.}(2006){Liu}, {van Paradijs}, \& {van den
  Heuvel}}]{catth}
{Liu}, Q.~Z., {van Paradijs}, J., \& {van den Heuvel}, E.~P.~J. 2006, \aap,
  455, 1165

\bibitem[{{Liu} {et~al.}(2007){Liu}, {van Paradijs}, \& {van den
  Heuvel}}]{cattw}
{Liu}, Q.~Z., {van Paradijs}, J., \& {van den Heuvel}, E.~P.~J. 2007, \aap,
  469, 807

\bibitem[{{Lowell} {et~al.}(2012){Lowell}, {Tomsick}, {Heinke}, {Bodaghee},
  {Boggs}, {Kaaret}, {Chaty}, {Rodriguez}, \& {Walter}}]{2012ApJ...749..111L}
{Lowell}, A.~W., {Tomsick}, J.~A., {Heinke}, C.~O., {et~al.} 2012, \apj, 749,
  111

\bibitem[{{Lutovinov} {et~al.}(2005{\natexlab{a}}){Lutovinov}, {Revnivtsev},
  {Gilfanov}, {Shtykovskiy}, {Molkov}, \& {Sunyaev}}]{luto}
{Lutovinov}, A., {Revnivtsev}, M., {Gilfanov}, M., {et~al.} 2005{\natexlab{a}},
  \aap, 444, 821

\bibitem[{{Lutovinov} {et~al.}(2005{\natexlab{b}}){Lutovinov}, {Revnivtsev},
  {Molkov}, \& {Sunyaev}}]{2005A&A...430..997L}
{Lutovinov}, A., {Revnivtsev}, M., {Molkov}, S., \& {Sunyaev}, R.
  2005{\natexlab{b}}, \aap, 430, 997

\bibitem[{{Lutovinov} {et~al.}(2013){Lutovinov}, {Revnivtsev}, {Tsygankov}, \&
  {Krivonos}}]{Lutovinov2013}
{Lutovinov}, A.~A., {Revnivtsev}, M.~G., {Tsygankov}, S.~S., \& {Krivonos},
  R.~A. 2013, \mnras, 431, 327

\bibitem[{{Lyubimkov} {et~al.}(1997){Lyubimkov}, {Rostopchin}, {Roche}, \&
  {Tarasov}}]{1997MNRAS.286..549L}
{Lyubimkov}, L.~S., {Rostopchin}, S.~I., {Roche}, P., \& {Tarasov}, A.~E. 1997,
  \mnras, 286, 549

\bibitem[{{Macomb} \& {Gehrels}(1999)}]{1999ApJS..120..335M}
{Macomb}, D.~J. \& {Gehrels}, N. 1999, \apjs, 120, 335

\bibitem[{{Mangano} {et~al.}(2008){Mangano}, {Romano}, {Sidoli}, {Sbarufatti},
  {Kennea}, {Vercellone}, {Burrows}, {Cusumano}, \&
  {Gehrels}}]{2008ATel.1727....1M}
{Mangano}, V., {Romano}, P., {Sidoli}, L., {et~al.} 2008, The Astronomer's
  Telegram, 1727, 1

\bibitem[{{Markwardt} {et~al.}(2008){Markwardt}, {Strohmayer}, \&
  {Swank}}]{2008ATel.1443....1M}
{Markwardt}, C.~B., {Strohmayer}, T.~E., \& {Swank}, J.~H. 2008, The
  Astronomer's Telegram, 1443, 1

\bibitem[{{Masetti} {et~al.}(2006{\natexlab{a}}){Masetti}, {Morelli},
  {Palazzi}, {Galaz}, {Bassani}, {Bazzano}, {Bird}, {Dean}, {Israel}, {Landi},
  {Malizia}, {Minniti}, {Schiavone}, {Stephen}, {Ubertini}, \&
  {Walter}}]{2006A&A...459...21M}
{Masetti}, N., {Morelli}, L., {Palazzi}, E., {et~al.} 2006{\natexlab{a}}, \aap,
  459, 21

\bibitem[{{Masetti} {et~al.}(2006{\natexlab{b}}){Masetti}, {Pretorius},
  {Palazzi}, {Bassani}, {Bazzano}, {Bird}, {Charles}, {Dean}, {Malizia},
  {Nkundabakura}, {Stephen}, \& {Ubertini}}]{2006A&A...449.1139M}
{Masetti}, N., {Pretorius}, M.~L., {Palazzi}, E., {et~al.} 2006{\natexlab{b}},
  \aap, 449, 1139

\bibitem[{{Mason} \& {Cordova}(1982)}]{1982ApJ...262..253M}
{Mason}, K.~O. \& {Cordova}, F.~A. 1982, \apj, 262, 253

\bibitem[{{Mauche} \& {Gorenstein}(1986)}]{MG1986}
{Mauche}, C.~W. \& {Gorenstein}, P. 1986, \apj, 302, 371

\bibitem[{{McClure-Griffiths} {et~al.}(2009){McClure-Griffiths}, {Pisano},
  {Calabretta}, {Ford}, {Lockman}, {Staveley-Smith}, {Kalberla}, {Bailin},
  {Dedes}, {Janowiecki}, {Gibson}, {Murphy}, {Nakanishi}, \&
  {Newton-McGee}}]{gass2009}
{McClure-Griffiths}, N.~M., {Pisano}, D.~J., {Calabretta}, M.~R., {et~al.}
  2009, \apjs, 181, 398

\bibitem[{{McKee} \& {Ostriker}(2007)}]{McKee2007}
{McKee}, C.~F. \& {Ostriker}, E.~C. 2007, \araa, 45, 565

\bibitem[{{McKee} \& {Ostriker}(1977)}]{MO1977}
{McKee}, C.~F. \& {Ostriker}, J.~P. 1977, \apj, 218, 148

\bibitem[{{McMillan}(2011)}]{massmodel}
{McMillan}, P.~J. 2011, \mnras, 414, 2446

\bibitem[{{Mereghetti} {et~al.}(2000){Mereghetti}, {Tiengo}, {Israel}, \&
  {Stella}}]{2000A&A...354..567M}
{Mereghetti}, S., {Tiengo}, A., {Israel}, G.~L., \& {Stella}, L. 2000, \aap,
  354, 567

\bibitem[{{Migliari} {et~al.}(2003){Migliari}, {Di Salvo}, {Belloni}, {van der
  Klis}, {Fender}, {Campana}, {Kouveliotou}, {M{\'e}ndez}, \&
  {Lewin}}]{2003MNRAS.342..909M}
{Migliari}, S., {Di Salvo}, T., {Belloni}, T., {et~al.} 2003, \mnras, 342, 909

\bibitem[{{Mineo} {et~al.}(2014){Mineo}, {Gilfanov}, {Lehmer}, {Morrison}, \&
  {Sunyaev}}]{Mineo2013}
{Mineo}, S., {Gilfanov}, M., {Lehmer}, B.~D., {Morrison}, G.~E., \& {Sunyaev},
  R. 2014, \mnras, 437, 1698

\bibitem[{{Mineo} {et~al.}(2012){Mineo}, {Gilfanov}, \& {Sunyaev}}]{mineo2012}
{Mineo}, S., {Gilfanov}, M., \& {Sunyaev}, R. 2012, \mnras, 419, 2095

\bibitem[{{Misiriotis} {et~al.}(2006){Misiriotis}, {Xilouris},
  {Papamastorakis}, {Boumis}, \& {Goudis}}]{Misiriotis2006}
{Misiriotis}, A., {Xilouris}, E.~M., {Papamastorakis}, J., {Boumis}, P., \&
  {Goudis}, C.~D. 2006, \aap, 459, 113

\bibitem[{{Morihana}(2012)}]{Morihana2012}
{Morihana}, K. 2012, \pasp, 124, 1132

\bibitem[{{Morihana} {et~al.}(2013){Morihana}, {Tsujimoto}, {Yoshida}, \&
  {Ebisawa}}]{morihana}
{Morihana}, K., {Tsujimoto}, M., {Yoshida}, T., \& {Ebisawa}, K. 2013, \apj,
  766, 14

\bibitem[{{Morrison} \& {McCammon}(1983)}]{Morrison}
{Morrison}, R. \& {McCammon}, D. 1983, \apj, 270, 119

\bibitem[{{Muno} {et~al.}(2004){Muno}, {Baganoff}, {Bautz}, {Feigelson},
  {Garmire}, {Morris}, {Park}, {Ricker}, \& {Townsley}}]{muno2004}
{Muno}, M.~P., {Baganoff}, F.~K., {Bautz}, M.~W., {et~al.} 2004, \apj, 613, 326

\bibitem[{{Muno} {et~al.}(2007){Muno}, {Baganoff}, {Brandt}, {Park}, \&
  {Morris}}]{sgr5}
{Muno}, M.~P., {Baganoff}, F.~K., {Brandt}, W.~N., {Park}, S., \& {Morris},
  M.~R. 2007, \apjl, 656, L69

\bibitem[{{Murakami} {et~al.}(2000){Murakami}, {Koyama}, {Sakano}, {Tsujimoto},
  \& {Maeda}}]{sgr3}
{Murakami}, H., {Koyama}, K., {Sakano}, M., {Tsujimoto}, M., \& {Maeda}, Y.
  2000, \apj, 534, 283

\bibitem[{{Nagata} {et~al.}(2003){Nagata}, {Kato}, {Baba}, {Nishiyama},
  {Nagayama}, {Nagashima}, {Kurita}, {Sato}, {Kato}, {Uemura}, {Yamaoka},
  {Monard}, {Ita}, {Matsunaga}, {Nakajima}, {Tamura}, {Nakaya}, \&
  {Sugitani}}]{2003PASJ...55L..73N}
{Nagata}, T., {Kato}, D., {Baba}, D., {et~al.} 2003, \pasj, 55, L73

\bibitem[{{Naik} \& {Paul}(2004)}]{2004A&A...418..655N}
{Naik}, S. \& {Paul}, B. 2004, \aap, 418, 655

\bibitem[{{Narayan} {et~al.}(1998){Narayan}, {Mahadevan}, {Grindlay}, {Popham},
  \& {Gammie}}]{narayan1998}
{Narayan}, R., {Mahadevan}, R., {Grindlay}, J.~E., {Popham}, R.~G., \&
  {Gammie}, C. 1998, \apj, 492, 554

\bibitem[{{Natalucci} {et~al.}(2000){Natalucci}, {Bazzano}, {Cocchi},
  {Ubertini}, {Heise}, {Kuulkers}, {in 't Zand}, \&
  {Smith}}]{2000ApJ...536..891N}
{Natalucci}, L., {Bazzano}, A., {Cocchi}, M., {et~al.} 2000, \apj, 536, 891

\bibitem[{{Negueruela} \& {Okazaki}(2001)}]{2001A&A...369..108N}
{Negueruela}, I. \& {Okazaki}, A.~T. 2001, \aap, 369, 108

\bibitem[{{Negueruela} {et~al.}(1999){Negueruela}, {Roche}, {Fabregat}, \&
  {Coe}}]{1999MNRAS.307..695N}
{Negueruela}, I., {Roche}, P., {Fabregat}, J., \& {Coe}, M.~J. 1999, \mnras,
  307, 695

\bibitem[{{Negueruela} {et~al.}(2006){Negueruela}, {Smith}, {Harrison}, \&
  {Torrej{\'o}n}}]{2006ApJ...638..982N}
{Negueruela}, I., {Smith}, D.~M., {Harrison}, T.~E., \& {Torrej{\'o}n}, J.~M.
  2006, \apj, 638, 982

\bibitem[{{Nobukawa} {et~al.}(2011){Nobukawa}, {Ryu}, {Tsuru}, \&
  {Koyama}}]{sgr11}
{Nobukawa}, M., {Ryu}, S.~G., {Tsuru}, T.~G., \& {Koyama}, K. 2011, \apjl, 739,
  L52

\bibitem[{{Oosterbroek} {et~al.}(2001){Oosterbroek}, {Barret}, {Guainazzi}, \&
  {Ford}}]{2001A&A...366..138O}
{Oosterbroek}, T., {Barret}, D., {Guainazzi}, M., \& {Ford}, E.~C. 2001, \aap,
  366, 138

\bibitem[{{Overbeck}(1965)}]{overbeck1965}
{Overbeck}, J.~W. 1965, \apj, 141, 864

\bibitem[{{Paerels} {et~al.}(2001){Paerels}, {Brinkman}, {van der Meer},
  {Kaastra}, {Kuulkers}, {den Boggende}, {Predehl}, {Drake}, {Kahn}, {Savin},
  \& {McLaughlin}}]{2001ApJ...546..338P}
{Paerels}, F., {Brinkman}, A.~C., {van der Meer}, R.~L.~J., {et~al.} 2001,
  \apj, 546, 338

\bibitem[{{Papitto} {et~al.}(2008){Papitto}, {Menna}, {Burderi}, {di Salvo}, \&
  {Riggio}}]{2008MNRAS.383..411P}
{Papitto}, A., {Menna}, M.~T., {Burderi}, L., {di Salvo}, T., \& {Riggio}, A.
  2008, \mnras, 383, 411

\bibitem[{{Parmar} {et~al.}(1989){Parmar}, {Gottwald}, {van der Klis}, \& {van
  Paradijs}}]{1989ApJ...338.1024P}
{Parmar}, A.~N., {Gottwald}, M., {van der Klis}, M., \& {van Paradijs}, J.
  1989, \apj, 338, 1024

\bibitem[{{Parmar} {et~al.}(2001){Parmar}, {Oosterbroek}, {Sidoli}, {Stella},
  \& {Frontera}}]{2001A&A...380..490P}
{Parmar}, A.~N., {Oosterbroek}, T., {Sidoli}, L., {Stella}, L., \& {Frontera},
  F. 2001, \aap, 380, 490

\bibitem[{{Pavlinsky} {et~al.}(2008){Pavlinsky}, {Sunyaev}, {Churazov},
  {Gilfanov}, {Vikhlinin}, {Hasinger}, {Predehl}, {Mitsuda}, {Kelley},
  {McCammon}, {Ohashi}, {den Herder}, {Ramsey}, {Gubarev}, {O'Dell}, \&
  {Fujimoto}}]{srg}
{Pavlinsky}, M., {Sunyaev}, R., {Churazov}, E., {et~al.} 2008, in Society of
  Photo-Optical Instrumentation Engineers (SPIE) Conference Series, Vol. 7011,
  Society of Photo-Optical Instrumentation Engineers (SPIE) Conference Series

\bibitem[{{Persic} \& {Rephaeli}(2007)}]{pr2007}
{Persic}, M. \& {Rephaeli}, Y. 2007, \aap, 463, 481

\bibitem[{{Pikel'Ner}(1968)}]{Pikelner1968}
{Pikel'Ner}, S.~B. 1968, \sovast, 11, 737

\bibitem[{{Ponti} {et~al.}(2013){Ponti}, {Morris}, {Terrier}, \&
  {Goldwurm}}]{Ponti2013}
{Ponti}, G., {Morris}, M.~R., {Terrier}, R., \& {Goldwurm}, A. 2013, in
  Advances in Solid State Physics, Vol.~34, Cosmic Rays in Star-Forming
  Environments, ed. D.~F. {Torres} \& O.~{Reimer}, 331

\bibitem[{{Ponti} {et~al.}(2010){Ponti}, {Terrier}, {Goldwurm}, {Belanger}, \&
  {Trap}}]{sgr7}
{Ponti}, G., {Terrier}, R., {Goldwurm}, A., {Belanger}, G., \& {Trap}, G. 2010,
  \apj, 714, 732

\bibitem[{{Porquet} {et~al.}(2003){Porquet}, {Rodriguez}, {Corbel}, {Goldoni},
  {Warwick}, {Goldwurm}, \& {Decourchelle}}]{2003A&A...406..299P}
{Porquet}, D., {Rodriguez}, J., {Corbel}, S., {et~al.} 2003, \aap, 406, 299

\bibitem[{{Pozdnyakov} {et~al.}(1983){Pozdnyakov}, {Sobol}, \&
  {Syunyaev}}]{pss1983}
{Pozdnyakov}, L.~A., {Sobol}, I.~M., \& {Syunyaev}, R.~A. 1983, Astrophysics
  and Space Physics Reviews, 2, 189

\bibitem[{{Predehl} \& {Schmitt}(1995)}]{predehl1995}
{Predehl}, P. \& {Schmitt}, J.~H.~M.~M. 1995, \aap, 293, 889

\bibitem[{{Ranalli} {et~al.}(2003){Ranalli}, {Comastri}, \& {Setti}}]{rcs2003}
{Ranalli}, P., {Comastri}, A., \& {Setti}, G. 2003, \aap, 399, 39

\bibitem[{{Rao} \& {Vadawale}(2012)}]{2012ApJ...757L..12R}
{Rao}, A. \& {Vadawale}, S.~V. 2012, \apjl, 757, L12

\bibitem[{{Rathborne} {et~al.}(2009){Rathborne}, {Johnson}, {Jackson}, {Shah},
  \& {Simon}}]{GRS2009}
{Rathborne}, J.~M., {Johnson}, A.~M., {Jackson}, J.~M., {Shah}, R.~Y., \&
  {Simon}, R. 2009, \apjs, 182, 131

\bibitem[{{Ratti} {et~al.}(2010){Ratti}, {Bassa}, {Torres}, {Kuiper},
  {Miller-Jones}, \& {Jonker}}]{2010MNRAS.408.1866R}
{Ratti}, E.~M., {Bassa}, C.~G., {Torres}, M.~A.~P., {et~al.} 2010, \mnras, 408,
  1866

\bibitem[{{Reig} {et~al.}(2005{\natexlab{a}}){Reig}, {Negueruela}, {Fabregat},
  {Chato}, \& {Coe}}]{2005A&A...440.1079R}
{Reig}, P., {Negueruela}, I., {Fabregat}, J., {Chato}, R., \& {Coe}, M.~J.
  2005{\natexlab{a}}, \aap, 440, 1079

\bibitem[{{Reig} {et~al.}(2005{\natexlab{b}}){Reig}, {Negueruela},
  {Papamastorakis}, {Manousakis}, \& {Kougentakis}}]{2005A&A...440..637R}
{Reig}, P., {Negueruela}, I., {Papamastorakis}, G., {Manousakis}, A., \&
  {Kougentakis}, T. 2005{\natexlab{b}}, \aap, 440, 637

\bibitem[{{Reig} {et~al.}(2011){Reig}, {Nespoli}, {Fabregat}, \&
  {Mennickent}}]{2011A&A...533A..23R}
{Reig}, P., {Nespoli}, E., {Fabregat}, J., \& {Mennickent}, R.~E. 2011, \aap,
  533, A23

\bibitem[{{Revnivtsev}(2003)}]{Revnivtsev2003}
{Revnivtsev}, M. 2003, \aap, 410, 865

\bibitem[{{Revnivtsev} {et~al.}(2009){Revnivtsev}, {Sazonov}, {Churazov},
  {Forman}, {Vikhlinin}, \& {Sunyaev}}]{Revnivtsev2009}
{Revnivtsev}, M., {Sazonov}, S., {Churazov}, E., {et~al.} 2009, \nat, 458, 1142

\bibitem[{{Revnivtsev} {et~al.}(2006){Revnivtsev}, {Sazonov}, {Gilfanov},
  {Churazov}, \& {Sunyaev}}]{Revnivtsev2006}
{Revnivtsev}, M., {Sazonov}, S., {Gilfanov}, M., {Churazov}, E., \& {Sunyaev},
  R. 2006, \aap, 452, 169

\bibitem[{{Revnivtsev} {et~al.}(2004){Revnivtsev}, {Churazov}, {Sazonov},
  {Sunyaev}, {Lutovinov}, {Gilfanov}, {Vikhlinin}, {Shtykovsky}, \&
  {Pavlinsky}}]{Revnivtsev2004}
{Revnivtsev}, M.~G., {Churazov}, E.~M., {Sazonov}, S.~Y., {et~al.} 2004, \aap,
  425, L49

\bibitem[{{Reynolds} {et~al.}(1997){Reynolds}, {Quaintrell}, {Still}, {Roche},
  {Chakrabarty}, \& {Levine}}]{1997MNRAS.288...43R}
{Reynolds}, A.~P., {Quaintrell}, H., {Still}, M.~D., {et~al.} 1997, \mnras,
  288, 43

\bibitem[{{Rieke} \& {Lebofsky}(1985)}]{Rieke}
{Rieke}, G.~H. \& {Lebofsky}, M.~J. 1985, \apj, 288, 618

\bibitem[{{Ritter} \& {Kolb}(2003)}]{caton}
{Ritter}, H. \& {Kolb}, U. 2003, \aap, 404, 301

\bibitem[{{Robitaille} \& {Whitney}(2010)}]{sfr2010}
{Robitaille}, T.~P. \& {Whitney}, B.~A. 2010, \apjl, 710, L11

\bibitem[{{Rolf}(1983)}]{rolf1983}
{Rolf}, D.~P. 1983, \nat, 302, 46

\bibitem[{{Roman-Duval} {et~al.}(2009){Roman-Duval}, {Jackson}, {Heyer},
  {Johnson}, {Rathborne}, {Shah}, \& {Simon}}]{romdutw}
{Roman-Duval}, J., {Jackson}, J.~M., {Heyer}, M., {et~al.} 2009, \apj, 699,
  1153

\bibitem[{{Roman-Duval} {et~al.}(2010){Roman-Duval}, {Jackson}, {Heyer},
  {Rathborne}, \& {Simon}}]{romdu}
{Roman-Duval}, J., {Jackson}, J.~M., {Heyer}, M., {Rathborne}, J., \& {Simon},
  R. 2010, \apj, 723, 492

\bibitem[{{Sala} \& {Greiner}(2006)}]{2006ATel..791....1S}
{Sala}, G. \& {Greiner}, J. 2006, The Astronomer's Telegram, 791, 1

\bibitem[{{Sazonov} {et~al.}(2006){Sazonov}, {Revnivtsev}, {Gilfanov},
  {Churazov}, \& {Sunyaev}}]{Sazonov2006}
{Sazonov}, S., {Revnivtsev}, M., {Gilfanov}, M., {Churazov}, E., \& {Sunyaev},
  R. 2006, \aap, 450, 117

\bibitem[{Schoonjans {et~al.}(2011)Schoonjans, Brunetti, Golosio, del Rio,
  Sole, \& C.~Ferrero}]{xraylib}
Schoonjans, T., Brunetti, A., Golosio, B., {et~al.} 2011, Spectromchimica Acta
  Part B: Atomic Spectroscopy, 66, 776

\bibitem[{{Shaw} {et~al.}(2009){Shaw}, {Hill}, {Kuulkers}, {Brandt},
  {Chenevez}, \& {Kretschmar}}]{2009MNRAS.393..419S}
{Shaw}, S.~E., {Hill}, A.~B., {Kuulkers}, E., {et~al.} 2009, \mnras, 393, 419

\bibitem[{{Shtykovskiy} \& {Gilfanov}(2005)}]{Shtykovskiy2005}
{Shtykovskiy}, P. \& {Gilfanov}, M. 2005, \mnras, 362, 879

\bibitem[{{Shtykovskiy} \& {Gilfanov}(2007)}]{sg2007}
{Shtykovskiy}, P.~E. \& {Gilfanov}, M.~R. 2007, Astronomy Letters, 33, 437

\bibitem[{{Shull} \& {van Steenberg}(1985)}]{shull1985}
{Shull}, J.~M. \& {van Steenberg}, M.~E. 1985, \apj, 298, 268

\bibitem[{{Sidoli} {et~al.}(1999){Sidoli}, {Mereghetti}, {Israel},
  {Chiappetti}, {Treves}, \& {Orlandini}}]{1999ApJ...525..215S}
{Sidoli}, L., {Mereghetti}, S., {Israel}, G.~L., {et~al.} 1999, \apj, 525, 215

\bibitem[{{Smith} \& {Dwek}(1998)}]{SmithDwek1998}
{Smith}, R.~K. \& {Dwek}, E. 1998, \apj, 503, 831

\bibitem[{{Snow} \& {McCall}(2006)}]{snow2006}
{Snow}, T.~P. \& {McCall}, B.~J. 2006, \araa, 44, 367

\bibitem[{{Spitzer}(1978)}]{ppim}
{Spitzer}, L. 1978, {Physical processes in the interstellar medium} (New York:
  Wiley-Interscience)

\bibitem[{{Stecker} {et~al.}(1975){Stecker}, {Solomon}, {Scoville}, \&
  {Ryter}}]{stecker1975}
{Stecker}, F.~W., {Solomon}, P.~M., {Scoville}, N.~Z., \& {Ryter}, C.~E. 1975,
  \apj, 201, 90

\bibitem[{{Steele} {et~al.}(1998){Steele}, {Negueruela}, {Coe}, \&
  {Roche}}]{1998MNRAS.297L...5S}
{Steele}, I.~A., {Negueruela}, I., {Coe}, M.~J., \& {Roche}, P. 1998, \mnras,
  297, L5

\bibitem[{{Stevens} {et~al.}(1997){Stevens}, {Reig}, {Coe}, {Buckley},
  {Fabregat}, \& {Steele}}]{1997MNRAS.288..988S}
{Stevens}, J.~B., {Reig}, P., {Coe}, M.~J., {et~al.} 1997, \mnras, 288, 988

\bibitem[{{Sunyaev} \& {Churazov}(1996)}]{sunchu}
{Sunyaev}, R.~A. \& {Churazov}, E.~M. 1996, Astronomy Letters, 22, 648

\bibitem[{{Sunyaev} {et~al.}(1993){Sunyaev}, {Markevitch}, \&
  {Pavlinsky}}]{sunyaev1993}
{Sunyaev}, R.~A., {Markevitch}, M., \& {Pavlinsky}, M. 1993, \apj, 407, 606

\bibitem[{{Sunyaev} {et~al.}(1999){Sunyaev}, {Uskov}, \& {Churazov}}]{suny2}
{Sunyaev}, R.~A., {Uskov}, D.~B., \& {Churazov}, E.~M. 1999, Astronomy Letters,
  25, 199

\bibitem[{{Tajima} {et~al.}(2010){Tajima}, {Blandford}, {Enoto}, {Fukazawa},
  {Gilmore}, {Kamae}, {Kataoka}, {Kawaharada}, {Kokubun}, {Laurent}, {Lebrun},
  {Limousin}, {Madejski}, {Makishima}, {Mizuno}, {Nakazawa}, {Ohno}, {Ohta},
  {Sato}, {Sato}, {Takahashi}, {Takahashi}, {Tanaka}, {Tashiro}, {Terada},
  {Uchiyama}, {Watanabe}, {Yamaoka}, \& {Yonetoku}}]{astroh}
{Tajima}, H., {Blandford}, R., {Enoto}, T., {et~al.} 2010, in Society of
  Photo-Optical Instrumentation Engineers (SPIE) Conference Series, Vol. 7732,
  Society of Photo-Optical Instrumentation Engineers (SPIE) Conference Series,
  773216

\bibitem[{{Takahashi} {et~al.}(2010){Takahashi}, {Mitsuda}, \&
  {Kelley}}]{Takahashi2010}
{Takahashi}, T., {Mitsuda}, K., \& {Kelley}, R. 2010, X-ray Astronomy 2009;
  Present Status, Multi-Wavelength Approach and Future Perspectives, 1248, 537

\bibitem[{{Tanaka}(2002)}]{tanaka2002}
{Tanaka}, Y. 2002, \aap, 382, 1052

\bibitem[{{Tanaka} {et~al.}(1994){Tanaka}, {Inoue}, \& {Holt}}]{Tanaka1994}
{Tanaka}, Y., {Inoue}, H., \& {Holt}, S.~S. 1994, \pasj, 46, L37

\bibitem[{{Tanaka} {et~al.}(1999){Tanaka}, {Miyaji}, \&
  {Hasinger}}]{tanaka1999}
{Tanaka}, Y., {Miyaji}, T., \& {Hasinger}, G. 1999, Astronomische Nachrichten,
  320, 181

\bibitem[{{Terrier} {et~al.}(2010){Terrier}, {Ponti}, {B{\'e}langer},
  {Decourchelle}, {Tatischeff}, {Goldwurm}, {Trap}, {Morris}, \&
  {Warwick}}]{sgr8}
{Terrier}, R., {Ponti}, G., {B{\'e}langer}, G., {et~al.} 2010, \apj, 719, 143

\bibitem[{{Torrej{\'o}n} {et~al.}(2010){Torrej{\'o}n}, {Negueruela}, {Smith},
  \& {Harrison}}]{2010A&A...510A..61T}
{Torrej{\'o}n}, J.~M., {Negueruela}, I., {Smith}, D.~M., \& {Harrison}, T.~E.
  2010, \aap, 510, A61

\bibitem[{{Torres} {et~al.}(2006){Torres}, {Steeghs}, {Garcia}, {McClintock},
  {Jonker}, {Kaplan}, {Casares}, {Greimel}, \&
  {Augusteijn}}]{2006ATel..784....1T}
{Torres}, M.~A.~P., {Steeghs}, D., {Garcia}, M.~R., {et~al.} 2006, The
  Astronomer's Telegram, 784, 1

\bibitem[{{Tr{\"u}mper} \& {Sch{\"o}nfelder}(1973)}]{trumper1973}
{Tr{\"u}mper}, J. \& {Sch{\"o}nfelder}, V. 1973, \aap, 25, 445

\bibitem[{{Tsygankov} \& {Lutovinov}(2005)}]{2005AstL...31...88T}
{Tsygankov}, S.~S. \& {Lutovinov}, A.~A. 2005, Astronomy Letters, 31, 88

\bibitem[{{Uchiyama} {et~al.}(2013){Uchiyama}, {Nobukawa}, {Tsuru}, \&
  {Koyama}}]{Uchiyama2013}
{Uchiyama}, H., {Nobukawa}, M., {Tsuru}, T.~G., \& {Koyama}, K. 2013, \pasj,
  65, 19

\bibitem[{{Ueda} {et~al.}(2003){Ueda}, {Akiyama}, {Ohta}, \& {Miyaji}}]{cxb1}
{Ueda}, Y., {Akiyama}, M., {Ohta}, K., \& {Miyaji}, T. 2003, \apj, 598, 886

\bibitem[{{Vainshtein} {et~al.}(1998){Vainshtein}, {Syunyaev}, \&
  {Churazov}}]{Vainshtein1998}
{Vainshtein}, L.~A., {Syunyaev}, R.~A., \& {Churazov}, E.~M. 1998, Astronomy
  Letters, 24, 271

\bibitem[{{Vallee}(1995)}]{Vallee1995}
{Vallee}, J.~P. 1995, \apj, 454, 119

\bibitem[{{Vall{\'e}e}(2008)}]{Vallee2008}
{Vall{\'e}e}, J.~P. 2008, \aj, 135, 1301

\bibitem[{{Verner} \& {Yakovlev}(1995)}]{Verner}
{Verner}, D.~A. \& {Yakovlev}, D.~G. 1995, \aaps, 109, 125

\bibitem[{{Wang} \& {Chakrabarty}(2004)}]{2004ApJ...616L.139W}
{Wang}, Z. \& {Chakrabarty}, D. 2004, \apjl, 616, L139

\bibitem[{{Werner} {et~al.}(2004){Werner}, {in't Zand}, {Natalucci},
  {Markwardt}, {Cornelisse}, {Bazzano}, {Cocchi}, {Heise}, \&
  {Ubertini}}]{2004A&A...416..311W}
{Werner}, N., {in't Zand}, J.~J.~M., {Natalucci}, L., {et~al.} 2004, \aap, 416,
  311

\bibitem[{{White} \& {Angelini}(2001)}]{2001ApJ...561L.101W}
{White}, N.~E. \& {Angelini}, L. 2001, \apjl, 561, L101

\bibitem[{{White} \& {van Paradijs}(1996)}]{1996ApJ...473L..25W}
{White}, N.~E. \& {van Paradijs}, J. 1996, \apjl, 473, L25

\bibitem[{{Wilson} {et~al.}(2002){Wilson}, {Finger}, {Coe}, {Laycock}, \&
  {Fabregat}}]{2002ApJ...570..287W}
{Wilson}, C.~A., {Finger}, M.~H., {Coe}, M.~J., {Laycock}, S., \& {Fabregat},
  J. 2002, \apj, 570, 287

\bibitem[{{Wolfire} {et~al.}(1995){Wolfire}, {Hollenbach}, {McKee}, {Tielens},
  \& {Bakes}}]{Wolfire1995}
{Wolfire}, M.~G., {Hollenbach}, D., {McKee}, C.~F., {Tielens}, A.~G.~G.~M., \&
  {Bakes}, E.~L.~O. 1995, \apj, 443, 152

\bibitem[{{Wolfire} {et~al.}(2003){Wolfire}, {McKee}, {Hollenbach}, \&
  {Tielens}}]{Wolfire2003}
{Wolfire}, M.~G., {McKee}, C.~F., {Hollenbach}, D., \& {Tielens}, A.~G.~G.~M.
  2003, \apj, 587, 278

\bibitem[{{Worrall} {et~al.}(1982){Worrall}, {Marshall}, {Boldt}, \&
  {Swank}}]{worrall}
{Worrall}, D.~M., {Marshall}, F.~E., {Boldt}, E.~A., \& {Swank}, J.~H. 1982,
  \apj, 255, 111

\bibitem[{{Yamauchi} \& {Koyama}(1993)}]{yako}
{Yamauchi}, S. \& {Koyama}, K. 1993, \apj, 404, 620

\bibitem[{{Zhang} {et~al.}(2010){Zhang}, {Torres}, {Li}, {Chen}, {Rea}, \&
  {Wang}}]{2010MNRAS.408..642Z}
{Zhang}, S., {Torres}, D.~F., {Li}, J., {et~al.} 2010, \mnras, 408, 642

\end{thebibliography}
\onecolumn
\begin{appendix}
\section{List of LMXB and HMXB sources responsible for peaks}\label{appb}
{The list of LMXBs and HMXBs \ct{that} are responsible for some of the
  brightest peaks in the scattered GRXE in the Galactic plane is given in
  Table \ref{tablepeaks}. A full list of sources with the corresponding data
  used in this paper is available at \url{http://www.mpa-garching.mpg.de/~molaro}}

\begin{table*}[htp]
\caption{List of sources responsible for peaks in the longitude and latitude profiles, ranked by 2-10keV luminosity.}
\label {tablepeaks}
 \begin{tabular}{ l | l|c | c | c | c | c | c | l | l | l }

Source & Type & $l$ ($^\circ$) & $b$ ($^\circ$) & distance  & $ L_{2-10{\rm keV}}$ & $ L_{17-60{\rm keV}}$  & $\Gamma$ & distance & $ L_{2-10{\rm keV}}$ & $ L_{17-60{\rm keV}}$ \\
 &  &  &   & (kpc) & (erg/s) & (erg/s) &  &  &  & \\
\hline\hline
 & & & & & & & & & & \\
4U 1516-569 & LM & 322.12&  0.037 &  9.2 &  3.31$\times 10^{38}$ &  5.91$\times 10^{35}$ &  6 & (4)  &(3) &   (1)\\ 
GRS 1915+105 & LM & 45.36 &  -0.22 &  11 &  2.96$\times 10^{38}$ &  4.88$\times 10^{37}$ &  3.03 & (4) & (3) &  (1) \\
GX 5-1 & LM & 5.079 &  -1.019 &  7.2 &  1.42$\times 10^{38}$ &  3.83$\times 10^{36}$ &  5.49 &  (7)  &  (3)&  (1) \\ 
GX 340+0 & LM & 339.59 &  -0.08 &  11 &  1.41$\times 10^{38}$ &  5.42$\times 10^{36}$ &  6.28 & (7) &(3)  &   (1)\\ 
GX 13+1 & LM & 13.52 &  0.106 &  7 &  4.32$\times 10^{37}$ &  8.03$\times 10^{35}$ &  4.89 & (5)  & (3)&  (1) \\ 
Cyg X-3 & HM & 79.85 &  0.7 &  7.2 &  2.43$\times 10^{37}$ &  1.01$\times 10^{37}$ &  3.17 &   (2)&(3) & (2)  \\ 
GRO J1744-28 & LM & 0.0445 &  0.3015 &  8.5 &  1.57$\times 10^{37}$ &  - &  - & (3) &(3)&  - \\

GX 354-0 & LM & 354.30 &  -0.15 &  5.3 &  7.17$\times 10^{36}$ &  1.89$\times 10^{36}$ &  3.19 & (4)  & (3)  &   (1)\\ 
Cen X-3 & HM & -67.9 &  0.33 &  5.7 &  6.09$\times 10^{36}$ &  2.45$\times 10^{36}$ &  5.24 &  (2) &(3) &  (2) \\ 
GRS 1758-258  & LM & 4.508  &  -1.362  & 8.5 & 5.90$\times 10^{36}$  &  5.67$\times 10^{36}$  &  1.92  & (3) & (3) & (1) \\

1E1740.7-2942 & LM & -0.873 & -0.105 & 8.5 & 5.03$\times 10^{36}$& 3.85$\times 10^{36}$ &  2.02& (12) & (3) & (11)\\ 
Cyg X-1 & HM & 71.34 &  3.07 &  1.86 &  4.01$\times 10^{36}$ &  3.90$\times 10^{36}$ &  1.97 &  (2) &(3) & (2)  \\ 
OAO 1657-415 & HM & -15.63 &  0.32 &  7.1 &  1.88$\times 10^{36}$ &  4.87$\times 10^{36}$ &  3.66 &  (2) &(3) & (2)  \\
4U 0115+63 & HM & 125.922 & 1.029 & 8 & 1.27$\times 10^{36}$ & 2.88$\times 10^{36}$ &  4.07 & (10) & (3) & (1) \\
1E 1145.1-6141 & HM & -64.5 &  -0.02 &  8.5 &  1.15$\times 10^{36}$ &  2.01$\times 10^{36}$ &  3.04 &   (2)&(3) & (2)  \\ 
4U 1908+075 & HM & 41.89 &  -0.81 &  7 &  1.06$\times 10^{36}$ &  9.66$\times 10^{35}$ &  2.86 &   (2)& (3) & (2)  \\ 
GX 301-2 & HM & -59.9 &  -0.03 &  3.5 &  9.31$\times 10^{35}$ &  3.15$\times 10^{36}$ &  5.83 &  (2) &(3) & (2)  \\
4U 1907+097 & HM & 43.74 &  0.47 &  5 &  8.98$\times 10^{35}$ &  4.39$\times 10^{35}$ &  5.15 &   (2)&(3) & (2)  \\  
A 1845-024 & HM & 30.4 & -0.381 & 10 & 8.42$\times 10^{35}$ & 8.21$\times 10^{34}$ & 2.57 & (7) & (3) & (1) \\ 
Sct X-1 & HM & 24.34 &  0.0657 &  10 &  7.50$\times 10^{35}$ &  - &  - & (3)  &(3) &  - \\ 
V0 332+53 & HM & 146.05 &  -2.194 &  7.5 &  1.03$\times 10^{35}$ &  1.11$\times 10^{37}$ &  5.09 & (10) &(3) &  (1) \\ 
1E 1743.1-2843 & LM & 0.25 &  -0.026 &  8 &  - &  4.54$\times 10^{35}$ &  3.36 &  (8)  & - & (1)\\ 
XTE J1810-189 & LM & 11.36 &  0.06 &  11.5 &  - &  3.35$\times 10^{35}$ &  2.21 &  (6)  & - & (1) \\ 
\end{tabular}
\tablebib{
(1)~\citet{Krivonos2012}; (2)~\citet{Lutovinov2013}; (3)~\citet{grimm}; (4)~\citet {2004MNRAS.354..355J}; (5)~\citet{1999MNRAS.306..417B}; (6)~\citet{2008ATel.1443....1M}; (7)~\citet{Grimm2002}; (8)~\citet{2003A&A...406..299P}; (9)~\citet{1999ApJ...525..215S}; (10)~\citet{1999MNRAS.307..695N}; (11)~\citet{Krivonos2007}; (12)~\citet{1996ApJ...473L..25W}.
}
\end{table*}
 \newpage
\section{Contribution of Sgr A$^{*}$}\label{app_sgrA}
{As discussed in \ct{Sect.} \ref{contrXrays}, the past activity of the \ct{currently} low quiescent source Sgr A$^*$ might have significantly contributed to the cumulative X-ray output of the Galaxy, and hence to the diffuse GRXE component. In Fig. \ref{sgr_A} we show the minimum luminosity required for the flux from this source to outshine the contribution of the entire XBs population at different positions on the Galactic plane, estimated as
\begin{equation}
L_{Sgr A^*} = 4 \pi R^2_{Sgr A^*} \times \sum_{i} \frac{L_{i}(\nu)}{4\pi R^2_i}
\end{equation}
Near the Galactic center, a luminosity of $\gtrsim 10^{37}$ ergs/s from Sgr
A$^*$ would be enough to become comparable with the illumination from X-ray
binaries. On the outskirts of the Galaxy\ct{,} on the other hand\ct{,} \changeRB{Sgr
A$^*$, or some other  ultra-luminous source near the Galactic center, can be ignored as long as its  luminosity is
$\lesssim 10^{39}-10^{40}$ ergs/s.}
}
\begin{figure*}[htp]
 \centering
  \includegraphics[trim = 20 0 20 10, width=9cm, height=8cm]{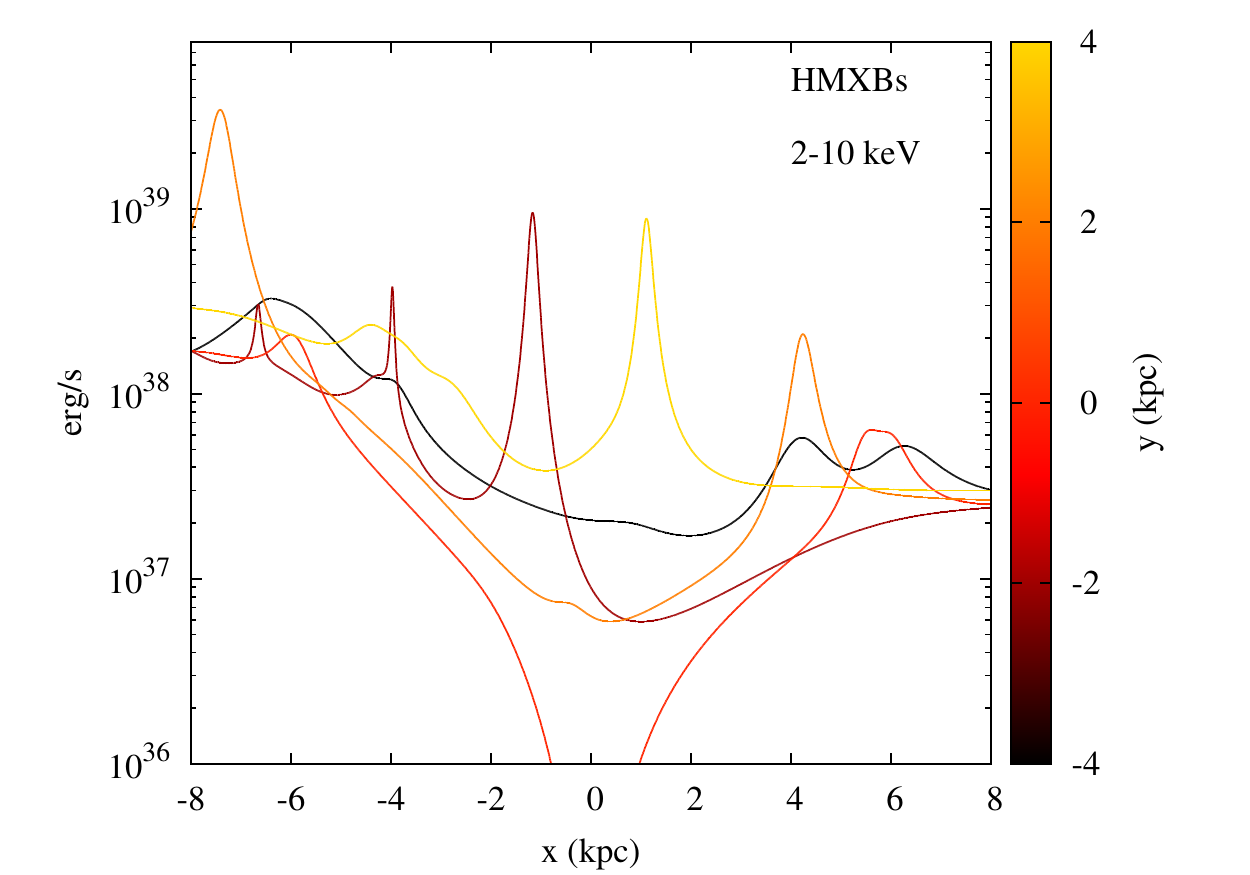} \includegraphics[trim = 20 0 20 10, width=9cm, height=8cm]{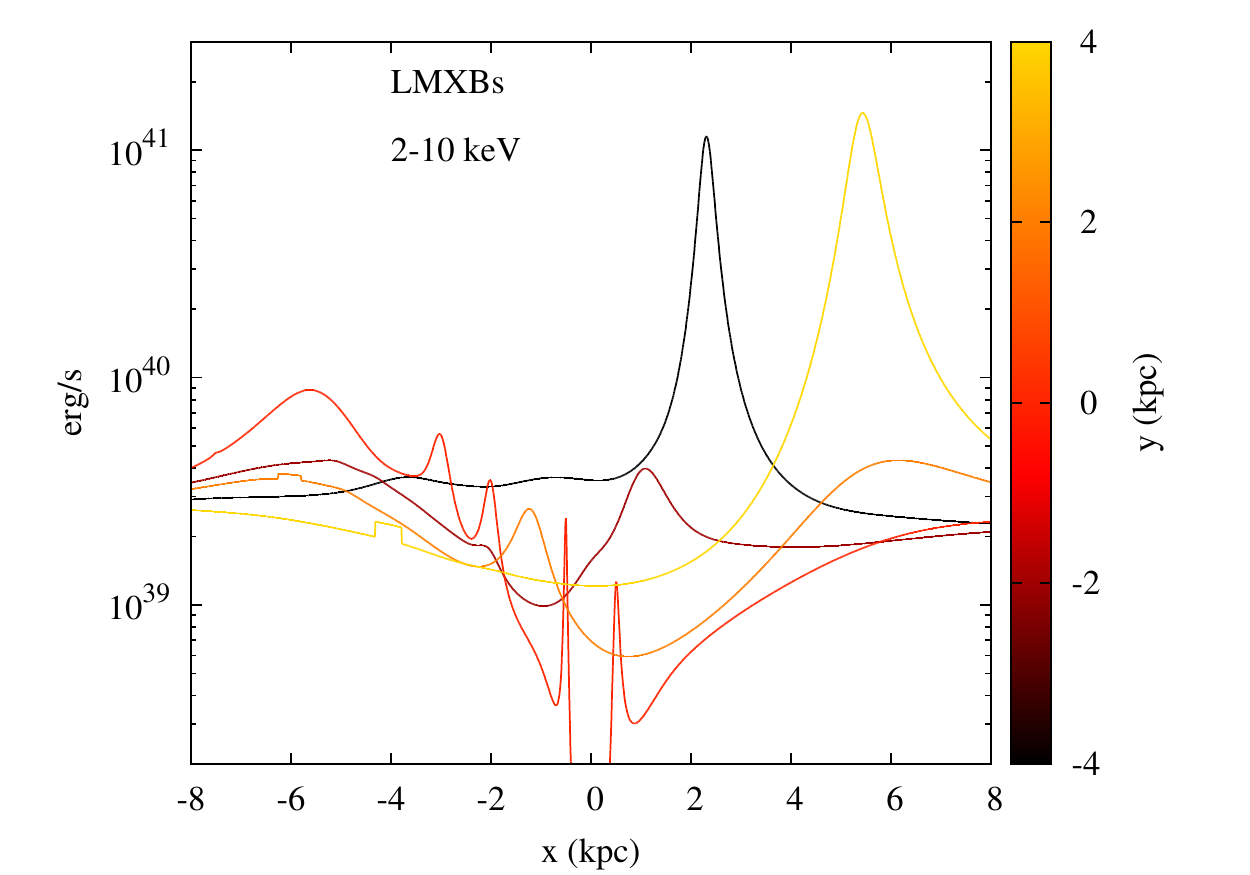} 
        \caption{Minimum Sgr A$^*$ luminosity required for the source's
          flux to be higher than the total HMXBs (\textit{left panel}) and
          LMXBs (\textit{right panel}) contribution at different positions
          on the Galactic plane. \changeRV{If we know the  star formation rate and the
mass of a galaxy, we can estimate the X-ray luminosity contributed by the  X-ray binaries and similarly find when the AGNs contribute
more than the X-ray binaries to the heating of the gas in \changeRR{other} galaxies with
AGNs.} }        
\label{sgr_A}
  \end{figure*}
\newpage

\section{ Scattering cross-sections}\label{appa}
In Fig. \ref{H2_energy_DCS} we illustrate the energy dependence of \changeRR{the scattering cross-section} for the case of H2 scattering. In Fig. \ref{ratio_H2_He_HI} and Fig.  \ref{Total_contr}  we highlight the additional effects \ct{caused by} bound
 electrons by showing the ratio of contributions from H2 and He with
 respect \ct{to the} free electrons. The energy dependence of
   Rayleigh scattering is clearly evident. At high energies, Rayleigh
   scattering operates only on \ct{a} very narrow range of scattering angles and
   the total scattering \ct{cross-section approaches the Klein-Nishina cross-section}. At low energies, coherent Rayleigh scattering results in the 
   enhancement of the cross section for a significant range of scattering
   angles\ct{,} resulting in a higher average scattering rate of \ct{low-energy}
   photons. The average scattered spectrum is therefore softer \ct{than} the
  average incident spectrum. {The contribution of molecular
    hydrogen as well as helium and heavier elements is compared in Table
    \ref{elements} in the Rayleigh scattering limit ($0^{\circ}$ scattering
    angle). The elements heavier than helium contribute $\lesssim
    10\%$ in this extreme case, and the actual contribution when suitably
    averaged over different scattering angles would be smaller \ct{than}
    the values in the table. 
}

\begin{table}[htp]
\centering
\caption{Elements \ct{according to} maximal contribution to Rayleigh scattering
  {if all hydrogen is in atomic form (or molecular form, given in parentheses).  $n_H$ is the
  total number density of hydrogen atoms in HI or H2 form. Solar
photospheric abundances from \citet{abundances} are assumed.}}
\label {elements}
 \begin{tabular}{ l | l }

Element & $Z^2\times n_Z/n_H$ \\
\hline \hline
& \\
HI (H2) & 1 (2) \\
 He & 0.340 \\
 O & 0.031 \\
 Fe & 0.021 \\
 C & 0.010 \\
 Ne & 0.008 \\
 Si & 0.006 \\
 Mg & 0.006 \\

 & \\
Total & 1.42 (2.42)\\
Other elements (total) & $\lesssim 0.01$ \\
\end{tabular}
\end{table}

\begin{figure}[htp]
\captionsetup{width=0.45\textwidth}
\centering
\begin{minipage}{.5\textwidth}
  \raggedright
  \includegraphics[trim = 30 40 0 50,height = 5cm, width = 9 cm]{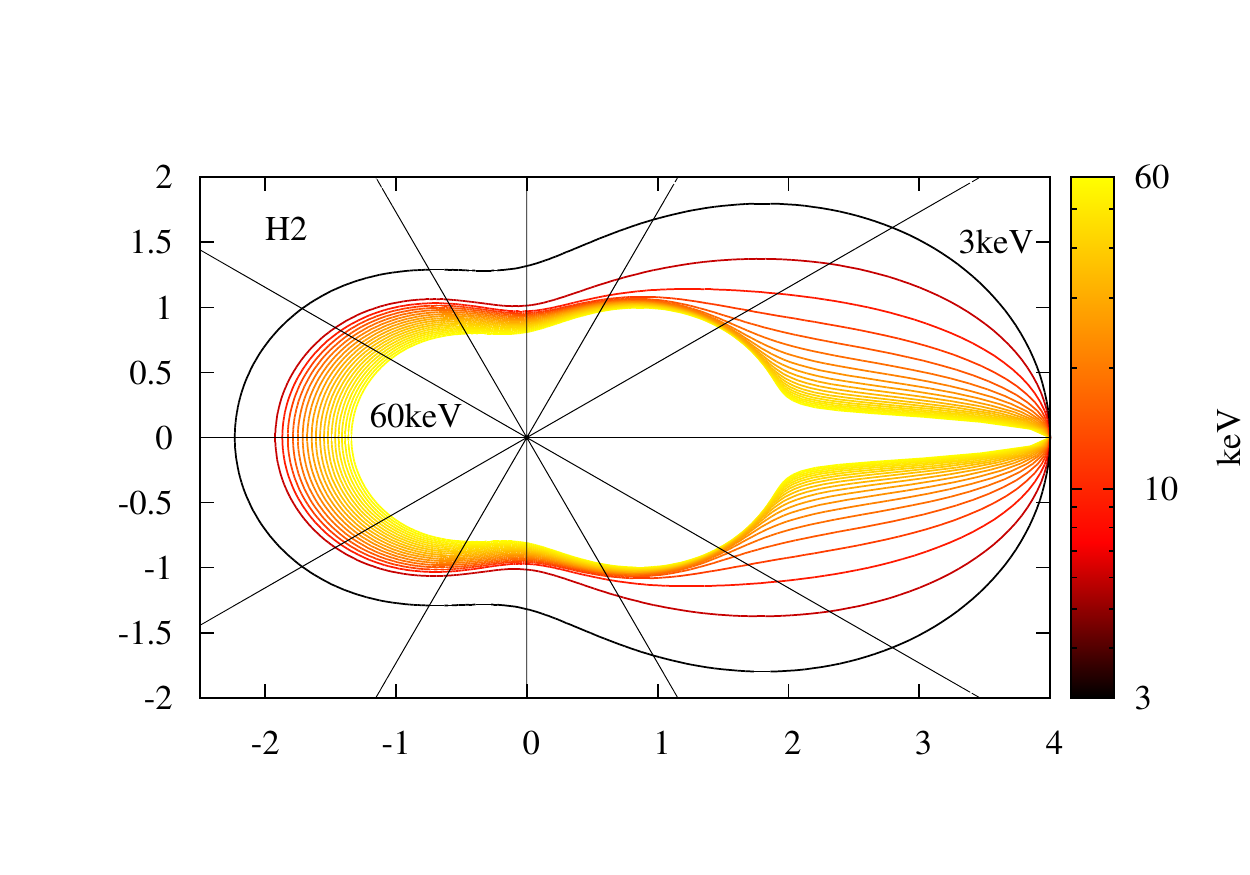}
 \raggedright \captionof{figure}{\changeRE{\ct{Cross-section} (Rayleigh+ Compton) in polar coordinates, $\sigma(\theta) e^{i\theta}$, where 
$\sigma$ is the amplitude of the \ct{cross-section} in units of $r_{\rm e}^2$ 
for the scattering angle $\theta$ and $r_{\rm e}$ is the classical 
electron radius}. The enhancement of Rayleigh scattering \ct{caused by} coherence effects is clearly visible at low energies. While the cross section remains constant with energy at angles close to zero, the contribution of Rayleigh scattering is shown to quickly decrease with increasing energy. Compton scattering is also suppressed \ct{by} relativistic effects.}
\label{H2_energy_DCS}
\end{minipage}%
\begin{minipage}{.5\textwidth}
 
\includegraphics[trim = 0 10 10 10,height = 6cm, width = 9 cm]{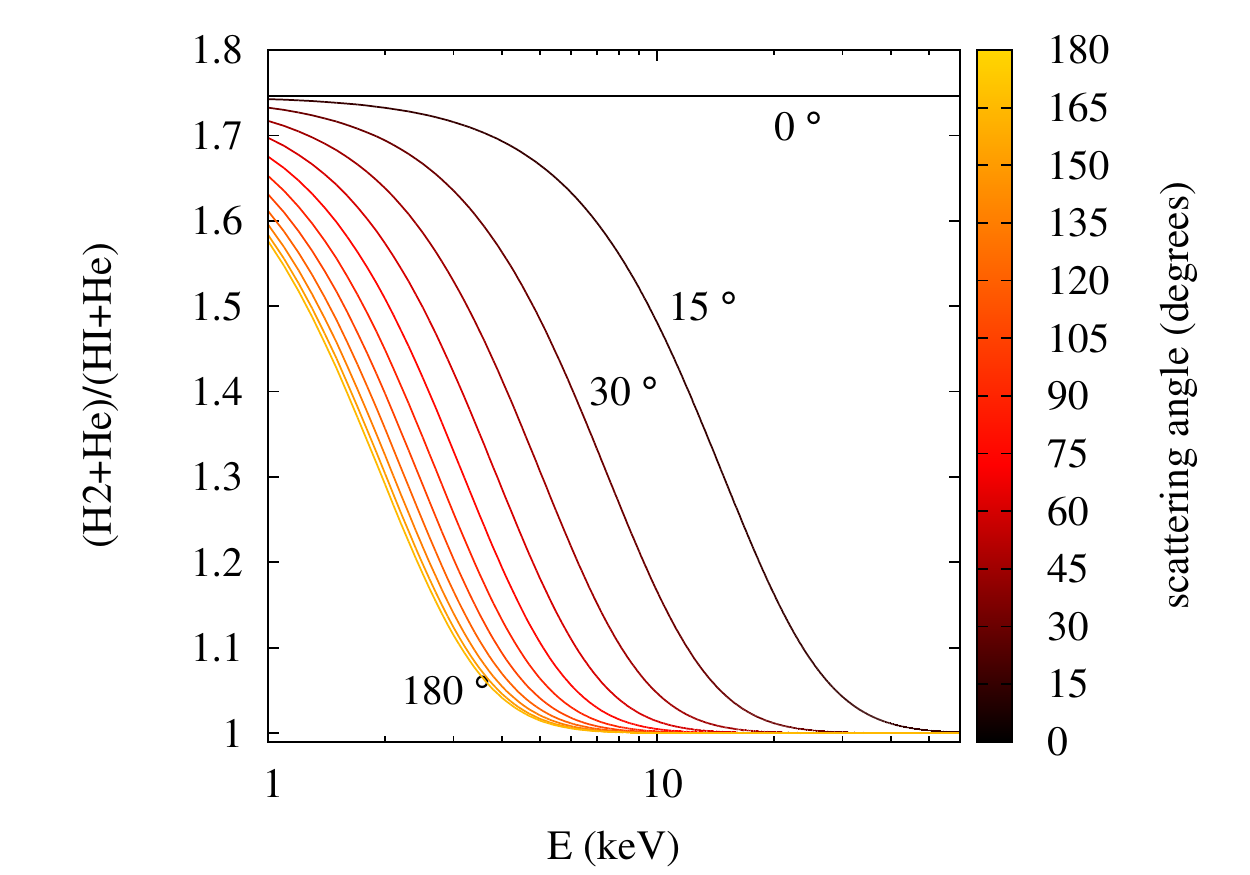}
  \raggedright \captionof{figure}{   Ratio of the \changeRR{Rayleigh + Compton} differential \ct{cross-section} of H2 + He to HI + He (where each element is weighted by relative abundance) as a function of energy for different scattering angles. The total cross section approaches that of unbound electrons as the importance of coherence effects decreases with energy, \ct{because of} the suppression of Rayleigh scattering.}
\label{ratio_H2_He_HI}
\end{minipage}
\end{figure}

\begin{figure*}[htp]
\centering
\includegraphics[trim = 20 10 20 10,height = 8cm, width = 9 cm]{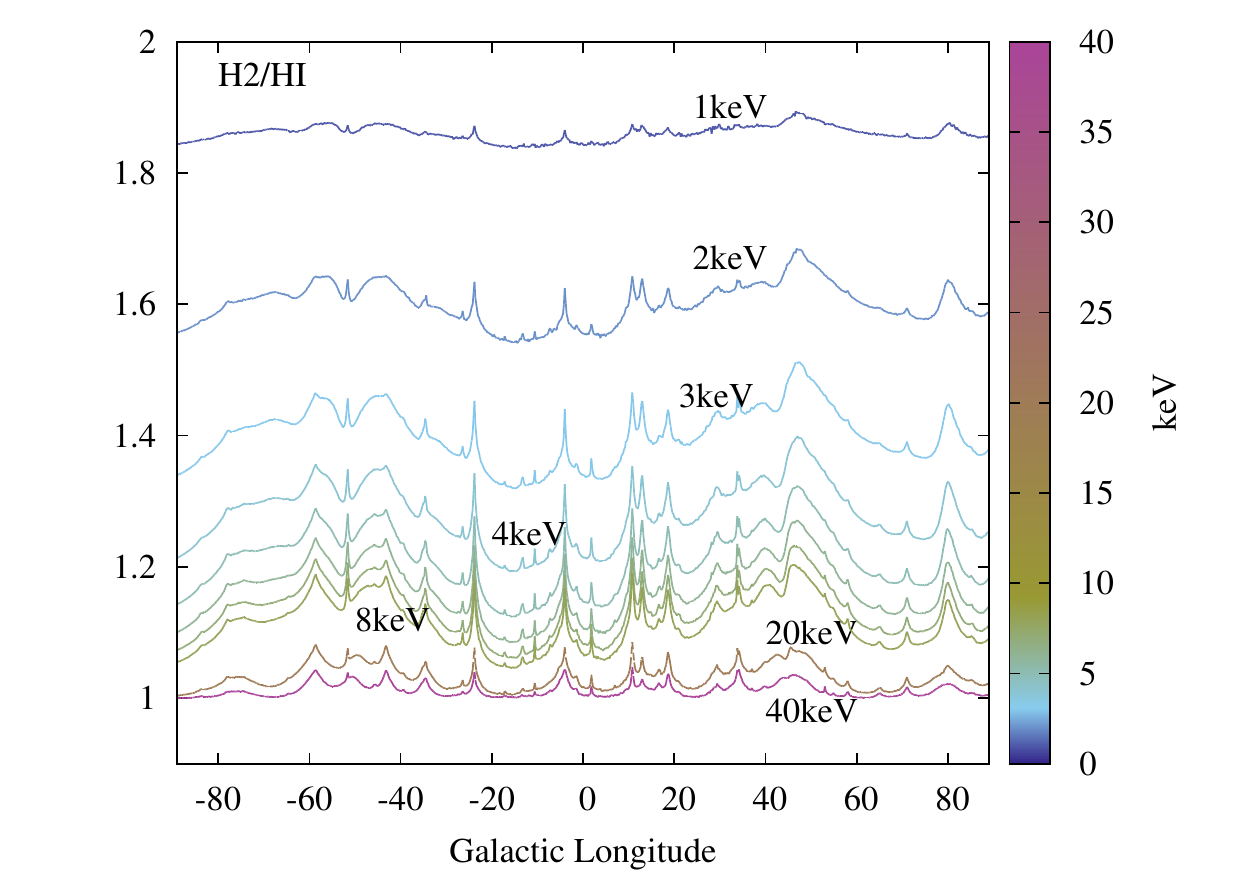}  \includegraphics[trim = 20 10 20 10,height = 8cm, width = 9 cm]{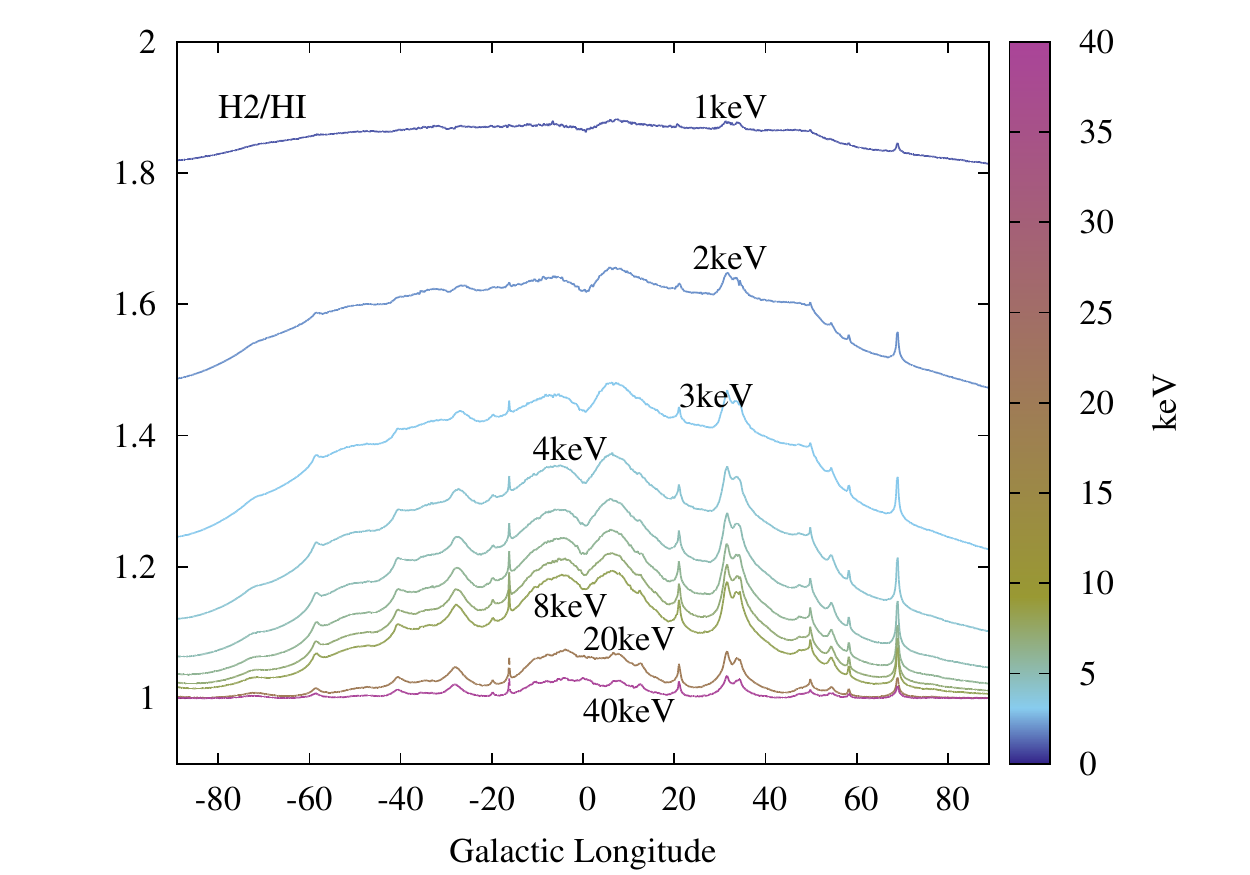}\\ 
\includegraphics[trim = 20 10 20 0,height = 8cm, width = 9 cm]{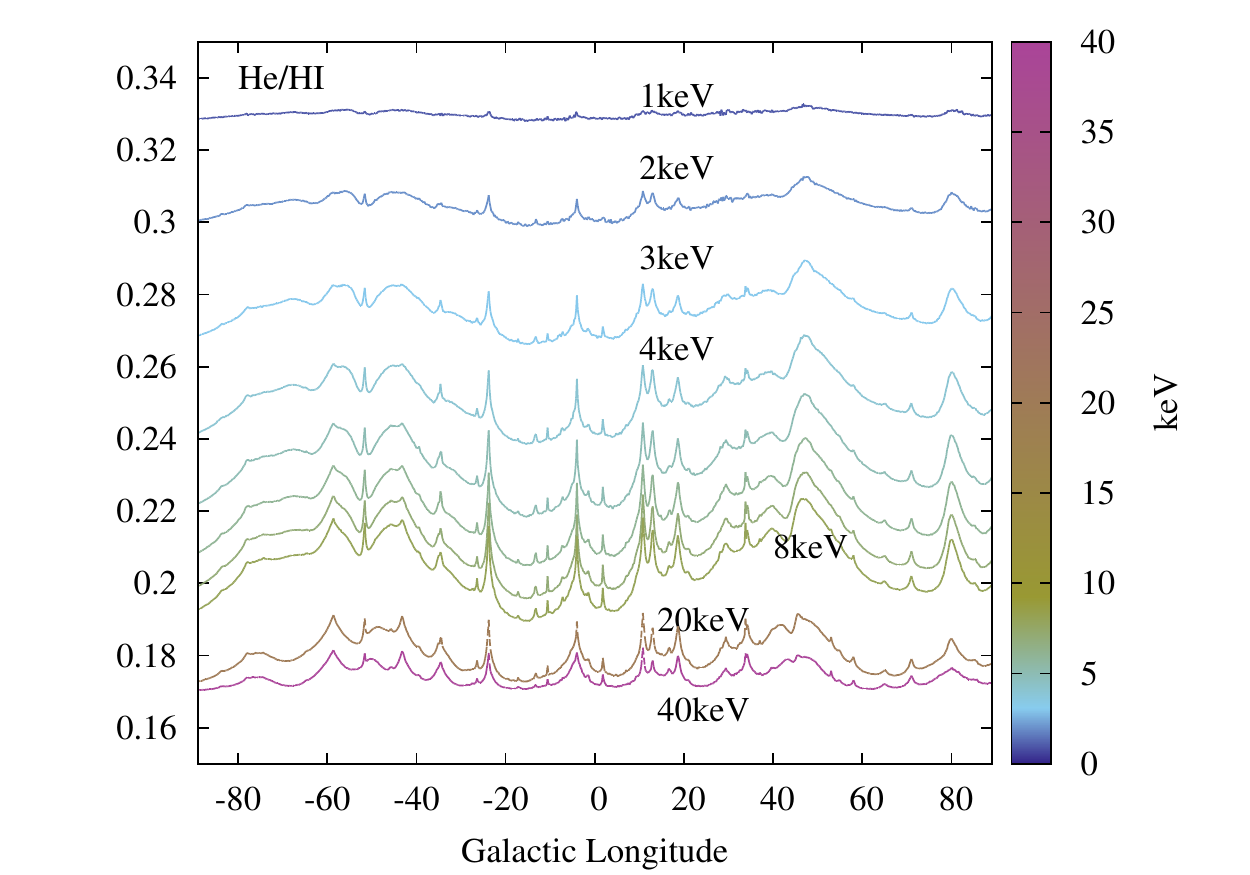} \includegraphics[trim = 20 10 20 0,height = 8cm, width = 9 cm]{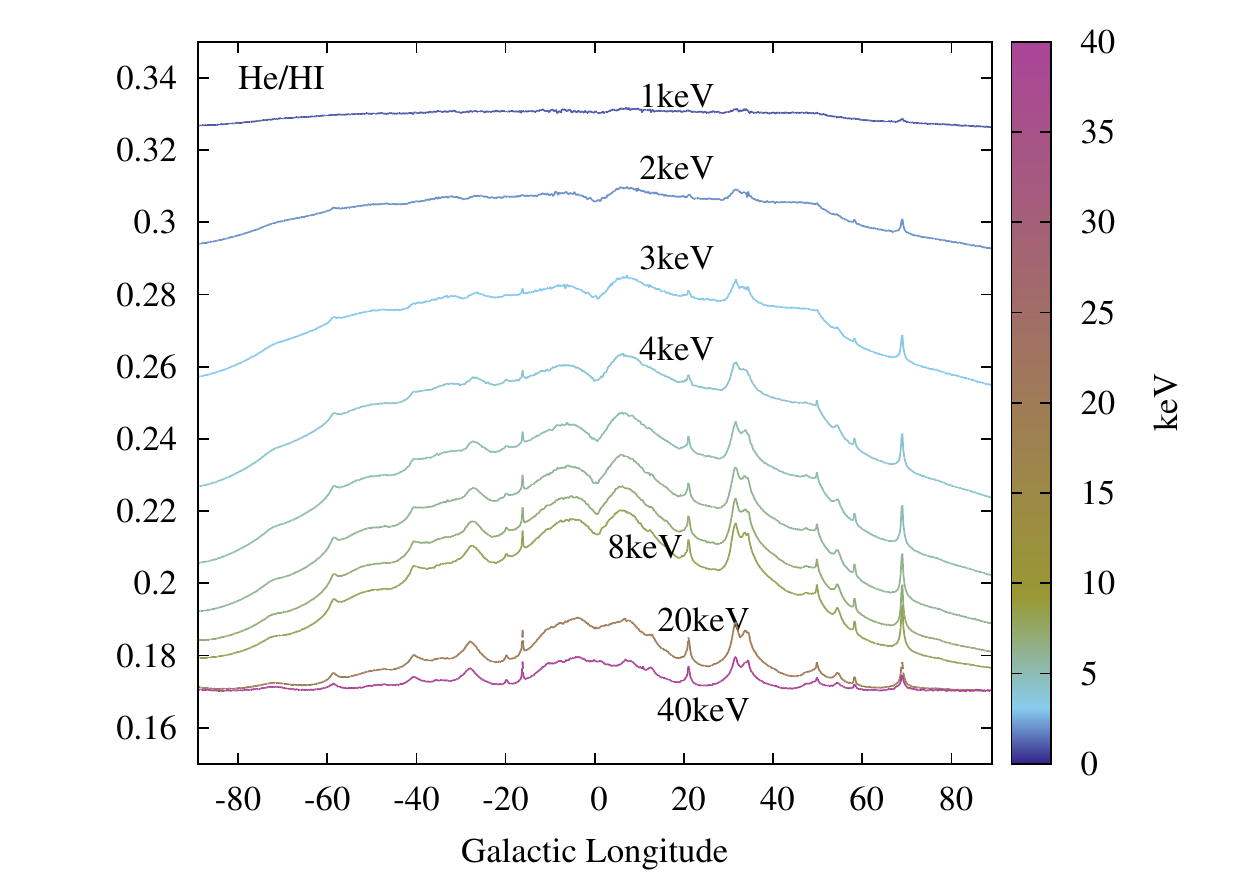} \\
\caption{Ratio of intensity from the \changeRR{Monte Carlo simulated HMXBs (\textit{left column}) and LMXBs (\textit{right column})} scattered along the Galactic plane ($b = 0$) by
  H2 (\textit{top row}) and He (\textit{bottom row}) to the intensity that
  would be scattered if all electrons were unbound. \changeRIV{The range of
    scattering angles over which the Rayleigh scattering dominates the
    scattering cross section depends on the characteristic size of
    \changeRV{the electron distribution in the atom or molecule}, which
    differs for different elements \changeRV{and molecules (see section
    \ref{sec:crosssec}). This leads to a \ct{nonlinear} dependence of the ratio
    of cross sections for different elements and molecules on the scattering
    angle.  At each longitude \ct{many different scattering angles contribute}, corresponding to the relative position of the
    X-ray sources w.r.t. the gas along the line of sight, resulting in the apparent longitudinal dependence of the ratio profiles.}}}
\label{Total_contr}
 \end{figure*}
 
\newpage
 
 \section{Additional references for distances and spectral parameters}\label{additional_ref}
 References for distances and spectral parameters of the sources used in
 the calculations not cited elsewhere in the paper are reported
 here. Please refer to our online \ct{catalog} for the full list of
 references associated to each source.
\citet{1982ApJ...262..253M}; 
\citet{1989ApJ...338.1024P}; 
\citet{1996ApJ...473L..25W}; 
\citet{1997ApJS..109..177C}; 
\citet{1997MNRAS.286..549L}; 
\citet{1997MNRAS.288...43R}; 
\citet{1997MNRAS.288..988S}; 
\citet{1998ApJ...495..435K}; 
\citet{1998MNRAS.297L...5S}; 
\citet{1999ApJ...525..215S}; 
\citet{1999ApJS..120..335M}.
\citet{1999MNRAS.306..417B}; 
\citet{1999MNRAS.307..695N}; 
\citet{2000A&A...354..567M}; 
\citet{2000ApJ...536..891N}; 
\citet{2000ApJ...543L.145D}; 
\citet{2000MNRAS.311..405K}; 
\citet{2001A&A...366..138O}; 
\citet{2001A&A...369..108N}; 
\citet{2001A&A...371.1018I}; 
\citet{2001A&A...380..490P}; 
\citet{2001ApJ...546..338P}; 
\citet{2001ApJ...561L.101W}; 
\citet{2001ApJS..133..187H}; 
\citet{2002A&A...389L..43I}; 
\citet{2002ApJ...569..903B}; 
\citet{2002ApJ...570..287W}; 
\citet{2003A&A...399..663K}; 
\citet{2003A&A...399..663K}; 
\citet{2003A&A...399..663K}; 
\citet{2003A&A...399..663K}; 
\citet{2003A&A...406..299P}; 
\citet{2003A&A...411L.487I}; 
\citet{2003MNRAS.342..909M}; 
\citet{2003PASJ...55L..73N}; 
\citet{2004A&A...416..311W}; 
\citet{2004A&A...418..655N}; 
\citet{2004ApJ...609..317H}; 
\citet{2004ApJ...616L.139W}; 
\citet{2004MNRAS.354..355J}; 
\citet{2005A&A...430..997L}; 
\citet{2005A&A...436..647F}; 
\citet{2005A&A...439..575I}; 
\citet{2005A&A...440..287I}; 
\citet{2005A&A...440..637R}; 
\citet{2005A&A...440.1079R}; 
\citet{2005A&A...444...15F}; 
\citet{2005ApJ...632..504C}; 
\citet{2005AstL...31...88T}; 
\citet{2006A&A...446L..17B}; 
\citet{2006A&A...449.1139M}; 
\citet{2006A&A...459...21M}; 
\citet{2006ATel..784....1T}; 
\citet{2006ATel..791....1S}; 
\citet{2006ATel..818....1K}; 
\citet{2006ApJ...638..982N}; 
\citet{2006ApJ...639.1033G}; 
\citet{2006ApJ...639L..31B}; 
\citet{2006MNRAS.365.1387C}; 
\citet{2007A&A...469L..27C}; 
\citet{2007ATel.1183....1L}; 
\citet{2007ATel.1207....1D}; 
\citet{2007ESASP.622..495D}; 
\citet{2008ATel.1443....1M}; 
\citet{2008ATel.1727....1M}; 
\citet{2008MNRAS.383..411P}; 
\citet{2009MNRAS.393..419S}; 
\citet{2010A&A...510A..61T}; 
\citet{2010A&A...510A..81C}; 
\citet{2010A&A...515L...1D}; 
\citet{2010MNRAS.404.1591D}; 
\citet{2010MNRAS.408..642Z}; 
\citet{2010MNRAS.408.1866R}; 
\citet{2011A&A...533A..23R}; 
\citet{2012ApJ...749..111L}; 
\citet{2012ApJ...757L..12R}; 
\citet{2013ApJ...764..185C}; 
\citet{2013ApJ...764..185C}; 
\citet{2013ApJ...768..183C};

\end{appendix}

\end{document}